\documentclass[fleqn]{2020SCGE}
\newcommand\msun{M_{\odot}}
\newcommand\arcsec{''}

\newcommand{\farcs}{$.\!\!^{\prime\prime}$}

\usepackage{hyperref}%PDF??????
\usepackage{multirow}
\usepackage{subfigure}
\usepackage{xspace}
\def\mosfit{{\tt MOSFiT}\xspace}
\begin{document}

\ensubject{subject}

%%%%%%%%%%%%%%%%%%%%%%%%%%%%%%%%%%%%%%%%%%%%%%%%%%%%%%%
%%% Authors do not modify the information below
%%% ????????????????
%%% ??????????, ????????????{}, ???????????????????
%Letter to the Editor??Article%??????
\ArticleType{Invited Review}%??Article
%\SpecialTopic{SPECIAL TOPIC: }%???????
%\Year{2023}
%\Month{January}
%\Vol{?}
%\No{?}
%\DOI{??}
%\ArtNo{000000}
%\ReceiveDate{?? ??, 2023}
%\AcceptDate{?? ??, 2023}
%\OnlineDate{January 1, 2016}
%%%%%%%%%%%%%%%%%%%%%%%%%%%%%%%%%%%%%%%%%%%%%%%%%%%%%%%

%%% title: ????
%%%   \title{title}{title for citation}
\title{Science with the 2.5-meter Wide Field Survey Telescope (WFST)} 
%%% Corresponding author: ???????
%%%   \author[number]{Full name}{{email@xxx.com}}
%%% General author: ???????
%%%   \author[number]{Full name}{}
\\

\author[1,2]{\\ Tinggui Wang}{{twang@ustc.edu.cn}}
\author[1,4]{Guilin Liu}{}
\author[1,4]{Zhenyi Cai}{}
\author[3]{Jinjun Geng}{} 
\author[3]{Min Fang}{} 
\author[3]{Haoning He}{}
\author[1,6]{\\ Ji-an Jiang}{}
\author[1,4]{Ning Jiang}{} 
\author[1,2]{Xu Kong}{{xkong@ustc.edu.cn}}
\author[3]{Bin Li}{} 
\author[3]{Ye Li}{}
\author[2,1]{Wentao Luo}{} 
\author[3]{Zhizheng Pan}{}
\author[3]{\\Xuefeng Wu}{}
\author[3]{Ji Yang}{}
\author[1,4]{Jiming Yu}{} 
\author[3]{Xianzhong Zheng}{xzz@pmo.ac.cn}
\author[1,2]{Qingfeng Zhu}{}
\author[1,4]{Yi-Fu Cai}{}
\author[3]{\\Yuanyuan Chen}{}
\author[3]{Zhiwei Chen}{}
\author[1,2]{Zigao Dai}{}
\author[1,2]{Lulu Fan}{}
\author[3]{Yizhong Fan}{}
\author[1,4]{Wenjuan Fang}{}
\author[1,4]{\\Zhicheng He}{}
\author[3]{Lei Hu}{}
\author[3]{Maokai Hu}{}
\author[3]{Zhiping Jin}{}
\author[3]{Zhibo Jiang}{}
\author[3]{Guoliang Li}{}
\author[3]{\\Fan Li}{}
\author[1,4]{Xuzhi Li}{}
\author[1,4]{Runduo Liang}{}
\author[1,4]{Zheyu Lin}{}
\author[3]{Qingzhong Liu}{}
\author[3]{Wenhao Liu}{}
\author[1,4]{\\Zhengyan Liu}{}
\author[3]{Wei Liu}{}
\author[3]{Yao Liu}{}
\author[3]{Zheng Lou}{}
\author[3]{Han Qu}{}
\author[2]{Zhenfeng Sheng}{}
\author[3]{Jianchun Shi}{}
\author[3]{\\Yiping Shu}{}
\author[1,4]{Zhenbo Su}{}
\author[3]{Tianrui Sun}{}
\author[3]{Hongchi Wang}{}
\author[1,4]{Huiyuan Wang}{}
\author[2,5]{Jian Wang}{} 
\author[1,4]{\\Junxian Wang}{}
\author[3]{Daming Wei}{}
\author[3]{Junjie Wei}{}
\author[1,4]{Yongquan Xue}{}
\author[3]{Jingzhi Yan}{}
\author[3]{Chao Yang}{}
\author[3]{Ye Yuan}{}
\author[1,4]{\\ Yefei Yuan}{}
\author[1,4]{Hongxin Zhang}{}
\author[3]{Miaomiao Zhang}{}
\author[3]{Haibin Zhao}{}
\author[1,4]{Wen Zhao}{}

%{{???@???.???}}%

%%% Author information for page head. ?š¹?§Ö????????
%%% ??????????????, ??????????author???
\AuthorMark{Wang}%\authorcr????????

%%% Authors for citation. ????????§Ö????????
%%% ??????????????, ??????????author???
\AuthorCitation{Wang et al}

%%% Address. ???
%%%   \address[number]{Address, City {\rm Postcode}, Country}
\address[1]{Department of Astronomy, University of Science and Technology of China, Hefei 230026, China}
\address[2]{Institute of Deep Space Sciences, Deep Space Exploration Laboratory, Hefei, 230026, China}
\address[3]{Purple Mountain Observatory, Chinese Academy of Sciences, Nanjing 210023, China}
\address[4]{CAS Key Laboratory for Research in Galaxies and Cosmology, Department of Astronomy, University of Science and Technology of China, Hefei 230026, China}
\address[5]{State Key Laboratory of Particle Detection and Electronics, University of Science and Technology of China, Hefei 230026, China}
\address[6]{National Astronomical Observatory of Japan, National Institutes of Natural Sciences, Tokyo 181-8588, Japan}

%\contributions{}%????????

\abstract{The Wide Field Survey Telescope (WFST) is a dedicated photometric surveying facility being built jointly by University of Science and Technology of China (USTC) and the Purple Mountain Observatory (PMO). It is equipped with a 2.5-meter diameter primary mirror, an active optics system, and a mosaic CCD camera with 0.73 gigapixels on the primary focal plane for high-quality image capture over a 6.5-square-degree field of view. The installation of WFST near the summit of Saishiteng Mountain in the Lenghu region is scheduled in summer 2023, and the operation is planned to start three months later. WFST will scan the northern sky in four optical bands ($u$, $g$, $r$ and $i$) at cadences from hourly/daily in the deep high-cadence survey (DHS) program, to semi-weekly in the wide field survey (WFS) program. During a photometric night, a nominal 30-second exposure in the WFS program will reach a depth of 22.27, 23.32, 22.84, and 22.31 (AB magnitudes) in these four bands, respectively, allowing for the detection of a tremendous amount of transients in the low-$z$ universe and a systematic investigation of the variability of Galactic and extragalactic objects. In the DHS program, intranight 90-second exposures as deep as 23 ($u$) and 24 mag ($g$), in combination with target of opportunity follow-ups, will provide a unique opportunity to explore energetic transients in demand for high sensitivities, including the electromagnetic counterparts of gravitational wave events, supernovae within a few hours of their explosions, tidal disruption events and fast, luminous optical transients even beyond redshift of unity. In addition, the final 6-year co-added images, anticipated to reach $g\simeq25.8$ mag in WFS or 1.5 mags deeper in DHS, will be of fundamental importance to general Galactic and extragalactic science. The highly uniform legacy surveys of WFST will serve as an indispensable complement to those of the Vera C. Rubin Observatory's Legacy Survey of Space and Time (LSST) that monitors the southern sky. } 

%%% Keywords. ?????


\maketitle

\tableofcontents%?????

%%%%%%%%%%%%%%%%%%%%%%%%%%%%%%%%%%%%%%%%%%%%%%%%%%%%%%%
%%% The main text. ???????
%???????????????????\cref{fig1}
%\twocolumn\onecolumn
%%%%%%%%%%%%%%%%%%%%%%%%%%%%%%%%%%%%%%%%%%%%%%%%%%%%%%%
\begin{multicols}{2}
\section{Introduction}\label{section_1}

Since the late 1950s, large surveys have played major roles in the development of virtually every domain of astronomy. 
The first large sky surveys in optical bands were conducted in 1950s through 1980s using the 1.2m Schmidt telescope of Palomar observatory on the northern hemisphere (Palomar Observatory Sky Surveys (POSS) I and II\cite{Reid1991}) and the UK Schmidt telescope at AAO and the ESO Schmidt telescope in Chile on the southern hemisphere. The Two-Micron All Sky Survey (2MASS), completed in 2001, employed three near-infrared bands and a pair of matched 1.3m diameter telescopes on both hemispheres (Arizona and Chile)\cite{Skrutskie2006}. These large-sky surveys have served as pools of significant discoveries in frontiers from the solar system to galaxies and quasars for dozens of years.

The Sloan Digital Sky Survey (SDSS)\cite{Gunn2006} is among the most ambitious and influential sky surveys in history. The dedicated 2.5-m aperture telescope employed by SDSS has mapped a quarter of the entire sky and has obtained spectra for millions of galaxies, quasars, and stars. In four phases of survey campaigns, SDSS has greatly advanced our understanding of the physics of galaxies, accreting supermassive black holes (quasars), the structure of the universe, and our own Galaxy. In addition to their initially designed science goals, the uniform and well-calibrated photometric and spectroscopic legacy data have engaged astronomers from virtually the entire astronomical community, leading to hundreds to thousands of scientific publications each year. Following the success of SDSS, imaging surveys in the southern sky were performed by the Dark Energy Survey (DES) Camera mounted on the 4-meter BLANCO telescope in Chile\cite{DES2005}. Compared to SDSS, DES detects 1.5 mag deeper over a sky area of 5000 square degrees, and 2.5 mag deeper over an area of 1000 square degrees with the ESO 4m survey telescope. The Large Sky Area Multi-Object Fiber Spectroscopic Telescope (LAMOST) has carried out the largest spectroscopic survey of stars in the Milky Galaxy\cite{2012RAA....12..723Z}. A high-sensitivity spectroscopic survey of galaxies and quasars in the northern sky is now ongoing, where the dark energy spectroscopic instrument (DESI) equipped on the Mayall 4m telescope is at work\cite{DESI2016}. 

Time domain surveys explore temporal changes of celestial objects, either intrinsically or extrinsically, by observing the sky repeatedly. These variations often contain crucial information to decipher the structure and nature of these variable sources. The blooming of time domain astronomy witnessed in the past decade has driven the technology development of wide-field survey facilities and the novel discoveries delivered by these facilities. The Catalina Real-Time Survey (CRTS)\cite{Drake2009ApJ...696..870D} searched for rare bright transients over a sky area of 33000 square degrees using 3 wide-field telescopes. The Palomar Transient Factory (PTF/iPTF)\cite{Law2009} and its successor, the Zwicky Transient Facility (ZTF)\cite{Bellm2019}, have monitored 3$\pi$ of the sky at a cadence of 3 days to a week, with complementary spectroscopy performed by follow-up telescopes. The Panoramic Survey Telescope and Rapid Response System (Pan-STARRS or PS)\cite{Kaiser2004}, the All-Sky Automated Survey for SuperNovae (ASAS-SN)\cite{Shappee2014}, and the Asteroid Terrestrial Impact Last Alert System (ATLAS)\cite{Tonry2018}, and Gaia, the global space astrometry mission, also conduct time-domain surveys and record transient sources. In general, time-domain surveys employ dedicated telescopes with apertures from a few tens of cm up to 1.3 m and large pixel sizes, with the exception of Pan-STARRS, which used two 1.8 m telescopes and gigapixel cameras. Currently, Pan-STARRS is largely dedicated to the search for near-earth asteroids (NEA). The limiting magnitudes in a single exposure for these surveys are in the range of 17.0 mag in the $V$ band for ASAS-SN to 21.8 mag in the $r$ band for Pan-STARRS.

At present, the demand for time domain surveys reaching fainter magnitude limits is growing due to the discovery of kilonovae, the electromagnetic emitter associated with the merger of neutron stars, and the increasing interest in high-redshift supernovae along with other transients with applications in cosmology and multi-messenger astronomy \cite{2017ApJ...848L..33A,2017ApJ...848L..17C}. The electromagnetic counterparts of gravitational wave sources detectable by the advanced LIGO/Virgo network in the upcoming five years will be typically 1-2 magnitudes fainter than the sensitivity limit of current major time domain surveys. These transient sources, moreover, are located in bright galaxies and therefore easily overwhelmed by starlight in the low-spatial-resolution images attained in current surveys. As for the southern sky, the Vera C. Rubin Observatory (VRO) with a flagship wide-field survey telescope of 8.4 meter aperture will be commission in the upcoming year, on which a 30 second exposure is expected to reach a single-epoch magnitude limit of 24.5 mag in $r$ band \cite{ivezic19}. However, no time-domain facilities are planned to be located on the northern hemisphere that is anticipated to reach a similar depth. The Wide Field Survey Telescope (WFST) is designed to fill this gap.

WFST has an aperture of 2.5 m and a field of view of about 3 $\deg$. It was designed to scan dynamic northern skies at depths 2 mag deeper than ZTF in $gri$, spatial resolutions of $\sim$1 arcsec, and daily time intervals. The telescope is expected to be installed on the Saishiteng Mountain  near Lenghu in summer 2023. High altitude and low water vapor result in a relatively high $u$ band efficiency, a distinct advantage among time-domain survey facilities targeting the northern sky. Regarding the site location, WFST and VRO are complementary in both longitude (158 degrees apart) and latitude (on the northern/southern hemisphere).

In this article, we describe the expected performance of WFST and its observation strategy in Section 2. Relevant time domain science, including supernovae, tide disturbance events (TDEs), multimessager astrophysics, and active galactic nuclei (AGNs), is covered in Sections 3.1 to 3.4, while topics on the Milky Way and asteroids are discussed in Sections 4 and 5, respectively. Section 6 presents the prospects for galaxy formation and cosmology, and Section 7 provides a summary of this paper.

\section{Expected Performance and Survey Strategy}\label{section_2}

WFST is a 2.5 m optical telescope with primary focus optics designed for a wide 3$^{\circ}$ field of view (FoV). The optical system consists of a primary mirror, five corrector lenses, an atmospheric dispersion compensator (ADC), and the filters of six optical bands ($u$, $g$, $r$, $i$, $z$ and $w$). Active optics (AO) is equipped to keep the telescope in a seeing-limited condition and to reduce primary-focus assembly (PFA) misalignment and primary mirror deformation. The scientific imaging array consisting of nine 9K$\times$9K CCDs (E2V CCD290-99) with a pixel size of 10$\mu$m will be installed in the primary focus plane, resulting in an effective FoV of about 6 square degrees.
The telescope will be located on the top of the Saishiteng Mountain near Lenghu (93$^{\circ}53^{\prime}$~E, 38$^{\circ}36^{\prime}$~N) at an altitude of 4200~m. The overall performance of the system will be presented in Section \ref{section_21}.

Two major sky surveys are planned in the $ugri$ bands. They will be described in Section \ref{section_22}. Targets of opportunity (TOO) follow-ups of high energy transients in multi-wavebands including the $z$-filter, and exploration of asteroids in the solar system with a sensitive $w$ band are considered. They will be described in subsections of Section 3. Additional small-scale programs can be scheduled for specific scientific objectives.

\subsection{Expected Performance of the System}\label{section_21}

The observing conditions of the site have been monitored for three years since 2018 \cite{2021Natur.596..353D}, giving a median value of seeing of 0\farcs75, an average night sky background brightness around 22.0 mag~arcsec${^{-2}}$ in V-band when the moon is below the horizon. The nightly observable time ranges from 5 hours in June to over 11 hours in January in each year. The clear time fraction is about 70\%, and the observation conditions in a significant number (337) of nights reach photometric requirements in the year 2021\footnote{http://lenghu.china-vo.org/sitecondition}. 
Taking into account a number of downgrade factors beyond the optical system's designed imaging performance, dome and atmospheric seeing, we expect overall image quality to be about $\sim 1$\arcsec~ assuming a median seeing of 0\farcs75. The averaged throughput is estimated to be 0.39, 0.72, 0.60, 0.56 and 0.33 for the $u$, $g$, $r$, $i$ and $z$ bands, respectively. 

	We estimate the limiting magnitudes of WFST based on the specification of the system design along with the relevant available data. We took a value of 22.0~mag~arcsec$^{-2}$ as the V-band sky background level, and adopted a model spectrum obtained from the SkyCalc code (version 2.0.9) developed by ESO astronomers. An airmass of 1.2 is assumed and aperture photometry is applied to estimate the limiting magnitudes of the system for point sources (Mag$_{\rm{30s}}$ or Mag$_{\rm{60s}}$) required to render a signal-to-noise ratio (SNR) of 5 for a 30 or 90 second exposure. We also computed the limiting magnitudes (Mag$_{\rm{50m}}$) of the images stacked from 100 30-second exposures with a total integration of 50 minutes. These results are reported in Table \ref{limiting_magnitude} \cite{2023arXiv230103068L}. 

\begin{table}[H]
\begin{center}
\begin{tabular}{c|c|c|c|c|c|c}
\hline
\textbf{Filter} & \em\textbf{u}& \em\textbf{g}& \em\textbf{r}& \em\textbf{i}& \em\textbf{z}& \em\textbf{w} \\
\hline
\textbf{Mag}$_{\textbf{\,sky}}$ & 22.51 & 22.33 & 21.39 & 20.65 & 19.71 & 21.42 \\
\hline
\textbf{Mag}$_{\textbf{\,30s}}$ & 22.27 & 23.32 & 22.84 & 22.31 & 21.38 & 23.47 \\
\hline
\textbf{Mag}$_{\textbf{\,90s}}$ & 23.17 & 24.04 & 23.51 & 22.96 & 22.04 & 24.10 \\
\hline
\textbf{Mag}$_{\textbf{\,50m}}$ & 24.82 & 25.85 & 25.36 & 24.83 & 23.90 & 25.99 \\
\hline
\end{tabular}
\end{center}
\caption{Site Sky Brightness and Limiting Magnitudes for 30s, 90s exposures and stacked 50 min exposures assuming airmass=1.2.}
\label{limiting_magnitude}
\end{table}

\subsection{Survey Strategy}\label{section_22}

In this subsection, we describe the two key programs planned for the WFST 6-year survey: the Wide-field Survey (WFS) program and the Deep High-cadence $u$-band Survey (DHS) program. The different designed survey modes, in terms of survey depth, area, and cadence, are commensurate with the primary scientific objectives of WFST. As part of the WFST 6-year survey, each program will occupy about 45\% of the total observing time. The remaining $\sim$ 10\% of the observing time (about 1,300 hours over 6 years) will be attributed to smaller campaigns for specific purposes, such as capturing time-critical targets and intensively scanning certain sky areas of particular interests (e.g. the Galactic plane).

The WFS program will cover an area of $\sim$ 8,000 deg$^2$ in the northern sky. It will employ four broad bands ($u, g, r, i$) with a single exposure of 30 seconds, leading to about 90 visits per pointing in each band over 6 years, if a clear night fraction of 70\% is assumed at the Lenghu site \cite{2021Natur.596..353D}. As for the purpose of long-term monitoring of specific targets (e.g. active galactic nuclei and variables), single-band visits will be evenly distributed in 6 years, i.e. 60 multiband visits (15 visits $\times$ 4 bands) per pointing per year, yielding yearly raw data of about 100 TB from the entire WFS fields. Observations for about 300 different pointings ($\sim$2,000 deg$^2$) with 60 visits per pointing will be executed throughout WFS during three months, leading to about 1,200 pointings in total every year. All of the $u$-band observations are scheduled in dark and gray nights, in view of the highly sky background-sensitive measurements planned in this band. To balance the efficiency and science goals of the survey and to optimize the homogeneity of the WFS visits, we will avoid consecutive observation in a single band, but we will observe in two bands every night, with the sole exception of the $u$ band. This strategy will result in a reasonable cadence and time span in characterizing multiband light curves for general purposes of time-domain research (e.g. transient classifications, variability statistics, and time-domain cosmology). Meanwhile, total integration in each band will reach $\sim 45$ min over 6 years, achieving deeper detection than any of the existing single-telescope surveys with comparable survey areas on the northern hemisphere.

In addition to WFS, we plan for the Deep High-cadence $u$-band Survey (DHS) program by virtue of the superior $u$-band imaging performance of WFST in time-domain investigations. DHS will routinely monitor a sky area of $2~\times \sim360$ deg$^2$ surrounding the equator every year (the ``Spring" and ``Autumn" fields; 6 months of observing per each). Considering the importance of $u$-band imaging and color information in revealing the nature of various energetic transient phenomena, for each 6-month campaign of DHS, we perform photometry in at least one more band besides $u$ in hour cadence in consecutive $\pm$ 7 days during every lunar cycle (starting from the new moon). Additionally, WFST will keep monitoring the same region in the $griz$ bands on the remaining nights of these 6 months. Such an innovative survey mode provides a unique opportunity to track transients right after their occurrences and to discover rare energetic explosive phenomena in the universe (e.g. early-phase supernovae, fast blue optical/ultraluminous transients, tidal disruption events, kilonovae, etc.). More details are deferred to \S \ref{section_31}--\S \ref{section_33}). WFST will also be combined with the next-generation Chinese space missions, e.g., EP \cite{2018SPIE10699E..25Y}; the Chinese Space Station Telescope (CSST) \cite{CSST21}) to be launched in the upcoming years, so that unprecedented synchronization of multiwavelength surveys between ground-based and space-borne wide-field survey facilities becomes feasible. By coordinating with EP and CSST, we will not only promptly identify optical counterparts of various high-energy astronomical events, but also attain real-time spectral energy distributions of various fast transients, by virtue of the anticipated synchronization and synergy.

Wide-field imaging is a mainstream tool employed in numerous fields of cutting-edge astronomy, whose success has been witnessed in numerous accomplished and ongoing wide-field survey projects. The prominent survey capability and high $u$ band sensitivity of WFST bring new opportunities to deep and wide exploration of the transient sky on the northern hemisphere, especially at blue optical wavelengths. The resultant large amount merger, accretion onto a newly formed compact object in a failed supernova, mergers of binary white dwarfs, and tidal disruption of stars by intermediate-mass or massive black holes (``IMBH TDE''; see \S \ref{section_323} for details). Recent studies find that FBUTs are usually accompanied by prominent emission in X-ray and radio wavelengths, indicating a compact object in the center of FBUTs. In view of the very low event rate in the local universe and the high UV luminosity of FBUTs, WFST DHS is expected to be the most promising survey project to accomplish a systematic investigation of this extreme transient phenomenon in the 2020s.

\subsubsection{Extreme Supernovae}\label{section_314}

The optical luminosity of superluminous supernovae (SLSNe) peaks at $\lesssim$ -21 mag \cite{gal-yam12}. Most SLSNe are 10 to 100 times brighter than typical CCSNe. The low event rate of SLSNe results in their first discovery as recent as 1999. After that, several SLSNe were occasionally found in the 2000s. In the last decade, over 100 SLSNe have been observed by unbiased transient surveys equipped with large-array CCD cameras. In the observational respect, the SLSNe population can be naturally divided into hydrogen-poor (SLSNe-I) and hydrogen-rich events (SLSNe-II). Most SLSNe-II emit narrow lines (SLSN-IIn), a feature similar to that of less luminous SNe IIn \cite{quimby07}. Therefore, they are interpreted as extreme cases of SNe IIn mainly driven by the interaction between the ejecta and dense CSM. SLSNe-I are less well understood, for which the dominating mechanism underlying their explosions is still under debate \cite{gal-yam19}.

A major open question about SLSNe is the energy source that powers these extremely luminous and long-lived events. Is a central engine necessarily required? If so, what kind of engine(s) (e.g., a magnetar, an accreting black hole, or both) are at work? The treatment of these questions requires samples consistently enlarged in high-cadence deep imaging surveys and intensive follow-up observations. The 6-year WFST WFS project will regularly monitor the northern sky at a cadence of a few days, so that SLSNe at $z\lesssim1$ will be detected with high completeness by virtue of their long-lasting and luminous light curves. SLSNe at high redshift is a potential focus of attention in the 2020s, not only of importance to the time-domain astronomy but also to tracking the star formation history in the high$z$ universe. Moreover, they may become useful distance estimators for cosmological measurements in the future. Taking advantage of the high UV luminosity of SLSNe and their higher event rate at higher redshift ($z \lesssim 2$), WFST, which has the superior sensitivity of the $u$ band and the properly designed telescope aperture, will be the most powerful telescope for searching SLSNe at $z > 0.5$ on the northern hemisphere. 

An extremely luminous type of SNe in theoretical prediction, known as pair-instability SNe (PISNe), remains elusive. PISNe are inferred to be the explosion of massive stars with zero-age main sequence (ZAMS) masses of about 130--260 M$_{\odot}$. The high temperature in the stellar cores of these massive stars causes a large amount of electron-positron pairs that, in turn, result in contraction of the core, followed by explosive oxygen burning that eventually unbinds these ultra-massive stars \cite{barkat67}. For stars with slightly lower ZAMS masses of 90--130 M$_{\odot}$, the progenitor may experience multiple non-destructive pair instability episodes that expel materials prior to the final core collapse. These pulses can lead to shell collisions that produce an SN-like transient. The succession of shell ejection may alternatively be followed by a PISN, of which the ejecta collide with the preceding ejected shells. This repetitive shell-collision system, with or without a final PISN, is called a pulsational pair-instability SN (PPISN \cite{woosley07}). PISNe and PPISNe are both extremely luminous SNe, but merely a few candidates have been reported due to the difficulty of yielding their massive progenitors in the low-$z$ universe. However, the large FoV and deep imaging capability of WFST will boost the sample of candidates for PISN / PPISN in the process of the planned 6-year WFST transient survey.

\begin{figure*}[!ht]
\centering
\includegraphics[width=0.75\linewidth]{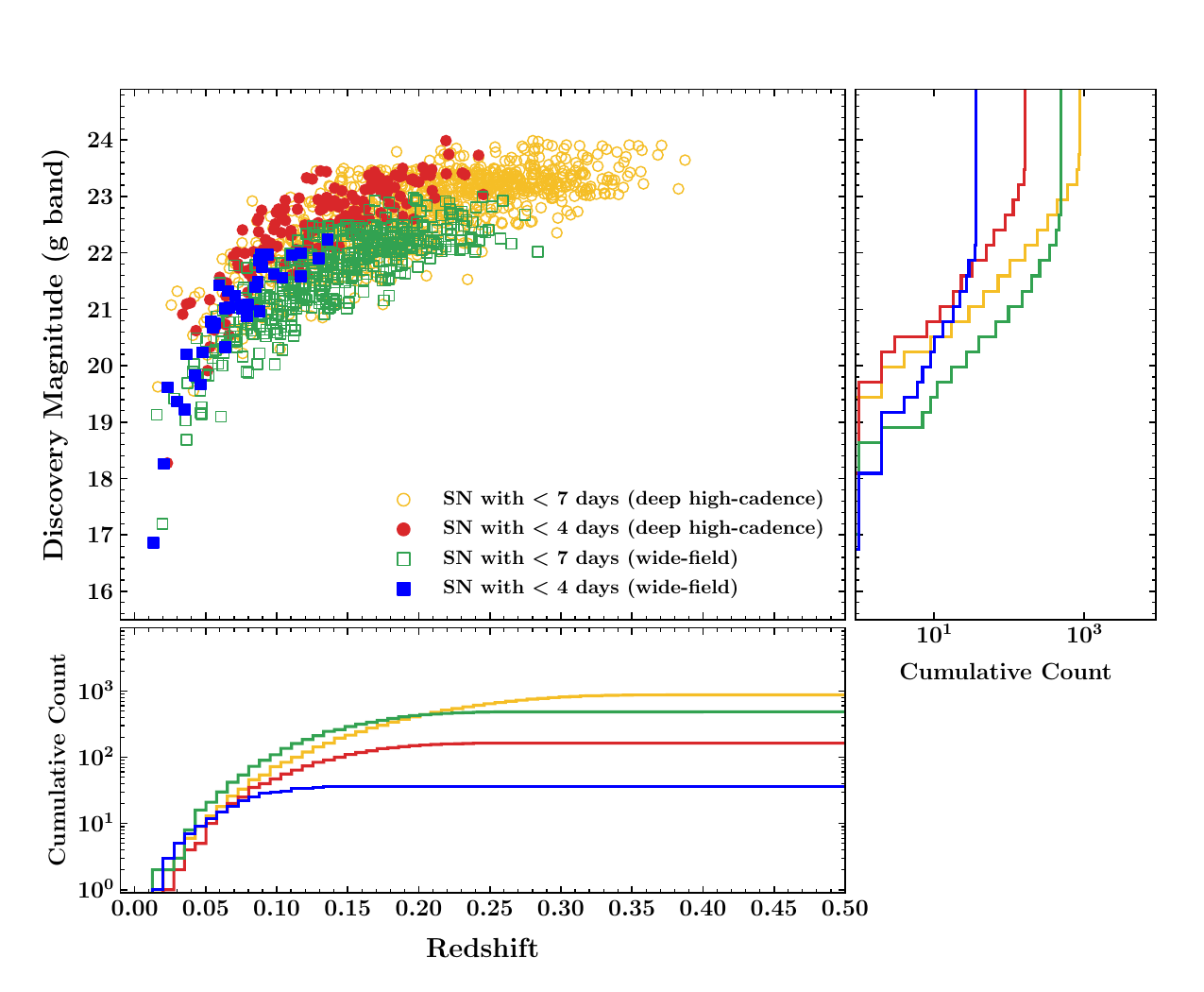}
\caption{Expected distribution of yearly SNe Ia on the discovery magnitude vs. redshift plane in WFST deep high-cadence (circles; 360 deg$^2$ daily) and wide-field (squares; 2000 deg$^2$ in 4 days) surveys.  SNe are divided into two groups according to the time of the second detection $t$: Open and solid symbols denote the SNe discovered with $t < 7$ days and $t < 4$ days, respectively. The right and bottom panels show cumulative counts in terms of discovery magnitude and redshift, respectively.} 
\label{fig:fig_sndiscovery}
\end{figure*}

\subsubsection{Cosmology and Gravitational Lensing}\label{section_315}

Two decades have passed since its discovery; the nature of dark energy remains a mystery. The recent measurement of $H_{0}$ from the local SN Ia distance ladder, calibrated to Cepheid variables, is in tension with the inference from the early universe using the cosmic microwave background (known as the ``Hubble tension"). At present, three large projects designed for measuring cosmological parameters with SNe Ia are underway or planned: the Hyper Suprime-Cam Subaru Strategic Program (HSC SSP \cite{miyazaki18}), the Vera C. Rubin Observatory Legacy Survey of Space and Time (VRO/LSST \cite{ivezic19}), and the Nancy Grace Roman Space Telescope. Nevertheless, none of these projects will construct an optimal SNe Ia sample at redshift below 0.3 to measure the cosmological parameters. Roman will find a very limited number of low-$z$ SNe, while the nominal-cadence survey strategies of HSC SSP and VRO/LSST will leave significant multiple/single-filter gaps in their low-$z$ SN light curves, which will downgrade the accuracy of SN Ia standardization. Hence, a large and unbiased WFST SN Ia sample to be observed with relatively high-cadence (i.e. $\lesssim3$ days) will further reduce the uncertainty in the measurement of dark energy density in the $0.1 < z < 0.3$ redshift bin, allowing for a precise comparison with the well-constrained measurements in the $z < 0.1$ bin. WFST SN Ia sample also promises to refine and extend SN Ia standardization models and to improve the constraints on the relationship between SN Ia distance measurements and the properties of their host galaxies.

Among the wide range of cosmological probes in the literature, SNe II are regarded as a promising independent method for deriving accurate distances and measuring cosmological parameters. Despite that SNe II display a large range of peak luminosities, several standardization methods have been developed, such as the expanding photosphere method, the standard candle method (SCM), the photospheric magnitude method, and the photometric color method (PCM). SCM is currently the most accurate and commonly used method to derive SN II distances, allowing one to construct a Hubble diagram with a $\sim$ 10\% dispersion in distance, suggesting that SNe II are potentially complementary and independent tools to constrain the nature of dark energy. Previous SN II Hubble diagrams based on SCM mainly focus on the low-$z$ universe ($z < 0.2$), where distinguishing different cosmic expansion histories is challenging. Therefore, measurements extending further back in time using SNe II at a higher redshift will be instrumental to distinguish cosmological models. With the expectation that thousands of WFST SNe II at $z>0.1$ will be found in the coming years, we will readily perform a direct comparison with the SN Ia measurements at $0.1 < z < 0.3$ and reveal the relevant implications.

Recent discoveries of strongly-lensed SNe have opened up a new frontier in the domains of cosmology and early-phase SNe. Strongly-lensed SNe are events where multiple light rays from a SN converge due to the gravity of an intervening object (e.g. a galaxy or a galaxy group or cluster), which results in multiple lensed SN images. A notable feature of such a system is the relative time delays among the lensed SN images due to the difference in light paths. The validation of time delays in strongly-lensed SN systems as an independent probe for the Hubble constant $H_0$ has been well recognized \cite{refsdal64}. Nevertheless, before the discovery of the first multiply-imaged SN in 2015 \cite{kelly15}, this ``time-delay cosmography'' technique had only been applied on strongly-lensed quasars, for which time delays are also measurable (\S \ref{section_345}). Up-to-date work achieved a 2.4\% precision measurement of $H_0$ from the combination of six strongly lensed quasars \cite{wong20}, demonstrating that the technique is a competitive and complementary approach. In comparison, the measurement of time delay is easier for lensed SNe, due to their characteristic light curves. Meanwhile, since SNe will eventually disappear, precise lens models are attainable by analyzing these systems without contamination from transients themselves. Eventually, strongly-lensed SNe are expected to provide more stringent constraints on $H_0$ than quasars \cite{ding21}. Strong-lensing time delays also offer a unique opportunity to probe SNe shortly after their explosions, in that once a lensed SN is found, follow-up observations can be scheduled well in advance to readily track the entire process of explosion.

To date, only five strongly charged SNe have been discovered. A deep wide-field imaging survey with WFST will substantially increase the sample size of strongly lensed SNe. According to Oguri et al.\cite{oguri10}, we expect to find over 20 strongly lensed SNe in the 6-year WFST WFS program. With dozens of WFST strongly lensed SNe in the 2020s in hand, we expect to embrace engaging opportunities in the frontiers of cosmology and early-phase SN study.

\subsubsection{Supernova Search with WFST}
\label{section_316}

The three key parameters of a transient survey are its depth, area, and cadence. The time-domain-related scientific output of the WFST surveys is optimized by properly coordinating these parameters. The weakness of most previous or ongoing transient surveys lies in the limited survey depth when small-aperture ($<$ 1.5m) telescopes are employed, or the low survey cadence in the case of large-aperture telescopes, hindering systematic investigations of the photometric behaviors of early-phase SNe and fast transients with faint brightness and fast-evolving light curves in minutes to a few days. Because of the specially designed large FoV and aperture of WFST, these objects of interest are expected to be efficiently discovered via WFST high-cadence deep-imaging surveys.

Here we present simulations of the one-year WFST survey in 3-day and 1-day cadences, corresponding to the wide-field and deep high-cadence surveys, respectively (Figure~\ref{fig:fig_sndiscovery}). Since we plan to obtain color information in each observable night, we simply assume in our simulations that the telescope monitors the same sky area in at least two bands (e.g. $u$ and $g$) every night. The clear night fraction, the influence of the moon phase and the visibility of the target have been taken into account \cite{2022Univ....9....7H}. To roughly demonstrate the SN detection efficiency of WFST, we focus on normal SNe Ia with well-established light curves and spectral templates. These SNe are stochastically generated at different redshifts based on the event rate derived from local SN Ia samples. SN Ia light curves are constructed through synthetic photometry using the Hsiao spectral template \cite{hsiao07}. In regard to the dispersion in the intrinsic luminosity of SNe Ia, we assume a uniform distribution of absolute magnitude at maximum light spanning a range of -18.5 to -19.5 mag. Finally, random foreground extinction from the Milky Way and the host galaxy is configured for each SN. 

In this simulation, an SN candidate that is detected at least twice on different nights is defined as a ``real'' SN detection. Figure~\ref{fig:fig_sndiscovery} shows the distribution of SNe Ia on the discovery magnitude versus the redshift plane based on two survey modes. Note that the time $t$ in the figure is defined as that of the second detection of an SN. As our main targets, SNe with $t < 4$ days (early-phase SNe; solid symbols) will be intensively observed by other observing facilities within the next few months to depict detailed multiband light curves and spectral evolution. The SNe Ia with $t < 7$ days (open symbols), mainly consisting of those for which a good coverage of multicolor light curves starting from $\sim$10--14 days before the peak is expected, will facilitate statistical investigations of the light-curve behaviors of SNe and the SN cosmology over a wide range of redshift. In the simulated observation of WFST for one year, we expect to discover more than 1000 SNe Ia at z $\lesssim$ 0.25 in $t < 7$ days, and particularly $\sim100$ early-phase SNe Ia at z $\lesssim$ 0.15 via WFST DHS. The number of early-phase SNe Ia is about three times larger than that discovered from WFST WFS, indicating the significance of a deep high-cadence survey for searching early-phase SNe (and other fast transients alike).

%%%%%%%%%%%%%%%%%%%%%%%%%%%%%%%%%%%%%%%%%%%%%%%%%%%
%%%%%%%%%%%%%%%%%%%%%%% TDE %%%%%%%%%%%%%%%%%%%%%%%
%%%%%%%%%%%%%%%%%%%%%%%%%%%%%%%%%%%%%%%%%%%%%%%%%%%

\subsection{Tidal Disruption Events}\label{section_32}

\subsubsection{Observational Status and Open Questions}\label{section_321}

A breakthrough in transient research during the past decade has been the detection of a rapidly growing number of tide disturbance events (TDE). A TDE occurs when a star occasionally wanders into the tidal sphere of a supermassive black hole (SMBH) residing in the center of a galaxy. The star will be tidally disrupted and partially accreted, producing a flash of electromagnetic radiation on timescales of months to years \cite{Rees1988}. The event rate is lower than that of a supernova by a factor of a few hundred, i.e. $10^{-4}-10^{-5}~\rm gal^{-1}~yr^{-1}$, placing TDEs in a class of rare transients.

Already theoretically predicted in the 1970s, TDEs were not identified until late in the 1990s from the archival ROSAT data as well as a few more subsequent events identified by XMM-Newton and Chandra, guided by the anticipation of a radiation peak in soft X-ray or extreme UV bands. These TDEs, however, were all found serendipitously from archival data, and synergetic information in other wavelength regimes is scarce. 
Thanks to a variety of wide-field optical surveys dedicated to time-domain surveys, an explosively growing number of TDEs have been found in the past decade (see recent review of \cite{Gezari2021}). In particular, the ZTF survey has boosted the rate of TDE discovery from $\lesssim$2 / year to $>$10 / year, opening up a new era of sample statistics \cite{vanVelzen2021}. At present, optical TDEs are being discovered in real time, and timely multi-wavelength follow-up observations therefore become feasible.

TDEs arouse broad interest in the community as a result of their distinctive scientific values. First, TDEs provide direct evidence for the existence of a SMBH in a quiescent galaxy beyond the current accessible regime that is based on stellar or gas dynamics, which is particularly useful in dwarf and distant galaxies. Even dormant intermediate-mass BHs (IMBHs) and SMBH binaries can be probed via TDEs. Moreover, TDEs serve as an ideal laboratory to scrutinize the accretion physics of SMBHs and tackle unsettled problems in AGNs by monitoring the entire life cycle of BH activity, or even by witnessing the formation of jets. The evolution of gas and the infrared and radio echoes of TDEs provide a novel tool to probe the sub-parsec environment of these distant quiescent SMBHs \cite{Jiang2021} inaccessible to other techniques. In the multi-messenger era, TDE is deemed an important astrophysical process as the origin of high-energy neutrinos\cite{Stein2021} (see details in \S \ref{section_334}).

As significant as the scientific value and advancement, many open questions about TDEs have yet to be answered. For instance, the TDEs found as yet exhibit an unexpected preference for post-starburst (or so-called ``E+A'') galaxies \cite{French2016}, which cannot be addressed by known selection effects. In addition, the observed total energy is one to two orders of magnitude lower than the theoretical prediction, leading to the puzzle of ``missing energy''. Also, the highly debatable origin of the bright optical-UV emission awaits more observational constraints. An associated issue is the connection between optically-selected and X-ray-selected TDEs, and the feasibility of constructing a simple model to unify them remains unclear. From an observational perspective, the mounting number of nuclear transients, both in normal and active galaxies, has raised a fundamental question: How to classify these transients (e.g. TDEs, turned-on AGNs, sporadic gas accretion, etc.) into different types of SMBH transient accretion events \cite{Zabludoff2021}? WFST, in synergy with other multi-wavelength/messenger time-domain facilities in the upcoming decade, offers an unprecedented opportunity to tackle these (and many other) challenging questions.  

\subsubsection{Demography of Dormant SMBHs Revealed by Large TDE Samples}\label{section_322}

As a direct probe of SMBHs, TDEs shed light on the distribution of mass (and even spin) of dormant SMBHs, which constitute the majority of SMBHs in the low-redshift universe. However, the sample size of known TDEs is insufficient as yet ($\lesssim$100 up to now \cite{Gezari2021}) to achieve a meaningful demography, an enlarged sample with improved completeness is indispensable.

The success of ZTF proves that high-cadence and multi-band observations during the same night provide critical color and evolution information that are remarkably beneficial to the TDE search. 
The observational feasibility is ensured by the fact that the TDEs exhibit an evidently bluer and more steady color, distinguishing themselves from the contaminating supernovae and the usual variable AGN \cite{vanVelzen2021}. 
WFST has the potential to surpass ZTF by taking advantage of improved depth, availability of the $u$-band, higher photometric accuracy, and high spatial resolution of imaging. In particular, as the optical band closest to the peak wavelength of the TDE SEDs, the employed $u$-band distinguishes WFST from the other facilities that will dominate the discovery and characterization of TDEs on the northern hemisphere. 

\begin{figure*}[htbp]
\centering
\subfigure[]{
	\includegraphics[scale=0.65]{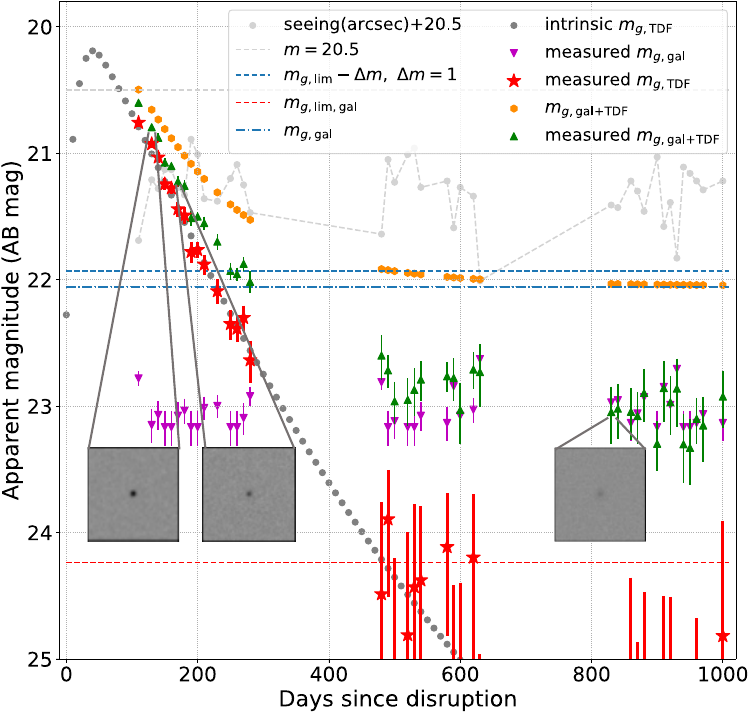}
}
\subfigure[]{
	\includegraphics[scale=0.25]{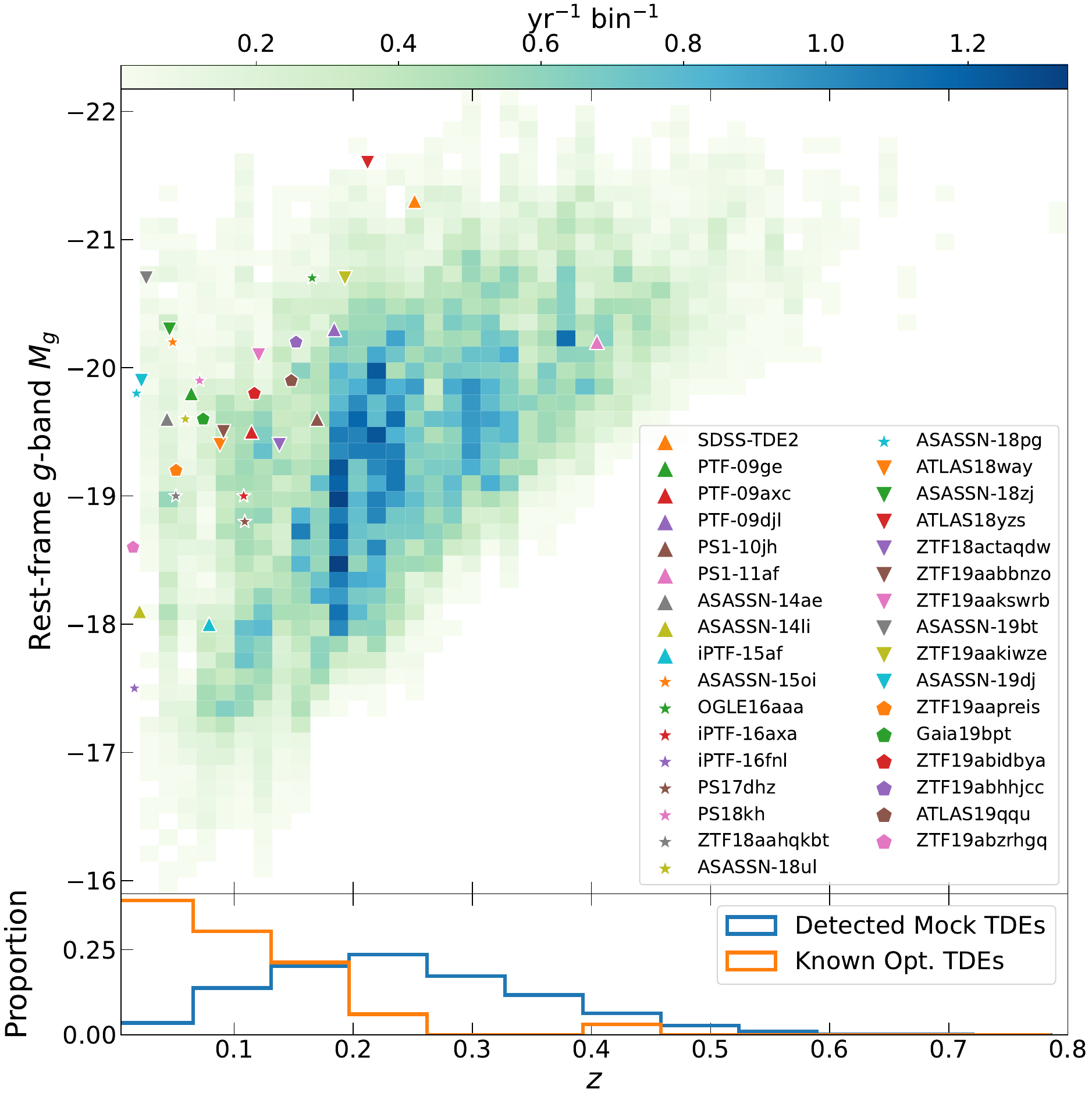}
}
\caption{Adapted from \cite{Lin2022}. (a): The $g$-band light curves and mock science images (insets) of a TDE at $z=0.253$ as an example of our mock observations.
(b) top: The absolute peak magnitude of the $g$ band ($M_g$) as a function of the redshift for the TDEs detected in our mock observations. 
The 33 optical TDEs summarized in Table~1 of \cite{vanVelzen2020} have also been overplotted for comparison.
(b) bottom: the histograms of the TDE redshift in the mock and known sample.} 
\label{tderesult}
\end{figure*}

In order to assess the TDE discovery capability of WFST, we performed mock observations taking site conditions, telescope parameters, and survey strategy into account. We start from billions of galaxies in the 440 deg$^2$ CosmoDC2 field \cite{2019ApJS..245...26K}, assign a TDE event rate to each galaxy as per its SMBH mass, and generate TDE light curves using the empirical model {\tt MOSFiT}. We assume a uniform survey strategy, in which the experimental 440 deg$^2$ field is scanned with 30-second exposures every 10 days in $u$, $g$, $r$, $i$ and $z$ bands, respectively. Also considered are the ``observation windows'' ($\sim180$ days per year) and the proportion of clear nights assumed to be 70\% (a clear night is defined as more than 4 hours of contiguous fully clear time \cite{2021Natur.596..353D}). In addition to a seeing distribution consistent with on-site monitoring, we adopt a sky background of 22.0 mag/arcsec$^2$ and readout noise of $10\ e^-$/pixel. 

In our definition, the discovery of a TDE satisfies the following minimum requirements (an example $g$-band light curve is given in Figure~\ref{tderesult}): 
1) the host galaxy is detectable in one band in the reference image; 
2) the excess in the galactic nucleus is significant in 10 epochs and 2 bands. After performing 1000 mock observations, we find that the combination of the $g$ and $r$ bands is the most effective. If we choose a more conservative strategy using the combination of the $u$, $g$, $r$, and $i$ bands so that comprehensive SED information is attainable (the$u$ band is particularly useful), then $29\pm6$ TDEs are detected in the CosmoDC2 field, equivalent to a detection rate of $532\pm100$ per year for the 8000~deg$^2$ WFS and $48\pm9$ per year for 720 deg$^2$ DHS\cite{Lin2022}. The real discovery power of DHS is even stronger given its depth and cadence advantages, i.e., boosted by a factor of $\gtrsim3$, while requiring serious challenge spectroscopic follow-ups for these faint sources.

In addition to enlarging the sample size of TDEs, WFST will substantially extend the redshift range to $z\sim0.8$, as suggested by our mock observations. After a planned 6-year survey, we expect to obtain a uniformly selected sample of thousands of TDEs. Combined with the host galaxy properties learned from WFST stacked images and CSST high-resolution images, this sample will allow for probing the occupations of SMBHs among different types of galaxies and constraining their mass functions in the local universe, a vital step towards deciphering the formation and growth history of SMBHs.

\subsubsection{Hunting for IMBHs through TDEs}\label{section_323}

SMBHs are believed to be the result of the growth of seeds that are significantly less massive. It is widely accepted that IMBHs lie in the mass range of $\sim10^2-10^5~\msun$, and were formed shortly after the formation of the first generation of galaxies. 
Investigations of IMBHs will undoubtedly advance our understanding of the BH family in the universe as a whole by bridging the gap between SMBHs in galactic nuclei and BHs of stellar masses in binaries. However, the paucity of unambiguously identified IMBHs and the poor understanding of their formation mechanism pose a major challenge \cite{Greene2020}. 

Until now, reported IMBH candidates have been exclusively noticed by their AGN features, yet their inactive counterparts have been largely overlooked. The stellar TDEs produced by IMBHs may provide a unique opportunity for uncovering the dormant IMBHs, which are tentatively invoked to explain the X-ray outburst in an off-centered massive globular cluster or an ultra-compact galaxy resulting from a minor merger \cite{Lin2018}. Besides normal (main-sequence) stars, white dwarfs (WDs) can be tidally disrupted by IMBHs, producing distinctive features. It has been proposed that thermonuclear explosions of WDs induced by the strong tidal compression of IMBHs will manifest themselves as optical transients similar to SNe Ia \cite{MacLeod2016}. Consequently, some WD TDEs have possibly been misclassified as normal SNe Ia in the past. Distinguishing between them solely through optical emission is challenging, but they are probably featured by distinctive signatures in other bands (e.g. X-ray emission from the accretion process in the WD TDE scheme).

As introduced in \S\ref{section_313}, the understanding of the physical mechanism underlying ultraluminous FBOT (peak bolometric luminosity $\gtrsim10^{44}$~erg s$^{-1}$), represented by AT~2018cow, remains controversial. IMBH TDEs have been suggested as a possible scenario, though an unusually long-lasting emission of highly super-Eddington accretion is required \cite{perley19}. The solution to the FBOT problem may involve a two-fold strategy: spotting them early and starting prompt observations in other wavelength regimes (e.g., X-ray, radio) and performing statistical analysis based on a large sample. However, to date, the number of AT~2018cow-like FBOTs remains in the single digits, so expanding the FBOT sample is of fundamental importance. If the IMBH-TDE scenario is correct, then the ultra-luminous FBOTs are likely the most efficient and direct probe of off-centered IMBHs. The defining blue ($g-r<-0.2$ at peak) and fast-evolving characteristics of FBOTs make them ideal targets for DHS in $u$-band (see details in~\S\ref{section_22}). In a deep survey of 720-$\rm deg^2$ field, we expect tens to hundreds of FBOTs per year (aware of the large uncertainty in the event rate), making WFST one of the most competitive facilities for the discovery of FBOTs.

TDEs with rapid ascending time ($t_{\rm rise}$) between FBOTs ($\sim3$ days) and usual TDEs ($\sim$ a month) are also potential ideal candidates for IMBHs, because $t_{\rm rise}$ is theoretically expected to correlate with the mass of the BH. The very recent discovery of a nuclear transient with a rising time of 13 days, AT~2020neh, can be plausibly explained by a main-sequence star tidally disrupted by an IMBH \cite{Angus2022} and is an exact demonstration of this strategy. WFST deep field is capable of unveiling more fast-rising optical TDE candidates like AT~2020neh, endowing us an opportunity to explore dormant IMBHs in the centers of dwarf galaxies.

\subsubsection{Other Opportunities}\label{section_324}

The rising phase of the BH light curve has not yet been explored sufficiently, but it provides valuable clues to BH and disrupted star mass and even to BH spin. Hitherto, ASASSN-19bt, which luckily falls in the TESS field, remains the sole TDE with consecutive sampling on a daily basis, allowing the light curve to be depicted before its peak \cite{Holoien2019}. In the WFST and LSST eras, the challenge of TDE research is distinguishing TDEs from other transients and coordinating limited follow-up observing resources for events with prominent scientific values as promptly as possible. Regular surveys at a cadence of days to weeks are not optimal for the discovery of TDEs at their early rising stage, while the advantages of the planned deep high-cadence field of WFST are distinct. Our estimation shows that the emission and color of about 10 TDEs will be measurable to WFST as early as (rest frame) 30 days before their peaks.

The overlap of the WFST timeline with that of the Einstein Probe \cite{2018SPIE10699E..25Y} is particularly interesting for the TDE study, because optical and X-ray campaigns have been playing a dominant role in the discovery of TDE. It remains enigmatic whether TDEs bright in optical and X-ray emission belong to distinctive populations or can be described in a unified picture, where the difference is due to orientation effects, dynamic evolution, or other effects. The weakness in combining the two wavelength regimes in previous TDE works is due to the shortage of dedicated time-domain surveys undertaken simultaneously in both bands. Optical TDEs unveiled in real time have been monitored in X-rays only for a short period since their discovery, yet revealed a complex relationship between X-ray and optical. The joint analysis of WFST and EP data promise to offer an unprecedented TDE sample with high-cadence light curves (or upper limits) and a solid foundation for deriving luminosity functions in the optical and X-ray bands. 

In addition to classical TDEs involving a star that plunges into the tidal radius, partial tidal disruption at a position barely beyond the tidal radius is also possible, in which case only the stellar envelope is stripped and ripped apart, leaving a compact naked core, which may be completely disrupted later\cite{2020ApJ...904..100R}. The event rate of partial TDEs is naturally expected to be higher than that of normal ones, but their lower luminosity poses a challenge to observations. Dozens of partial TDEs are probably detectable by the ZTF survey every year \cite{Chen2021}, but have been overlooked. The power of WFST to detect weak optical emission allows us to anticipate the discovery of a significant number of partial TDEs, but the success of this strategy likely hinges on distinguishing them from other massive nuclear transients. The partial TDE scenario is also a proposed explanation for the intriguing periodic optical flares found in galactic nuclei \cite{Payne2021}, and a potential source for low -frequency gravitational waves. 

The IR echoes of TDEs have been proven to be effective in tracing the (sub)parsec environment of SMBHs in normal galaxies, which are otherwise extremely difficult to probe\cite{Jiang2021}. The statistics of environmental differences between quiescent and active galaxies is instrumental to revealing the triggering and fueling mechanism of AGN. However, the construction of a panorama is hindered by the strong preference of the known TDE hosts for post-starburst galaxies and thus by the absence of star-forming and passive galaxies. WFST will help construct a TDE sample with enhanced completeness 
by detecting a remarkable amount of optically-weak TDEs,
and the analysis of dust and gas echoes based on a virtually unbiased sample will become realistic. Once completed, the upshot will be a major step towards an in-depth understanding of the pc-scale environment of SMBHs in various types of galaxies, which will ultimately facilitate the construction of a panoramic picture of the SMBH activity.

\subsection{Multi-messenger Events}\label{section_33}

Stellar transients result from a variety of processes in stellar evolution, including the explosive death (e.g. SNe and Gamma-Ray Bursts or GRBs), the violent behaviors of the compact remnants of the explosion (e.g. pulsars and possibly Fast Radio Bursts or FRBs), as well as processes related to the merger of binaries (e.g. Gravitational Wave Events or GWEs). Among these transients, SNe and GRBs are possibly neutrino-related events. In this section, we discuss the observation plans of stellar transients with WFST.

%%%%%%%%%%%%%%%%%%%%%%%%%%%%%%%%%%%%%%%%%%%%%%%%%%%%%%%%
%%%%%%%%%%%%%%%%%%%%%%% Kilonova %%%%%%%%%%%%%%%%%%%%%%%
%%%%%%%%%%%%%%%%%%%%%%%%%%%%%%%%%%%%%%%%%%%%%%%%%%%%%%%%

\subsubsection{Gravitational Wave Events}\label{section_331}
The observations of GW170817 \cite{2017PhRvL.119p1101A}, GRB 170817A \cite{2017ApJ...848L..14G, 2017ApJ...848L..15S} and AT2017gfo \cite{2017ApJ...848L..33A,  2017ApJ...848L..17C} have opened up a new era of multi-messenger GW astronomy. Electromagnetic (EM) counterparts of GWE are of fundamental importance to extreme relativistic physics and redshift measurement of standard sirens. In this subsection, we discuss the prospects of WFST in the search for optical counterparts of GWE.

\paragraph{Kilonovae}\label{section_3311}
During the coalescence of binary neutron star (BNS) and some neutron star-black hole (NSBH) binaries, neutron-rich ejecta are released through shocks at the contact interface, tidal interactions, and disk outflows. Rapid neutron capture ($r$-process) nucleosynthesis renders heavy elements to form and decay in these ejecta \cite{1982ApL....22..143S}, powering a rapidly evolving and roughly isotropic thermal transient ``kilonova'' \cite{2010MNRAS.406.2650M}. \par

The observations of AT2017gfo along with GW170817/GRB 170817A have confirmed that BNS mergers produce kilonovae. 
Detection of kilonovae will help to locate the source, thus determining the redshift of GW events, to explain the origin of heavy elements in universe, to probe the nature of ejecta and merger remnants, and to constrain the NS equation of state (EoS). Thus, we simulate 10,000 BNS mergers spread over the redshift range of 0 to 0.2 to characterize the WFST detection capability of kilonova. \par

A binary neutron star merger, if the merger remnant is a strongly magnetized millisecond pulsar (or millisecond magnetar), is believed to result in a kilonova along with an afterglow brighter than those from the decay of radioactive heavy elements and the interaction of a relativistic jet with its ambient medium \cite{2018ApJ...861..114Y, 2018ApJ...856L..33G, 2021ApJ...918...52L}. Observations of such transients have posed new constraints on the EoS for dense neutron star matter, showing that the EoS therein is probably highly stiff. In parallel, the inconsistency between the Hubble constant determined from SNe Ia and that from the Cosmis Microwave Background (CMB), or the so-called ``Hubble constant tension'', is currently a focus of cosmological research. The electromagnetic signals together with the gravitational waves from a binary neutron star merger promise to help resolve this problem by providing an independent and unique probe of the Hubble constant \cite{2020Sci...370.1450D}. \par 

During their dynamical time, BNS mergers eject neutron-rich matter through shocks at the contact interface and tidal interactions in the equatorial planes. In general, the tidal ejecta have a sufficiently low electron fraction $Y_e\lesssim0.25$ along with the production of heavy nuclei. These ejecta are lanthanide-rich, with a high opacity and known as ``red'' components. Polar ejecta have a larger electron fraction ($Y_e\gtrsim0.25$) due to the effects of $e^\pm$ captures and neutrino irradiation. These ejecta are known as the ``blue'' components due to the lack of heavy nuclei synthesis and the bluer colors. After the BNS merger, an accretion disk is formed around the central remnant NS or BH, whereas the disk loses a fraction of its mass because of the neutrino-heated winds and spiral density waves. In this case, the electron fraction and opacity of these ejecta lie between those of the ``red'' and ``blue'' components, which are therefore known as the ``purple'' components.\par

For NS (double NS or BH-NS) mergers, the binary chirp mass is among the measured parameters best determined from GW signals, while the type and mass ratio of the two companions are poorly constrained. As the ejecta properties of the kilonova are sensitive to the type of merger and the mass ratio, they are useful for diagnosing the progenitor. The construction of more relevant samples will help us fill the gap between NSs and BHs \cite{2019ApJ...887L..35B}. In a double NS merger, a possible remnant includes a stable NS, a supermassive NS supported by solid-body rotation, or a hypermassive NS supported by differential rotation, or a collapsing system that promptly evolves into a BH, depending on the EoS and total mass of the double NS system \cite{2017ApJ...850L..19M}. 

Using the mass distribution of Galactic double NSs and EoS from the constraints of GW170817/AT2017gfo, we calculate the mass and velocities of the three components following \cite{2017CQGra..34j5014D, 2018ApJ...869..130R, 2014MNRAS.441.3444M}. We also derive the kilonova light curves from these samples employing the Modular Open Source Fitter for Transients (\mosfit), and calculate their GW signals and the expected signal-to-noise ratio (SNR) if they are detected by the second generation (2G) GW detector network. Hereafter, we denote the network of advanced LIGO-Livingston/Handford and advanced Virgo as LHV, and the network of LHV, LIGO-India and KAGRA as LHVIK. \par

\begin{figure*}[htbp]
\centering
\subfigure[u]{
	\includegraphics[width=8cm]{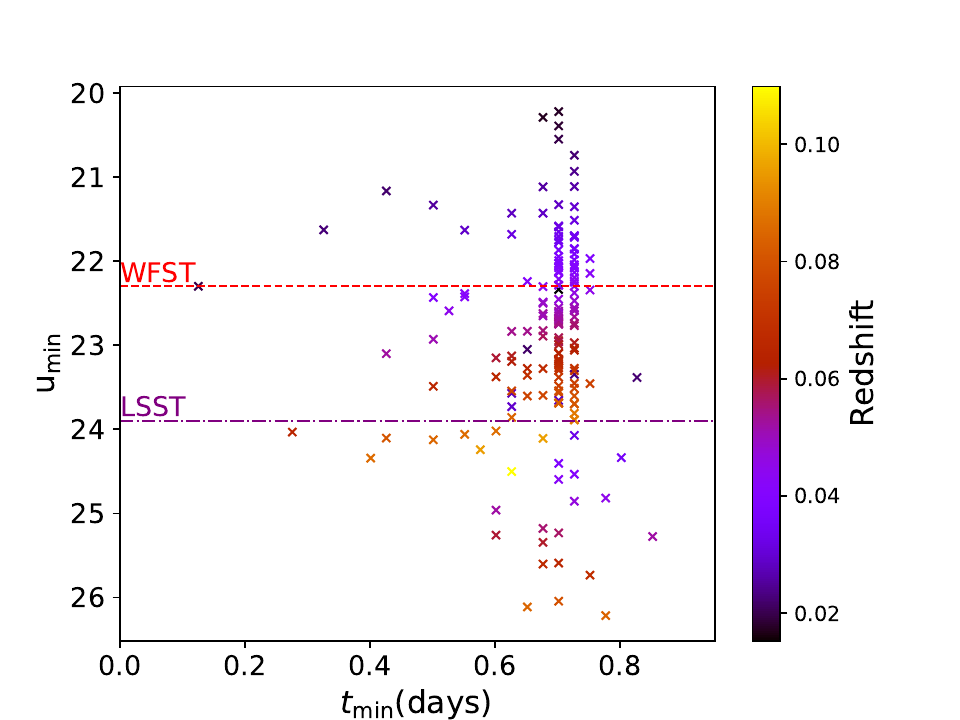}
}
%\hspace{0.5pt}
\subfigure[g]{
	\includegraphics[width=8cm]{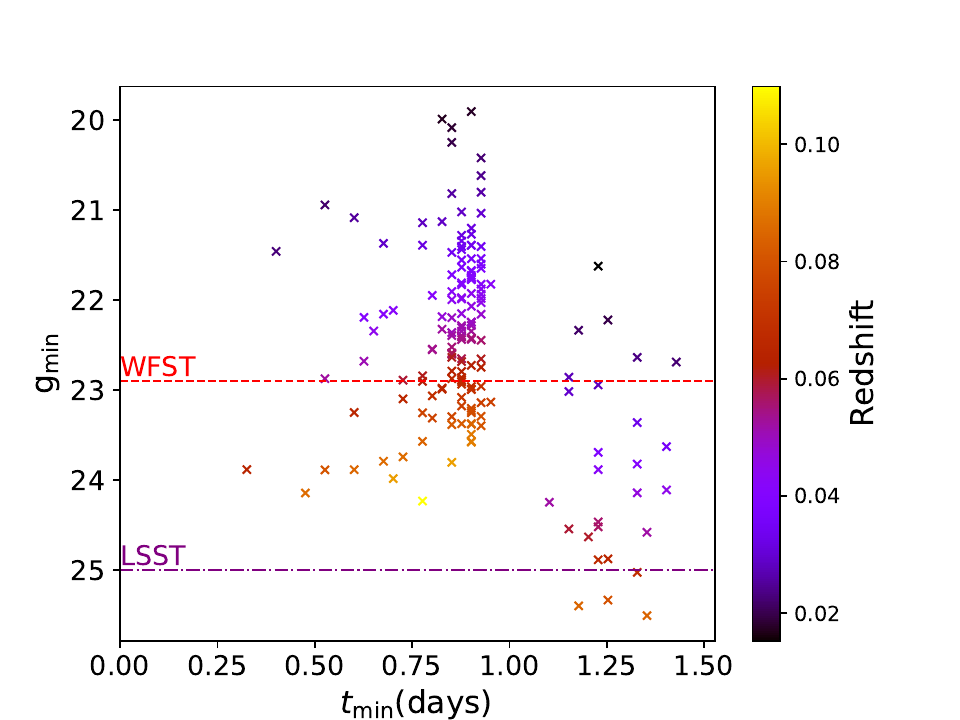}
}
%\hspace{0.5pt}
\subfigure[r]{
	\includegraphics[width=8cm]{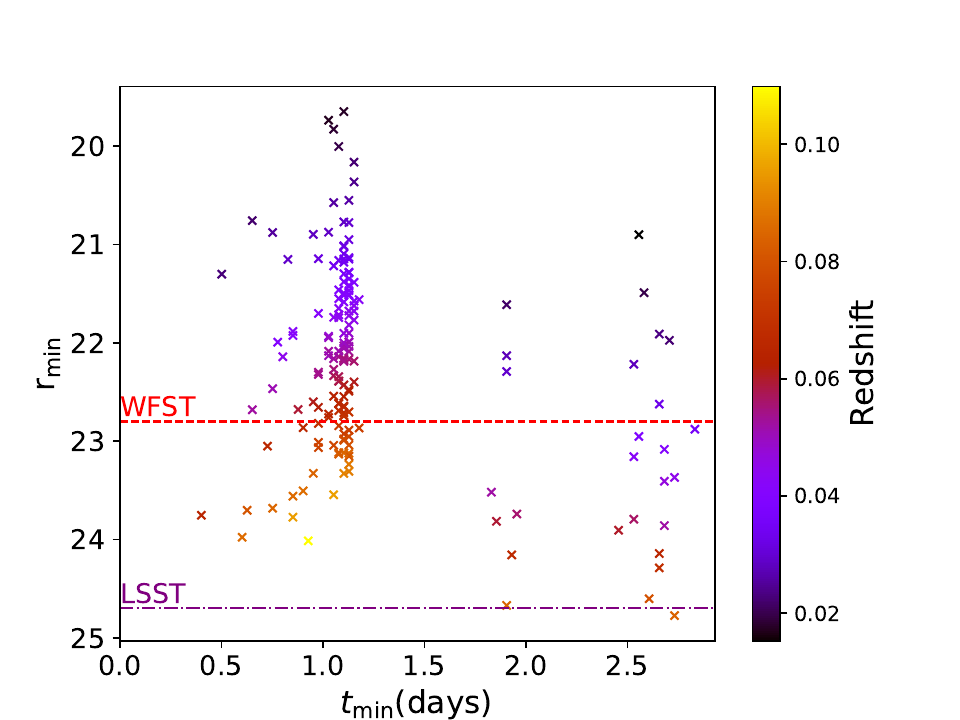}
}
\hspace{0.5pt}
\subfigure[i]{
	\includegraphics[width=8cm]{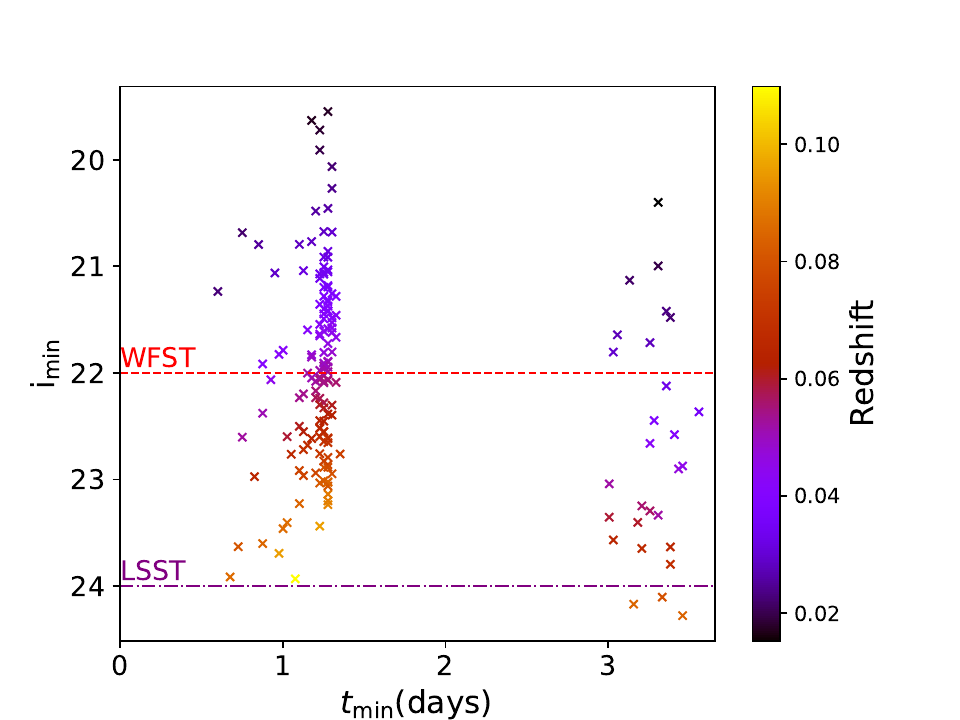}
}
%\hspace{0.5pt}
%\subfigure[z]{
%	\includegraphics[width=8cm]{figs_GWE/LHV_mag_z.pdf}
%}
\caption{The distributions of  magnitudes of kilonovae at their maximum brightness and the corresponding times for BNS mergers with GW SNR$>10$. Colors depict redshifts of sources. Four panels (a-d) represent the results of $u$-, $g$-, $r$-, $i$-bands, respectively.}
\label{figure_kilonova}
%\vspace{0.8cm}
\end{figure*}

In Fig. \ref{figure_kilonova}, we show the magnitude of the kilonovae at their peak luminosity and the corresponding time for BNS mergers detectable by LHV with $\mathrm{SNR}>10$. The two dashed lines in each panel depict the single-visit depth of a 30s exposure for WFST and LSST. The redshift limit of LHV is $\sim 0.12$, while WFST can observe kilonovae at a maximum redshift of $\sim0.06$ in the $r$ band. As shown in the $i$ band panel, the time at which peak luminosity is reached is concentrated around 1--3 days, a consequence of the fact that the fraction of ``red''/``blue'' components is strongly influenced by the mass ratio. For BNS with unequal masses, the less massive NS is tidally disrupted before contact, and shock production and the ``blue'' component are suppressed. The ``red'' component has a larger opacity and it takes the photons therein more time to diffuse, so the kilonova dominated by the ``red'' component reaches the maximum luminosity at a later time. Hence, $i$-band observations allow for a deeper understanding of the color evolution of kilonovae and the nature of ejecta. In Fig. \ref{figure_kilonova}, we further note that the luminosity of the $u$-band reaches its maximum within a few hours. Current AT2017gfo observing campaigns lack $u$-band imaging, and a quick WFST search in the $u$-band facilitates investigations of the kilonova evolution within the first few hours.\par

Assuming a local BNS merger rate of $80-810\ \mathrm{Gpc}^{-3}\ \mathrm{yr}^{-1}$ \cite{2021ApJ...913L...7A}, a follow-up area of $\sim50\%$ of the whole sky, and that a fraction of $\sim70\%$ are observable nights, we report the amount of BNS mergers per year with observable GW and kilonova signals in Table \ref{table_kilonova}. For WFST and LHV, the rate of multi-messenger detections is $\sim1-13$ per year in the $g$- and $r$-bands, and is slightly lower in the $u$ and $i$, but the $z$-band is likely unusable in a kilonova search due to the relatively low sensitivity. We plan to focus on campaigns for $u$- and $g$-bands ($u$ in particular) in the first few hours of our kilonova search and then switch gears to $r$- and $i$-bands, especially when we optimize the search efficiency for red kilonovae \cite{Liu2023}.

%**************
\begin{table*}
\begin{center}
\begin{tabular}{|c|c|c|c|c|c|c|}
\hline
\multicolumn{2}{|c|}{}&$u$&$g$&$r$&$i$&$z$\\
\hline
\multirow{2}{*}{WFST}
&LHV&0.6-5.8&1.1-11.6&1.3-12.9&0.8-8.5&0.2-2.4\\
\cline{2-7}
&LHVIK&0.8-7.9&1.8-18.2&2.0-25.2&1.1-11.8&0.3-2.9\\
\hline
\multicolumn{7}{|c|}{}\\
\hline
\multirow{2}{*}{LSST}
&LHV&1.5-15.3&1.8-17.9&1.8-18.0&1.8-18.0&1.6-16.8\\ 
\cline{2-7}
&LHVIK&2.9-29.0&3.8-37.9&3.8-39.0&3.8-39.0&3.2-32.7\\
\hline
\end{tabular}
\end{center}
\caption{Number of BNS mergers per year with observable GW signals and kilonova.}

\label{table_kilonova}
\end{table*}

\paragraph{Gamma-ray bursts and afterglows}\label{section_3312}
For high-redshift events, the expected WFST detection of EM counterparts is the short-GRBs (sGRBs) and their afterglows. However, GRB emission is beamed, i.e. the $\gamma$-ray radiation is emitted in a narrow cone more or less perpendicular to the plane of the inspiral. Hence, only a small fraction of BNS mergers are expected to have produced observable GRBs and afterglows. In a previous work \cite{2021ApJ...916...54Y}, we calculated the detection rate of BNS mergers observable by GW detectors, X-ray and $\gamma$-ray facilities (EP; GECAM; Swift-BAT; SVOM-ECLAIRS; Fermi-GBM), and optical telescopes (WFST, LSST) hunting for their afterglows. We simulated $10^7$ BNS mergers in the redshift range of 0--0.3 and assumed a Gaussian-shaped jet profile for all of them \cite{2002ApJ...571..876Z}, which is supported by that of GW170817/GRB 170817A. \par

\begin{figure*}[htbp]
\centering
\subfigure[LHV]{
	\label{LHV_flux7_scatter}
	\includegraphics[width=8cm]{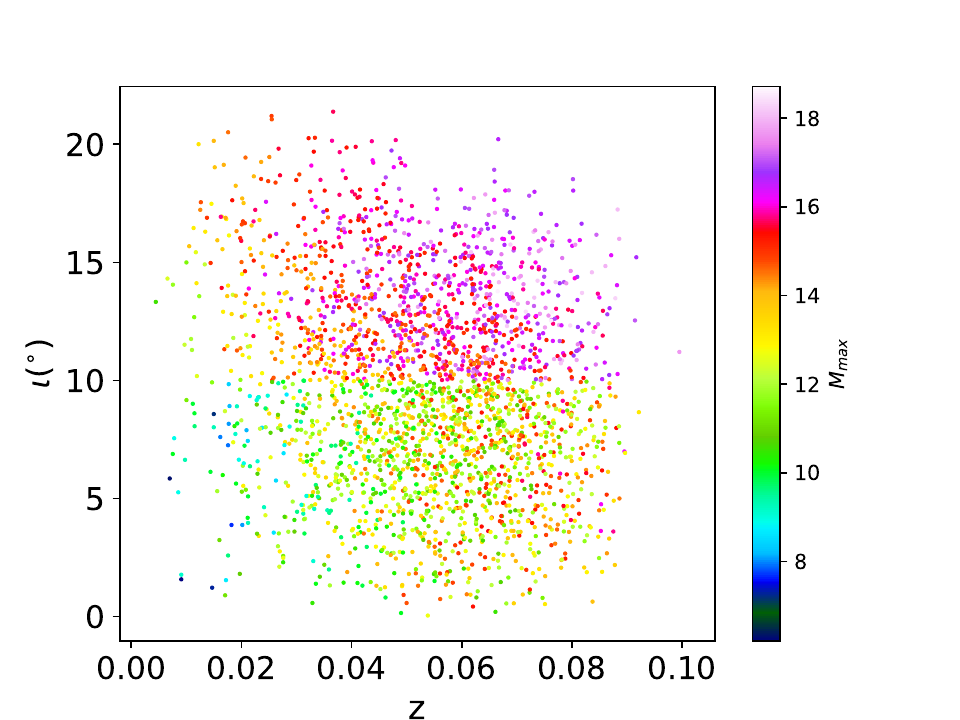}
}
\subfigure[LHVIK]{
	\label{LHVIK_flux7_sactter}
	\includegraphics[width=8cm]{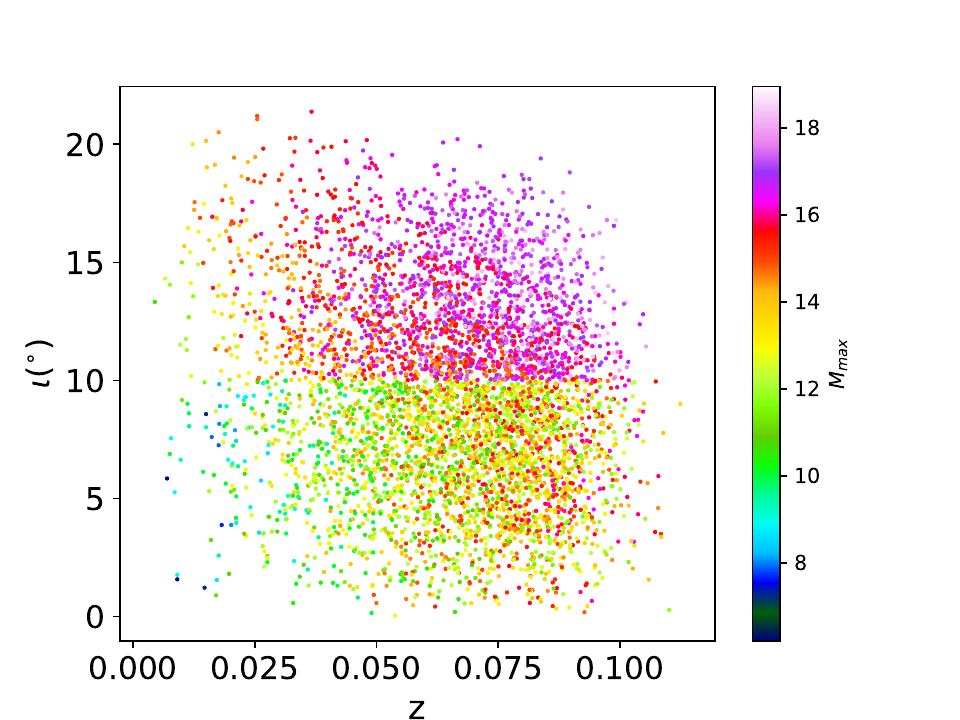}
}
\caption{The distributions of inclination angle, redshift of BNS samples and their afterglow fluxes, which can be triggered by GW detectors and GECAM. The colorbars show their $r$-band magnitude of the afterglows with $\theta_j=10^{\circ}$ \cite{2021ApJ...916...54Y}. (a) and (b) are for GW detectors with and without LIGO-India and Kagra together. }
\label{flux_scatter}
\end{figure*}

In Table \ref{table_GRB}, we list the rate of multi-messenger detections per year. For the case of LHV, this rate is 0.042-0.425 per year when Swift-BAT is involved and is 0.072-0.731 per year if SVOM-ECLAIRS is at work. For the case of GECAM and Fermi-GBM, the rate is a few times higher due to their significantly larger survey areas. Despite its better sensitivity, the EP result is slightly worse than Swift-BAT due to its smaller survey area. When Kagra and LIGO-India are added, LHVIK renders a rate about twice higher than that of LHV. Here, we select the BNS samples that can trigger both GW interferometers and $\gamma$-ray detectors, adopt the GECAM result as fiducial, and summarize the distribution of the BNS redshift and inclination angles ($\iota$) in Fig. \ref{flux_scatter}. \par

After that, we employ standard afterglow models \cite{1998ApJ...497L..17S} to estimate the afterglow magnitudes in the $r$ band. When the Lorentz factor $\gamma$ drops below the half-opening angle $\theta_j$ of a jet, the jet materials begin to spread sideways; such a phenomenon is known as ``jet break''. For an on-axis observer, the light curve consists of two power-law segments connected at the jet-break time; as for an off-axis observer, the light curve reaches a peak after the jet-break time and displays a power-law decline ever since. For off-axis GRB samples, we can calculate the peak magnitudes of afterglows in $r$-band; but for the on-axis case, the afterglows decay with time in a power-law manner, rendering $r$ unattainable from their light curve. In the latter case, we adopt the $r$-band magnitude at the jet-break time instead. $r$ values are exhibited by the colorbars in Fig. \ref{flux_scatter}. Our work shows that the afterglows under consideration are detectable by WFST. After accounting for the fractions of observable area and time, we find that the joint observation rate of sGRBs and afterglows is less than $\sim$ 2 per year, remarkably lower than that of kilonova. Therefore, our WFST search programs for GW EM counterparts will be focused on kilonovae.

\begin{table*}
\begin{center}
\begin{tabular}{|c|c|c|c|c|c|}
\hline
&Swift-BAT&SVOM-ECLAIRS&GECAM&Fermi-GBM&EP\\
\hline
LHV&0.042-0.425 &0.072-0.731&0.278-2.820&0.198-2.001&0.029-0.297\\
\hline
LHVIK&0.084-0.856&0.146-1.474&0.553-5.598&0.394-3.985&0.058-0.593\\
%\hline
\hline
\end{tabular}
\end{center}
\caption{The expected detection rates (in unit of year$^{-1}$)  of multi-messenger sources of BNS mergers via synergy of ground-based gravitational wave detectors and various $\gamma$/X-ray large field telescopes.}
\label{table_GRB}
\end{table*}

\paragraph{Optical counterparts of other GW events}

Kilonovae and optical afterglows from BH-NS mergers are another type of multi-messenger sources that we expect to discover with WFST. The two events, GW200105 and GW200115, followed by several other candidates, GW190426, GW190917, GW191219, GW190814, and GW200210, were discovered during the third observation time (O3) of the LIGO Scientific Collaboration and Virgo Collaboration (LVC). Unfortunately, no electromagnetic counterpart was identified. Several works made efforts to explain the lack of EM identification in theory (e.g., \cite{2011.02717}). The EoS of NSs, the spin of BH, and the mass ratio of the binaries have been found to significantly influence the kilonova luminosity function and the EM detection through their parameter distributions. In the case of a primary BH with a high-spin distribution and its NS companion being less massive with a stiff EoS, the NS is expected to be disrupted by the BH in nearly every case, powering a bright kilonova and an afterglow. In optimal estimation, WFST will detect this kind of optical counterparts at a rate of around $\mathcal{O}(1)$ per year \cite{2011.02717}.

Binary black hole (BBH) mergers also produce EM radiation in some special cases, e.g. the BH has the electric charge or the BBH resides inside the accretion disk of a galaxy. The event GW190521 is possibly the first multi-messenger observation of a BBH event. The detection of an electromagnetic signal has been reported as ZTF19abanrhr by the Zwicky Transient Facility (ZTF) in a sky area consistent with that initially reported by the LVK in an early warning, rendering it a candidate counterpart to GW190521 \cite{Graham2020PhRvL.124y1102G}. A flare peaking at $\sim$50 days after the trigger of GW caused a flux elevation of $\sim$0.3 mag that sustained for $\sim$50 days, assuming a typical bolometric correction factor for quasars. The EM flare is consistent with the expectations for a kicked BBH merger residing in the accretion disk of an active galactic nucleus, which potentially has paramount implications in interpreting GWs from compact mergers, forecasting future counterparts, and measuring the Hubble constant. EM campaigns as follow-up observations of GW alerts are planned to monitor AGN at multiple cadences, from days to weeks, to optimize the efficiency of searching for EM counterparts in the AGN channel.

It is challenging to quantify the detection rate of the optical counterparts for these GW events as a result of the perplexing parameter dependence. For WFST, the GW-triggered target-of-opportunity observations are instrumental in demystifying the formation and evolution of these events.

%%%%%%%%%%%%%%%%%%%%%%%%%%%%%%%%%%%%%%%%%%%%%%%%%%%
%%%%%%%%%%%%%%%%%%%%%%% GRB %%%%%%%%%%%%%%%%%%%%%%%
%%%%%%%%%%%%%%%%%%%%%%%%%%%%%%%%%%%%%%%%%%%%%%%%%%%

\subsubsection{Gamma-ray Bursts}\label{section_332}
Gamma-ray bursts (GRBs), the most energetic stellar explosions in the Universe, are relativistic beaming of jet emission towards the observer. The jet is launched by a compact central engine, being either a BH or a rapidly rotating and highly magnetized NS. No thorough consensus of GRB jet properties (e.g., jet composition, emission radius) exists as yet. The temporal/spectral evolution of the prompt/afterglow emission brings up the primary clues to investigating the GRB jets. A statistically significant sample of GRB prompt/afterglow light curves is fundamental to pinning down the jet properties, necessitating wide-field surveys of the optical counterparts of GRBs. 
 
\paragraph{The Early Optical Afterglow}\label{section_3321}
Multi-wavelength observations of GRB afterglows in the past years has led to the construction of the standard external shock scenario \cite{Piran93,Meszaros93}, in which the interaction between the blast waves and the surrounding medium heats up the ambient electrons to emit broadband afterglows in the form of synchrotron radiation. In observations, the optical afterglow typically commences at a time of $~10^3$~s after the GRB trigger, mainly because of the difficulty of timely optical follow-ups after a GRB is detected. Therefore, the early stage (within $10^3$~s) of a GRB afterglow, namely the early optical afterglow, is often missed. A wide-field survey of the GRB optical afterglow promises to expand the sample of early optical afterglows and improve our understanding of GRB jets. Late-stage optical afterglows are crucial in constraining the structure of the relativistic jet launched from the central engine and the density of the ambient environment\cite{Alexander17,Lazzati18}, early optical afterglows, in parallel, are a unique probe to unravel the composition of the jets and to clarify whether baryons or magnetic fields play a dominant role therein \cite{Piran99,Dai98,ZhangYan11}.

\begin{figure*}
    \centering
    % \begin{subfigure}
    \includegraphics[scale=0.32]{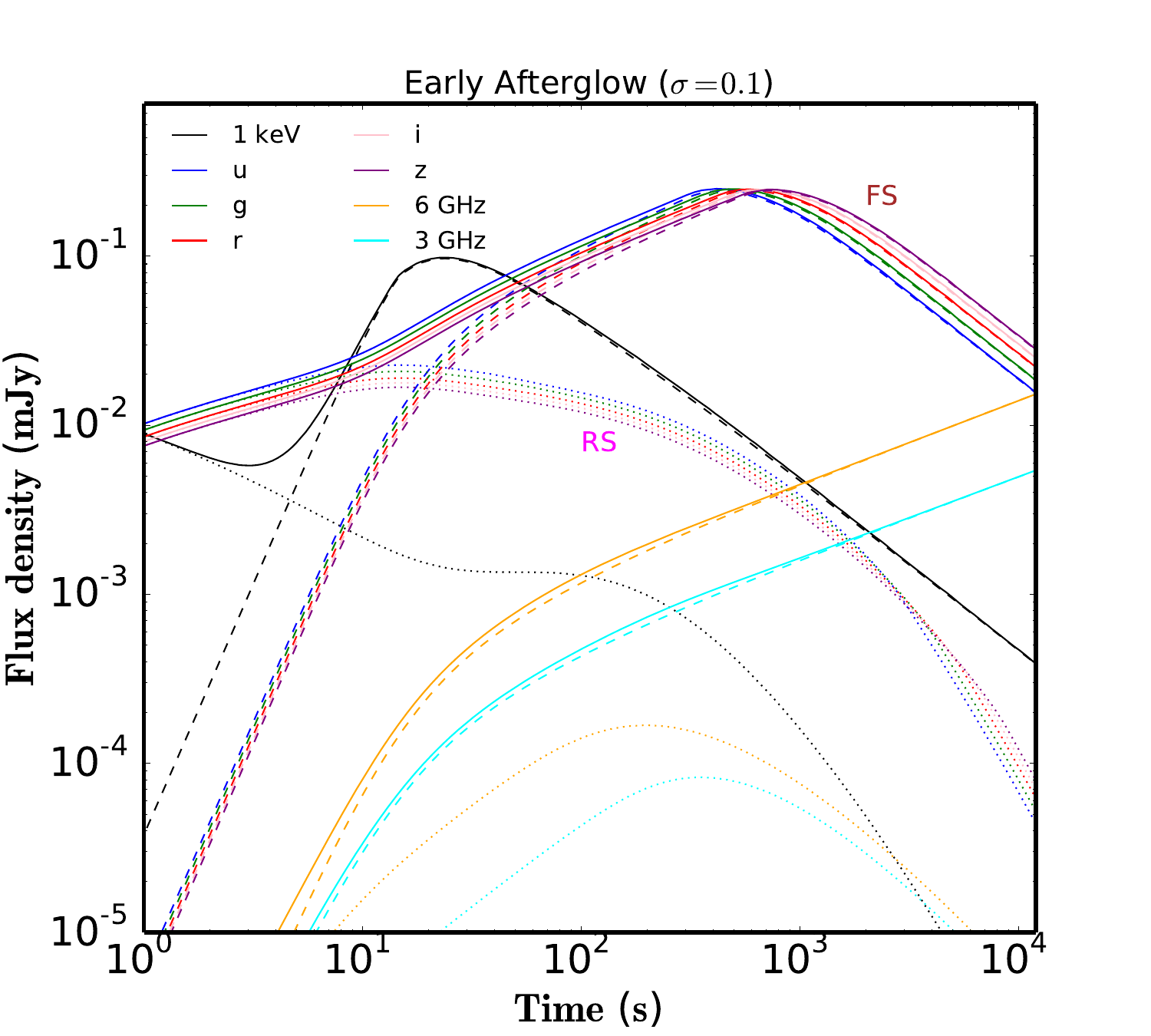}
    % \end{subfigure}
    % \begin{subfigure}
    \includegraphics[scale=0.32]{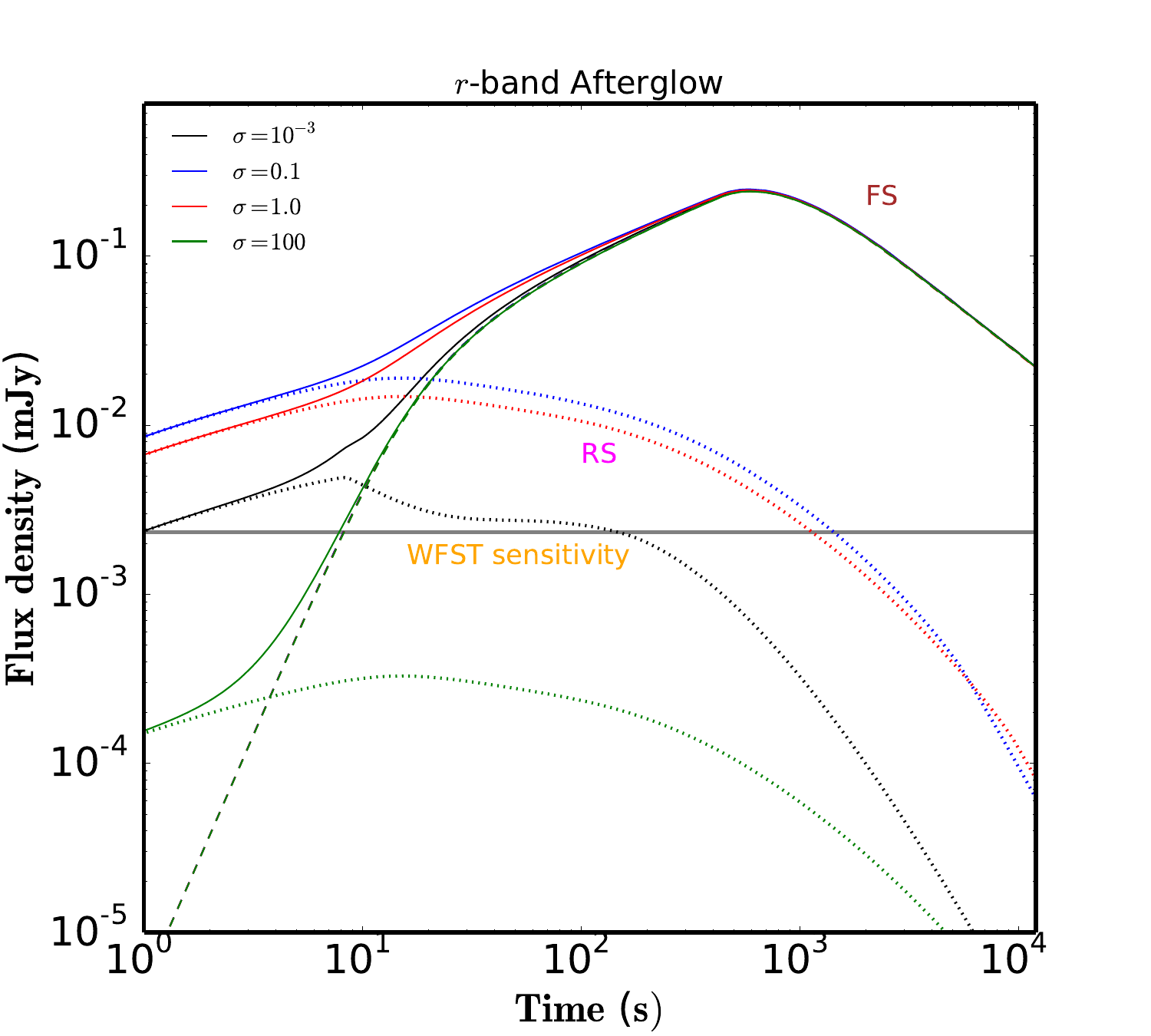}
    % \end{subfigure}
    \caption{Left: Multiwavelength afterglows of an FS-RS system with $\sigma = 0.1$ as an example of predicted observations of a GRB jet at a redshift of $z = 1$. The initial values of the jet parameter are $E_{\rm K,iso} = $~ erg (isotropic equivalent kinetic energy), $n = 1$~ cm$^{-3}$ (circumburst density), $\Gamma_2 = 200$ (bulk Lorentz factor of the FS),
    and $\Gamma_4 = 10^4$ (bulk Lorentz factor of magnetized ejecta),
the microphysical parameters of the FS are $\varepsilon_{\rm e} = 0.1$ and $\varepsilon_{\rm B} = 0.01$. The dashed and dotted lines present emissions from the FS and the RS, respectively. The solid lines are the total flux.  
    Right: The $r$-band lightcurves of FS-RS systems with different values of $\sigma$. The other parameters used are the same as those for the left panel. The grey horizontal line exhibits the sensitivity of WFST with an exposure of 30 s.
    }
\label{fig:Early-Afterglow}
\end{figure*}

When a jet interacts with its surrounding medium, two shocks develop simultaneously, one propagating outward into the external medium (the ``forward shock''; FS) and the other traveling backward into the jet (the ``reverse shock''; RS). Consequent bright optical flashes of the RS in the early episode are predicted theoretically \cite{Meszaros97,Meszaros99,Sari99,ZhangB03,Fan04}, though the early optical afterglows of a few GRBs have shown evidence for an additional emission component arising from a strong RS \cite{Vestrand14,Troja17}. Using a series of numerical methods to solve the dynamics of an FS–RS system proposed in previous work \cite{Beloborodov06,Uhm11,Geng14,Geng16,AiZhang21}, we relate the contribution of RS emission in the early afterglow to the magnetization parameter of the GRB jet, i.e., $\sigma = B_0^2/(4 \pi \rho_0 c^2)$, where both the magnetic field $B_0$ and the fluid density $\rho_0$ are defined in the comoving frame of the fluid. 
A set of numerical multi-wavelength light curves from the FS-RS system are given in Figure \ref{fig:Early-Afterglow}. The emerging RS emission renders early-stage light curves that deviate from those produced in the simple external shock scenario. Meanwhile, our results show that the RS emission is a significant contribution for ejecta with $\sigma$ over the range of 0.1--1, and is dominated over by the FS emission otherwise. This is because at an early stage, the weak magnetic field inhibits synchrotron radiation for $\sigma \ll 1$, whereas the strong magnetic field acts as a relaxant that weakens the RS itself for $\sigma > 1$. Therefore, observations of a substantial sample of early afterglows will constrain $\sigma$ of GRB jets with statistical significance.   

In our WFST surveys, the sensitivity limit lies safely below the early RS flux of a typical GRB, and the FoV can cover the uncertain region of the GRB location within several pointings, demonstrating the WFST's capability to capture early afterglows. When a GRB trigger notice is reported by a space-borne wide-field $\gamma$-ray detector (e.g. Fermi, GECAM \cite{GECAM20} or SVOM\cite{SVOM16}), a timely follow-up to the burst with a relatively small localization uncertainty in the gamma-ray may detect optical signals as promptly as possible. Fermi/GBM report $\sim$300 GRBs per year on average, of which at least 10\% reside within the WFST survey area (with site conditions and the fraction of observable nights taken into account). We plan to observe the targets with a position uncertainty of less than 10 degrees (corresponding to a fraction of $\sim$ 37\%) following the first notice of Fermi. With an exposure of 30 seconds for each pointing, our simulation shows that, for these target candidates, the possibility of spotting the rising phase of the early afterglow is $\sim$22\%. As a result, we expect WFST to capture golden early afterglows $\sim$ 2--3 per year. As a more optimistic consideration, the SVOM satellite that will be commissioned in 2024 is expected to report $\sim 70$ GRBs with a localization error of $\sim$ 10 arc minutes, resulting in a higher WFST detection rate of $\sim$7 golden early afterglows per year.

\paragraph{High-redshift Gamma-Ray Bursts}\label{section_3322}
Thanks to the combination of their extreme brightness with the spectroscopy of the optical afterglows, GRBs are detectable up to a high redshift, as already demonstrated by the cases of GRB 090423 at $z\sim8.2$ \cite{2009Natur.461.1258S,2009Natur.461.1254T} and GRB 090429 at $z\sim9.4$ \cite{2011ApJ...736....7C}. 
As bright beacons in the deep Universe, GRBs are viewed as a complementary, and to some extent unique, probe to the early Universe. 
Statistical analysis of high-redshift GRBs may shed light on cosmic expansion/dark energy, the cosmic star formation rate, Population III stars, the reionization epoch, the metal enrichment history, among other themes of fundamental importance (for a review, see \cite{2015JHEAp...7...35S,2015NewAR..67....1W}).

During its 18.5 years of operations, \emph{Swift} only detected 9 GRBs at $z>6$, although the redshift in 6 cases is spectroscopic, leaving others photometric. Despite the paucity of confirmed high$z$ GRBs in the \emph{Swift} era, theoretical models predict that bursts at $z>6$ represent more than 10\% of the entire population, implying that GRBs are efficient in sampling high$z$ objects \cite{2008MNRAS.385..189S,2015MNRAS.448.2514G}.
A prerequisite to further exploiting the potential of GRBs as a cosmological probe is the construction of a larger sample of high-$z$ GRBs. The optimal strategy to detect the largest possible amount of high-$z$ GRBs is to design a facility operated on soft X-rays with high sensitivity \cite{2015MNRAS.448.2514G,2015JHEAp...7...35S}. Similarly, the Einstein Probe (EP) to be operated in the 0.5--4 keV energy band reaching an unprecedentedly high sensitivity of $10^{-10}$ erg ${\rm s^{-1}}$ ${\rm cm^{-2}}$ in an exposure of $10$ s is expected to detect $\sim20$ GRBs $\rm yr^{-1}$ $\rm sr^{-1}$ at $z\geq6$, or $\sim6$ GRBs $\rm yr^{-1}$ $\rm sr^{-1}$ at $z\geq8$ \cite{2018SSPMA..48c9505W}. Once high-$z$ GRBs are detected, the first and foremost issue is to measure their redshift, but the optical
afterglows of GRBs fade so rapidly that a few hours later they commonly become too faint to permit accessing the redshift.
We expect WFST to contribute to the process of prompt identification of high-$z$ candidates that deserve deep spectroscopy in near-IR by endowing follow-up multicolor images that facilitate photometric redshift estimates. In the EP era, the combination of fast optical photometry using WFST and subsequent deep spectroscopic measurements using larger ground-based telescopes will enable a highly efficient pipeline-wise identification of GRBs at $z>6$.

%%%%%%%%%%%%%%%%%%%%%%%%%%%%%%%%%%%%%%%%%%%%%%%%%%%
%%%%%%%%%%%%%%%%%%%%%%% FRB %%%%%%%%%%%%%%%%%%%%%%%
%%%%%%%%%%%%%%%%%%%%%%%%%%%%%%%%%%%%%%%%%%%%%%%%%%%

\subsubsection{Fast Radio Bursts}\label{section_333}

Fast radio bursts (FRBs) are millisecond-duration cosmological radio transients\cite{lorimer2007}, of which some repeat, but others apparently do not\cite{spitler2016, 2021arXiv210604352T}. As of 2021, hundreds of FRBs have been reported \cite{2021arXiv210604352T}, of which 18 with their host galaxies identified have been localized within arcseconds \cite{2020Natur.581..391M, 2021arXiv210801282B}. The comparison between host galaxies and subgalactic environments has shown that the surrounding environment of FRBs is similar to that of core collapse SNe (CCSNe), type Ia SNe, and short-duration GRBs (SGRBs), but dissimilar to that of long-duration GRBs (LGRBs) and superluminous SN (SLSNe) \cite{2020ApJ...895L..37B,liye2020}, indicative of an association of the progenitors of FRBs with those of CCSNe or SGRBs. This association is (at least partially) confirmed by the discovery of FRB 200428, an FRB from the Galactic magnetar SGR1935+2154 \cite{2020Natur.587...59B, 2020Natur.587...54C, 2020Natur.587...63L} in association with a supernova remnant (SNR). Hence, whether all FRBs originate from CCSNe-associated magnetars, or, to be more specific, whether repeating FRBs and apparently non-repeating FRBs have the same origin, are the most appealing questions awaiting to be addressed. We anticipate that WFST surveys will help tackle these themes from the respect of their host galaxies and optical counterparts.

\paragraph{Host Galaxy}\label{section_3331}

The similarity of the host galaxy and subgalactic environments hints at an association of FRBs with other transients. As mentioned above, the number of identified FRB host galaxies as yet is 18, hosting 7 repeating FRBs and 11 apparently non-repeating FRBs. The limited sample size severely hinders in-depth investigation, rendering the proposed models on the mechanism and origin of repeating and non-repeating FRBs indiscernible. 
Localizing FRBs to arcsecond precision requires wide-field radio arrays as powerful as the Australian Square Kilometer Array Pathfinder (ASKAP). The Square Kilometer Array (SKA), the Five-hundred-meter Aperture Spherical Radio Telescope (FAST) and the Canadian Hydrogen Intensity Mapping Experiment (CHIME) promise to deliver a detection rate $\sim$100 yr$^{-1}$, if an yearly observing time comparable to that of ASKAP is assumed. 
The deep imaging of WFST on the northern hemisphere will set signposts for scrutinizing the FRB host galaxies.
To assess the possibility of distinguishing between repeating and non-repeating FRB host galaxies, we enlarge the FRB host galaxy sample size to 72 by resampling the known FRBs \footnote{https://frbhosts.org/} and perform Kolmogorov–Smirnov tests on the host properties of repeaters vs. non-repeaters, including the stellar mass, the star formation rate (SFR), the specific star formation rate (sSFR) and the galactocentric offset of the FRBs. As a result, we find the probability of repeaters/non-repeaters drawn from the same sample to be less than 0.05. Hence, we conclude that an FRB host galaxy sample made available by the deep imaging of WFST and an enlarged FRB sample with arcsecond localization from future radio telescope arrays will allow for distinguishing the repeating and non-repeating FRBs, if they originate differently.

\paragraph{Optical Counterparts}\label{section_3332}

As elusive as the engine and the emission mechanism of FRBs, a number of models have predicted multi-wavelength counterparts \cite{2014ApJ...792L..21Y, 2019ApJ...878...89Y
} detectable in future WFST surveys. Mechanisms producing FRBs, curvature radiation, or maser, may also produce prompt optical radiation with a millisecond duration similar to those of FRBs. During their propagation towards the Earth, the FRB photons may be inverse Compton scattered by high-energy electrons into optical bands. If the electrons are from the magnetosphere of a magnetar, or if the FRB is produced by maser, then the duration of this optical signal is similar to that of FRBs; but if the FRB is surrounded by SNRs filled with high energy electrons, the optical counterpart may last, instead, thousands of seconds \cite{2019ApJ...878...89Y}. Furthermore, when an outflow accompanies the FRB, a phenomenon evidenced by a pair of X-ray counterparts detected in the Galactic FRB 200428 \cite{2021NatAs...5..378L}, the interaction between the outflow and the interstellar medium (ISM) produces optical afterglows. Dependent on the energy of the FRBs, the time scale of the optical afterglows is on the order of an hour \cite{2014ApJ...792L..21Y}.

Theoretical models predict that the optical-to-radio flux ratio $\eta_{\nu}=f_{\rm opt}/f_{\rm radio}$ of FRBs ranges from $<$ 10$^{-11}$ to 0.1 \cite{2019ApJ...878...89Y, 2020ApJ...896..142B, 2020ApJ...897..146C}, and the optical radiation most detectable by WFST results from the inverse Compton scattering of FRBs inside a neutron star's magnetosphere or an SNR, which typically yields $\eta_{\nu}=5 \times 10^{-5}$ and $10^{-4}$, respectively. 
Assuming an FRB to last 1 millisecond, the FRB fluence function from CHIME observations leads to the flux function $N(>f_{\rm radio})=818^{+229}_{-210}(\frac{f_{\rm radio}}{5\rm \ Jy})^{-1.4}$ sky$^{-1}$ day$^{-1}$.
WFST detection rate of an optical counterpart of FRB is thus estimated by
$N=N_{\rm FRB}(>f_{\rm opt}/\eta_{\nu})*{\rm FOV}$, where $f_{\rm opt}= t_{\rm FRB,o}f_{\rm opt,30}/t_{\rm obs}$ for counterparts with duration $t_{\rm FRB,o}<30s$, $f_{\rm opt,30}$ is the 30 s exposure $r$-band detection limit of WFST, and a 7 deg$^2$ FoV is applied. As a result, the event rate of the ms optical counterpart produced by magnetospheric IC is estimated to be 0.02 yr$^{-1}$, while the optical counterpart lasting for hours produced by FRB-SNR IC is 200 day$^{-1}$ in an ideal case. It should be noted that $\eta_{\rm nu}=10^{-4}$ used here is largely an upper limit with significant uncertainty, and the fraction of FRBs that are surrounded by SNRs is unknown. Moreover, an optical counterpart with a duration of an hour is often difficult to confirm, because normal surveys only record one observing point and coordinated radio observations are required to complete the confirmation. The result of WFST surveys will have profound implications for FRBs, because unambiguous detection of their optical counterparts will open up a new window for this frontier, whereas no detection also provides constraints for the present models \cite{2020ApJ...897..146C, 2021Univ....7...76N}. 

In addition, other transients probably associated with FRBs include CCSNe (if the origin is young magnetars produced by CCSNe), gravitational wave signals and SGRBs/kilonovae (if the origin is magnetars produced by merger of compact stars). The data archive produced by WFST surveys will be a valuable legacy for future exploration of the FRB-transient association. 

%%%%%%%%%%%%%%%%%%%%%%%%%%%%%%%%%%%%%%%%%%%%%%%%%%%%%%%%%%%
%%%%%%%%%%%%%%%%%%%%%%% Neutrino CP %%%%%%%%%%%%%%%%%%%%%%%
%%%%%%%%%%%%%%%%%%%%%%%%%%%%%%%%%%%%%%%%%%%%%%%%%%%%%%%%%%%

\subsubsection{Optical Counterparts of High-energy Neutrinos}\label{section_334}

When particles are accelerated in an astronomical object (e.g. by terminal shocks), the interaction between the accelerated cosmic rays and the surrounding matter or target photons often produces high-energy neutrinos and photons.
The electromagnetic counterparts of high-energy neutrinos are instrumental in the identification of candidate neutrino sources, the determination of the distance to these sources, the exploration of their properties, and our understanding of the acceleration and radiation mechanisms therein, highlighting the necessity of searching for electromagnetic counterparts or transients in coincidence with neutrinos temporally and spatially.

To date, high-energy neutrinos have been detected by large neutrino telescopes settled in water (ANTARES \cite{ANITA2009}, Baikal-GVD \cite{Baikal2021}) and ice (IceCube \cite{IceCube1999}), and by the Auger surface detector and ANITA at high altitude \cite{ANITA2009}.
The IceCube neutrino observatory, the largest neutrino detector to date, detected TeV-PeV astrophysical neutrinos in 2013 \cite{IceCube2013}, of which the origin remains under debate. Since 2016, the IceCube neutrino observatory has been releasing public real-time alerts on single muon neutrino-induced track events with a highly possible astrophysical origin via the Astrophysical Multi-messenger Observatory Network (AMON) and the Global Cycling Network (GCN). The IceCube neutrino alerts include ``gold type'' and ``bronze type'' notices with a chance of astrophysical origin greater than 50$\%$ and $30\%$ and the detection rates are about 12 yr$^{-1}$ and 16 yr$^{-1}$,  respectively. The uncertainty in anchoring the direction of neutrinos ranges from $0.2^\circ$ to $0.75^\circ$.

In their optical real-time follow-up (OFU) program, the IceCube team delivers real-time alerts to the Robotic Optical Transient Search Experiment (ROTSE) and the Palomar Transient Factory (PTF) \cite{Law2009,Rau2009} to start a search for optical counterparts, and the triggered observations are supplemented by a retrospective search in the Pan-STARRS1 wide field survey data \cite{Kaiser2004,Magnier2013}. Consequently, electromagnetic instruments all over the world point to the direction of the neutrino events and conduct follow-up observations in energy bands and messengers ranging from radio, optical, X-ray to GeV/TeV photons and gravitational waves, whose results are then reported on the GCN. Follow-up GeV, X-ray and optical observations of alert neutrinos have revealed BL Lacs, flat-spectrum radio quasars (FSRQs), and TDEs, among others \cite{IceCube2018MultiMessenger,Franckowiak2020,Stein2021}. 

As the neutrino events detected on the southern hemisphere are highly contaminated by muon backgrounds, the alerts released by IceCube are due to neutrinos from the northern hemisphere or the vicinity of the equator, for which IceCube has higher sensitivity. Residing on the northern hemisphere and possessing a sufficient FoV to cover the area of angular uncertainty for most neutrino events detectable by IceCube in a single exposure, WFST will serve as an ideal follow-up optical facility.
Meanwhile, the optical time-domain surveys by WFST will discover more SNe, FBOTs, TDEs, GRBs and AGNs, allowing cross-identification between the detected neutrinos (real-time or archival) and WFST's legacy data. WFST surveys also promise to help identify neutrino sources and further constrain the acceleration mechanism of cosmic rays, the radiation mechanism of neutrinos, and other properties of the sources of scientific interest.

\paragraph{Blazars}\label{section_3341}
Blazars are characterized by their relativistic jets driven by SMBHs with the direction aligned with the observer's line of sight. Blazars will make up an important part of the WFST targets, as will be discussed in Section \ref{section_34}. These jets may accelerate cosmic rays to high energy, and the interaction between energetic cosmic rays and target photons or matter in or near the acceleration sites may produce high-energy neutrinos and photons. Therefore, blazars have been proposed to be high-energy neutrino sources \cite{Stecker1991AGN}.
 
On September 22, 2017, the IceCube Observatory reported a track-like neutrino event (IceCube-170922A) as energetic as about 300 TeV. Follow-up observations found that this event was spatially and temporally associated with the optical-TeV active blazar TXS 0506+056 \cite{IceCube170922A} with a significance of 3$\sigma$. The optical follow-up observations were performed by observatories around the world, including ASASSN, the Liverpool Telescope, the Kanata Telescope, the Kiso Schmidt Telescope, the Southern African Large Telescope (SALT), the Subaru telescope, and the VLT/X-shooter. The spectra, light curve, and polarization were obtained, while the redshift was constrained by optical spectroscopy from the Liverpool, Subaru and VLT telescopes before the determination made by the Gran Telescopio Canarias (GTC). This was the first time the association between neutrinos and point sources was revealed at a high significance level. The potential association between the activity of TXS 0506+056 and the neutrino event renders it a promising candidate source of high-energy neutrinos. A 3.5-$\sigma$ excess of high-energy neutrino events with respect to the atmospheric background was later identified in the direction of TXS 0506+056  prior to the IceCube-170922A alert \cite{IceCube2018MultiMessenger}.
The blazar-neutrino association supports the scenario that AGNs can accelerate highly energetic cosmic rays and produce neutrinos during photohadronic or hadronuclear interactions \cite{IceCube170922A,IceCube2018MultiMessenger}.

In addition the follow-up of real-time neutrino triggers, in a sample of muon track neutrino events that happened between April, 2012 and May, 2017, 11 significant neutrino flares have been found to be associated with 10 AGN counterparts, including FSRQs, BL Lacs and radio galaxies \cite{Osullivan2019}.
Furthermore, 9 blazars are in possible association with single high-energy neutrino events, as per an analysis of both archival and alert neutrino events \cite{Franckowiak2020}. 

\paragraph{GRBs and SNe}\label{section_3342}

GRB/SN jets are believed to accelerate cosmic rays and produce high-energy neutrinos through interactions of cosmic rays with target photons or the surroundings \cite{Waxman1997GRB}.  
Neutrinos may also be produced when shock-accelerated cosmic rays interact with matter and photons during the shock breakout phase of SNe. WFST's capability to detect early phase SNe will help pin down the exploding time of SNe and probe the association between SN SBOs and neutrinos.

Alternatively, if these jets fail to break out through the stellar envelope (e.g. in red/blue supergiant stars), neutrinos and gamma-rays are produced in the interaction between accelerated protons and thermal photons in the jets choked in the thick stellar envelope or the extended material. The duration of the central engine may be longer than that of long GRBs \cite{Xiao2014,He2018SNIIneutrinos}. Since neutrinos and gamma-rays are produced inside the stellar envelope, the source is opaque to gamma-ray photons but transparent to neutrinos. Hence, the lack of association between the observed GRBs and IceCube neutrinos, as well as the tension between the diffuse gamma-ray observations and neutrino observations, can be explained. 
Because a Type {\sc ii} SN is predicted to explode a few hours after the neutrino emission, once an SN spatially associated with neutrinos is spotted, we can trace back to measure the SN explosion time using the observed SN light curve, and measure the time interval between the neutrino burst and the SN explosion.

Furthermore, as discussed in Section \ref{section_31}, some subclasses of SNe are powered by the interaction between the ejecta and the CSM or the companion (e.g. SNe Ia-CSM, SNe IIn, FBOTs, and SLSNe).
The terminal shocks produced by the ejecta-CSM interaction can accelerate cosmic rays to high energies. The cosmic-ray-CSM interaction may result in high-energy neutrinos, rendering the above subclasses of SNe possible optical counterparts of high-energy neutrinos. 

IceCube runs a real-time program to search for muon-neutrino doublets or multiplets. To keep the atmospheric background under control, two or more muon neutrinos detected within a time interval of 100 seconds and within an angular distance of $<3.5^{\circ}$ are required to trigger a doublet or multiplet alert. In March 2012, a neutrino doublet alert was triggered: A Type IIn SN PTF12csy at a distance of approximately 300 Mpc was found to be 0.2$^{\circ}$ away from the neutrino alert direction (with an error radius of 0.54$^{\circ}$), and the a posteriori significance of the chance detection of the neutrino doublet and the SN was 2.2$\sigma$ \cite{IceCube2015SN}. However, the SN was at least 169 days old and no long-term neutrinos signal was found throughout the year, suggesting that the doublet probably was not correlated with the SN. On February 17, 2016, the IceCube real-time neutrino search identified a triplet with three muon neutrino candidates arriving within 100 s of each other, with a probability of detecting at least one triplet from an atmospheric background of 32\%. However, no likely electromagnetic counterpart was detected\cite{IceCube2017multiplet}.
The above multiplet alert was selected under the assumption that the duration of neutrino bursts from transients (e.g. GRBs or CCSNe) is shorter than 100 s, a typical duration of long GRBs. However, as mentioned above, in the chocked-jet models or the interaction-powered SNe, the duration of neutrino bursts may be longer.

The detection of early-phase SNe by WFST will help to pin down the exploding time of SNe readily, allowing us to search for SNe associated with neutrinos in the WFST archival data by assuming a certain time lag between the SNe explosion and neutrinos. Investigations of associations between GRBs/SNe and neutrinos will provide more clues on progenitor stars and the radiation mechanisms.

\paragraph{TDEs}\label{section_3343}
TDEs may generate a relativistic jet or outflow that accelerates cosmic rays to high energies. Neutrinos may be produced when cosmic rays interact with target photons or matter. In a systematic search for optical counterparts to high-energy neutrinos with ZTF \cite{Stein2021}, TDE AT2019dsg was found to be associated with a $\sim 0.2$ PeV neutrino IC191001A with a probability of chance of about $0.2\%-0.5\%$. AT2019dsg was discovered by ZTF six months before the detection of IC191001A, and was later classified as a TDE by ePESSTO+ based on its optical spectrum. As mentioned in Section \ref{section_324}, being significantly more sensitive than ZTF, WFST promises to capture faint TDEs at earlier stages to construct a TDE sample with higher completeness, and to discover more candidate associations between TDEs and neutrinos that will facilitate in-depth investigations.

%%%%%%%%%%%%%%%%%%%%%%%%%%%%%%%%%%%%%%%%%%%%%%%%%%%
%%%%%%%%%%%%%%%%%%%%%%% AGN %%%%%%%%%%%%%%%%%%%%%%%
%%%%%%%%%%%%%%%%%%%%%%%%%%%%%%%%%%%%%%%%%%%%%%%%%%%

\subsection{Active Galactic Nuclei}\label{section_34}

Residing in the centers of active galaxies, luminous quasars, or active galactic nuclei (AGNs) in general, are the manifestations of gas accretion onto massive black holes (BHs) and are believed to play a key role in regulating the evolution of massive galaxies. Although the accretion-BH scenario of the central engine of AGNs has been established since the discovery of quasars over sixty years ago, many fundamental questions remain unresolved. For instance, how do SMBHs acquire their gas? What mechanism is responsible for their variability over a wide range of wavelengths? Are their activities triggered in a persistent or episodic mode, and what are the conditions at work in either case?

\begin{figure}[H]
\centering
\includegraphics[width=0.45\textwidth]{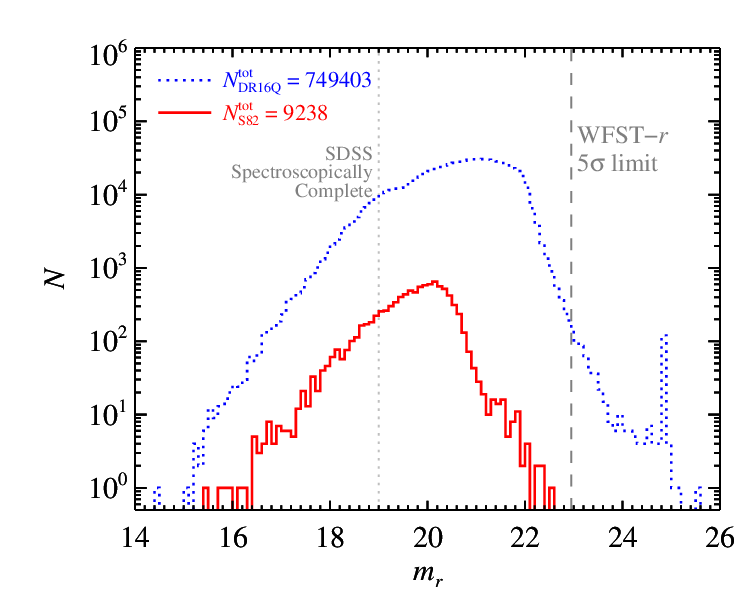}
\caption{Distributions of the apparent $r$-band magnitudes for spectroscopically confirmed quasars in the Stripe 82 (S82; red solid histogram; \cite{MacLeod2010}) and in the SDSS 16th data release quasar catalog (DR16Q; blue dotted histogram; \cite{Lyke2020ApJS..250....8L}). Note that only quasars with physical $r$-band magnitudes are used here. Shown for comparison are the spectroscopically complete limit of $\sim 19$ mag for SDSS quasars (light-gray dotted vertical line) and the WFST $r$-band 5$\sigma$ detection limit of $\sim 22.9$ mag in a single 30-sec exposure (gray dashed vertical line).}
\label{fig:sdss_s82_dr16q_magr}
\end{figure}

The ongoing and upcoming intensive time-domain surveys are instrumental in deciphering the mysteries about AGNs, which are predominantly spatially unresolvable. Illustrated in Figure~\ref{fig:sdss_s82_dr16q_magr}, the current SDSS survey has spectroscopically confirmed nearly $\sim 0.75$ million AGNs over $\sim 10,000~{\rm deg}^2$ primarily in the northern sky (SDSS DR16Q; \cite{Lyke2020ApJS..250....8L}), while only $\sim 1\%$ of them in the well-known Stripe 82 (S82) region of $\sim 290~{\rm deg}^2$ on the southern Galactic hemisphere along the celestial equator have decade-long light curves in five bands ({\em u, g, r, i, z}), which are mapped 8 times on average in a 2-to-3 months duration per year between 2000 and 2008 \cite{MacLeod2010}. Later on, there
have been several completed or ongoing time-domain surveys over the SDSS footprints, which are, however, largely with insufficient sensitivity to detect the majority of faint SDSS quasars and/or with fewer passbands than SDSS. For instance, the CRTS survey conducted in 2005--2013 covered $\sim 26,000~{\rm deg}^2$ in a single broad $V$ band and reached typical detection limits of $\sim 19 - 20$ mag (CRTS DR2; \cite{Drake2009ApJ...696..870D}). In 2009--2013, deeper PTF / iPTF surveys in $g$- and $R$-bands reached a depth of $R \sim 21.0$ mag (PTF DR3; \cite{Rau2009}). The ongoing ZTF survey has been releasing {\em g, r, i} images with a depth of $r \sim 20.5$ mag since March 2018 (ZTF DR8; \cite{Masci2019PASP..131a8003M}). The 3$\pi$ sterodian survey conducted by the Pan-STARRS1 (PS1) team between June 2009 and March 2014 in five passbands ($g_{\rm P1},\;r_{\rm P1},\;i_{\rm P1},\;z_{\rm P1},\;y_{\rm P1}$) reaches a 5$\sigma$ depth of $r_{\rm P1} \sim 21.8$ mag (PS1 DR2; \cite{Chambers2016arXiv161205560C}).
The footprint of WFST WFS is expectedly enclosed by but comparable to the SDSS one, while that of WFST DHS would contain and be larger than the famous SDSS S82 by a factor of $\sim 2.5$.
Considering the same five passbands as SDSS and a 5$\sigma$ detection limit of $r \sim 22.9$ mag in a 30-sec single-epoch exposure (Figure~\ref{fig:sdss_s82_dr16q_magr}), we {\bf thus } expect the WFST DHS and WFS surveys to provide decade-long light curves in three to four passbands (probably excluding $z$ band) for nearly all SDSS quasars, of which a significant amount is not observable by LSST on the southern hemisphere. Furthermore, the WFST surveys will extend the preexisting light curves to several decades for the quasars located in S82 and in the ten medium deep fields of PS1, contributing a highly valuable WFST legacy to the AGN community.

These new decades-long light curves will allow the physical origin of AGN variability to be explored both over longer timescales and towards the fainter end where BH masses lower than currently accessible are found. The increase of time baseline will lead to an increasing possibility of identification of new types of rare AGN associated events.
Thanks to the upcoming deep and high-resolution WFST images, constructing a sample of considerably close AGN pairs is foreseeable by virtue of the unique AGN colors, such that inspection of the triggering mechanism of AGN activity is made possible.
In addition, the long-term variability as well as the deep WFST stacked images (to a depth of $r \sim 25$ mag) will be of service in identifying and characterizing quasar candidates fainter than the completeness limit of SDSS spectroscopy. These quasar candidates will then be readily observable targets for subsequent major spectroscopic programs (e.g. LAMOST-II and MUST) that explore even fainter AGNs at high redshift with lower BH masses with the ultimate goal of attaining a panoramic view of the BH growth and its co-evolution with galaxy and tracking down the cosmological evolution of the intergalactic medium and the large-scale structure of the universe. Several relevant science cases are elaborated below.

\subsubsection{Physical Origin of AGN Optical Variability}\label{section_341}

The variability of AGN in optical is suspected to be driven by X-ray reprocessing \cite{Krolik1991}, accretion disk turbulence \cite{Cai2018}, or corona heating \cite{SunM2020b}, but the physical origin remains largely unclear. Hitherto, no self-consistent physical model has been validated by all relevant observations because of the perplexing accretion physics involved and the large observational uncertainties. The decades-long light curves from the WFST legacy survey will help improve the observational precision by conducting single-band and interband measurements of the variation.

\paragraph{Correlations}

In general, AGN variability appears aperiodic or even stochastic in single-band observations \cite{Ulrich1997ARA&A..35..445U}, although it can be described by a characteristic timescale and a long-term variation amplitude on a statistical basis \cite{Kelly2009}. Hence, scrutinization of the correlations between these two and other observational or physical parameters of AGNs (e.g., wavelength, redshift, BH mass, bolometric luminosity, Eddington ratio, metallicity, X-ray loudness, radio loudness, and the strength of emission lines) promises to shed light on the mysteries about AGN structure and accretion physics.
In particular, the correlation between the BH mass and the slope of variation amplitude to wavelength is a promising alternative method for BH mass estimation (M. Y. Sun et al. in preparation). A recently proposed approach to measure the density of gaseous outflow based on the variability of broad absorption lines also hinge on accurate measurement of the AGN light curves \cite{He2019NatAs...3..265H}. Accurate measurement of the time scale and the variation amplitude is therefore a primary goal of AGN science in the time-domain era. WFST legacy survey extending the preexisting quasar light curves to several decades will help pin down accurately the variation timescale up to a number of years.

\paragraph{Coordination and timelags}

\begin{figure*}
\centering
\includegraphics[width=0.4\textwidth]{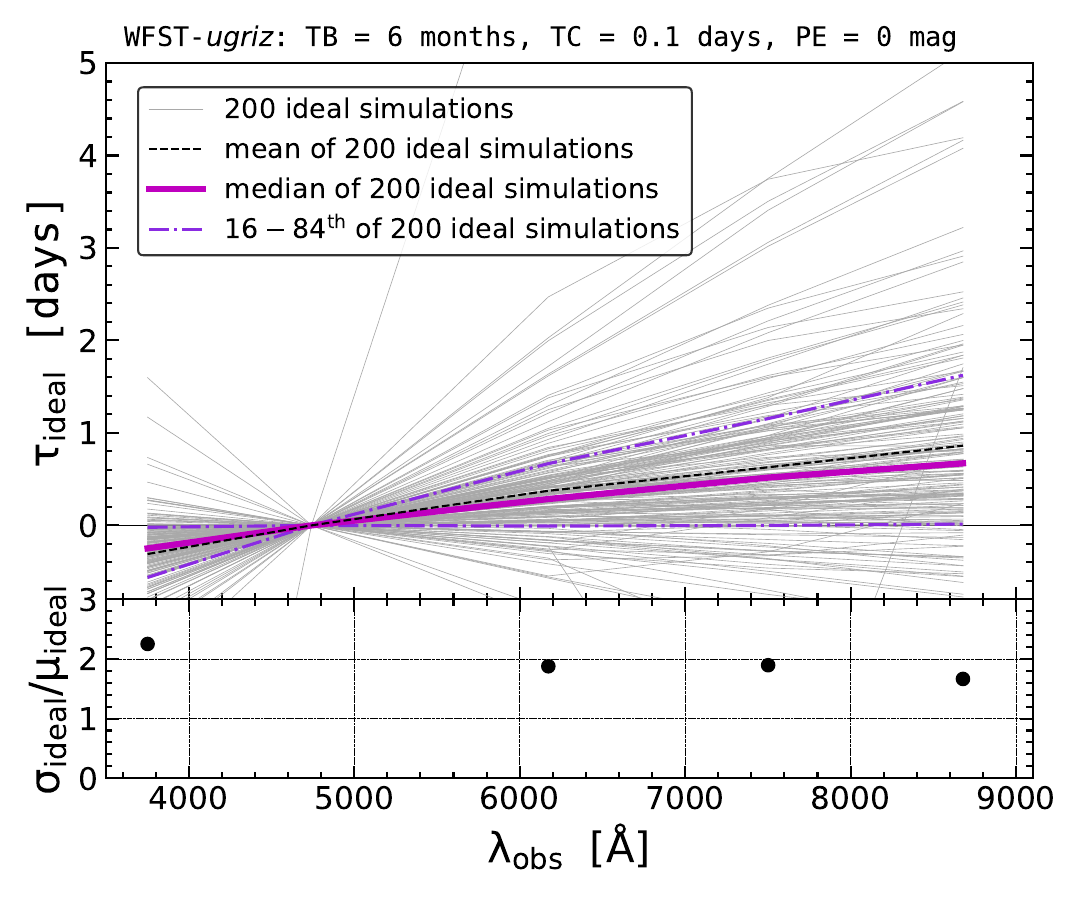}
\includegraphics[width=0.4\textwidth]{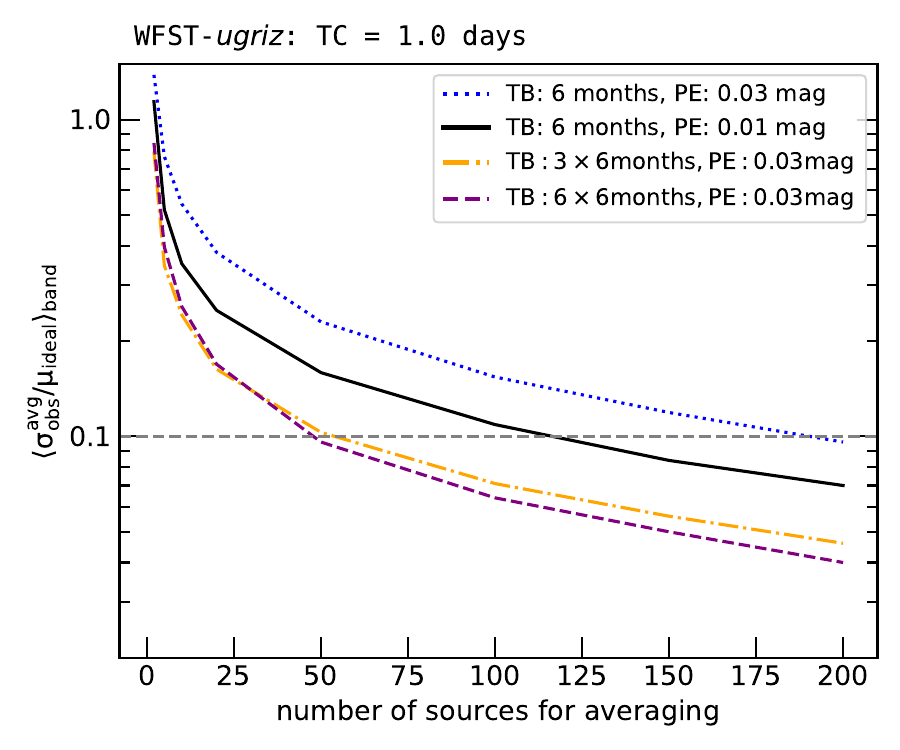}
\caption{Left panel: relative to WFST-$g$ band, the intrinsic inter-band timelag as a function of wavelength implied by the disk turbulence model \cite{Cai2020} for AGNs akin to NGC 5548 observed yearly in WFST-$ugriz$ passbands, assuming a temporal baseline of 6 months per year, a temporal cadence of 0.1 day, and without photometric error (Z. B. Su et al. 2023, in preparation). 
In the top subpanel, the thin gray solid lines show the results of individual simulations, while the median/mean and 16\%-84\% percentile ranges are shown accordingly. In the bottom subpanel, the intrinsic uncertainty of the individual timelag in each band is quantified as the ratio of the corresponding dispersion $\sigma_{\rm ideal}$ to the mean timelag $\mu_{\rm ideal}$.
Right panel: considering a real temporal cadence of 1 day in the WFST DHS and photometric errors of 0.01-0.03 mag, the observed uncertainties of the mean timelag decrease significantly with increasing the number of sources used in averaging and especially with increasing the cumulative observations from one year to three and six years.}
\label{fig:lag_wavelength}
\end{figure*}

Despite the fact that single-band quasar light curves appear stochastic, inter-band variations sometimes demonstrate well-established coordination, where brightening or dimming in phase across optical to UV wavelengths (and even X-ray bands) is seen. In addition, variations at longer wavelengths lag behind those at shorter wavelengths, a phenomenon termed the inter-band timelag. Uncorrelated variations \cite{SouH2022} and the failure to recover lags for the vast majority of AGNs seen in the Dark Energy Survey fields are also reported \cite{Yu2020}. Regardless of the complication, the inter-band timelags derived from optical continuum variations of AGNs are used to estimate the size of accretion disks \cite{Homayouni2019}, under the assumption that the inter-band timelags are closely related to the difference of light travel time among different disk regions irradiated by the central X-ray corona. However, this assumption is under debate as the role of X-ray reprocessing is challenged by multiple observations \cite{Zhu2018}. A new mechanism for the interband timelag has been proposed according to the thermal disk turbulence scenario \cite{Cai2018}. As seen in the left panel of Figure~\ref{fig:lag_wavelength}, for AGNs similar to NGC 5548 observed in the five WFST passbands, the disk turbulence model predicts an intrinsic dispersion of the inter-band timelag as a function of wavelength in seeming consistency with current preliminary observations. Using a tentative survey strategy described in~\S\ref{section_22}, $\lesssim$ 10\% - 30\% uncertainty of the measured time delay is easily achievable by averaging hundreds of AGNs with comparable BH mass and luminosity, even if the first year data are used only (Figure~\ref{fig:lag_wavelength}, right panel). Significant accuracy of the time-lag measurement is expected as a result of the 6-year data accumulation.

The $\sim 700$ square degree deep drilling fields frequently monitored by WFST are significantly larger than those in the PS1 Medium Deep Survey and those planned for LSST, and WFST is expected to make appealing progress in AGN research. In addition to evaluating the time lags between different wavelengths, these deep-dwelling fields will offer a unique opportunity to investigate the true variable SEDs as well as the timescale-dependent color variation of AGNs, of which the latter is deemed as a new path to probe and test accretion disk physics in the era of time-domain astronomy \cite{Sun2014}, demonstrating the potential of the WFST legacy survey to improve our understanding of AGN variability physics.

\subsubsection{Particular AGN Variability}\label{section_342}

Although most AGNs display stochastic variations, persistent monitoring of AGNs in the current time-domain era has led to the emergence of previously known types of AGN variability with enigmatic physical origins. 

Extremely variable (EV) AGNs are those that vary by $> 1$ mag on a time scale of decades \cite{Rumbaugh2018ApJ...854..160R}, in contrast to normal AGNs with a typical variation of $\sim 0.2$ mag on a similar time scale. The physical origin of EV AGN is under debate, but a universal mechanism underlying extreme and normal variations has been suggested \cite{Ren2021arXiv211107057R}. Intriguingly, $> 20\%$ of EV AGNs are spectroscopically confirmed as rare changing-look (CL) AGNs \cite{MacLeod2019ApJ...874....8M}. CL AGNs are featured by dramatic emergence or disappearance of broad emission lines on a timescale of decades, which pose challenges to the standard thin-disk theory. Although most CL AGNs are intrinsically related to changes in the accretion rate \cite{Sheng2017ApJ...846L...7S}, the cause of such a change is not yet known. Furthermore, the timescale and frequency of CL AGNs may place constraints on the episodic and net lifetimes of AGNs and are instrumental in probing the AGN triggering mechanism and the accretion process. Complemented with archival data, the WFST surveys will facilitate the construction of decades-long light curves and the characterization of EV and CL AGNs. 

From nearly a million quasars from the CRTS survey, Graham et al. \cite{Graham2017MNRAS.470.4112G} identified 51 events showing strange major flares atop of the normal stochastic quasar light curves. Their physical origin remains unclear, though micro-lensing by stars in the foreground galaxies is a possibility \cite{Lawrence2016MNRAS.463..296L}, and a more appealing proposed mechanism is associated with explosive stellar-related activity in the accretion disk, such as SNe, TDEs, or mergers of stellar-mass BHs \cite{Graham2017MNRAS.470.4112G}. Remarkably, the ZTF survey has potentially detected an event of binary BH merger in the accretion disk of an AGN in accordance with a reported gravitational wave event \cite{Graham2020PhRvL.124y1102G}. Nearly two magnitudes deeper than the ZTF campaigns, the WFST survey promises to significantly increase the number of detected extraordinary events as a basis for in-depth investigation of their nature.

Periodically varying quasars are considered supermassive binary BH candidates (SMBHB), of which several have been reported \cite{Graham2015MNRAS.453.1562G}. Recently, from a sample of $\sim 9000$ color-selected quasars in a $\sim 50~{\rm deg}^2$ sky area of the PS1 Medium Deep Survey, Liu et al. \cite{Liu2019ApJ...884...36L} identified 26 SMBHB candidates with more than 1.5 cycles of variation. WFST surveys will help verify these SMBHB candidates and identify new candidates, if deep fields larger than those of PS1 are monitored.

Notably, the decades-long light curves delivered by the WFST survey will benefit the search for peculiar AGNs with monotonically increasing/decreasing variations, minimal variations over a long timescale, and true turn-on/turn-off AGNs, potentially a crucial step towards revealing the triggering mechanism of AGNs.

\subsubsection{Low-luminosity AGNs and IMBHs}\label{section_343}

Low-luminosity AGNs in dwarf galaxies are of particular interest because they practically offer the opportunity to identify candidates of intermediate-mass black holes (IMBHs) that bridge the mass gap between SMBHs and stellar-mass BHs. IMBHs in the local universe, as relics/analogs of SMBH seeds in the early universe, are essential for investigating the seed formation mechanisms and the co-evolution of BHs and galaxies. However, IMBHs with supportive observational evidence remain scarce to date, making the increase in sample size a pressing demand (see \cite{Greene2020} for a review). 

A challenge in finding low-luminosity AGNs hosted by dwarf galaxies is posed by the weak AGN signal that is easily overwhelmed by the star-forming activity when conventional methods (e.g. optical spectroscopy, X-ray or radio mapping) are employed. Variability proves to be an effective tool for distinguishing real AGNs from star-forming galaxies and has resulted in the discovery of a considerable number of IMBH candidates in dwarf galaxies, including star-forming ones largely overlooked previously. Recently, the characteristic timescale of optical variability was found to correlate with the BH mass \cite{Burke2021}, paving the way for identifying IMBH candidates by mass estimation purely based on photometric variability. The high-resolution images to be obtained by WFST will significantly alleviate the dilution of stellar light from host galaxies, in contrast to current time-domain optical surveys. Reliable photometry of these weak AGNs will thus become accessible, allowing for detection of active IMBH candidates not only in isolated dwarf galaxies, but also in close dwarf companions of large galaxies, or even in the stripped cores of dwarf galaxies inside a massive galaxy. To distinguish the AGN from stellar activity, the properties of light curves, wide band SED, color variations, and galaxy location are used. In combination with a daily cadence in high-cadence fields, these photometric measurements promise to help construct an appreciable sample of IMBH candidates with BH mass estimates. 

\subsubsection{Off-nucleus AGNs}\label{section_344}

Observationally, off-nucleus AGNs are featured by the spatial offset and are physically connected to nearby companion galaxies. According to the standard framework of hierarchical structure formation, a galaxy merger is naturally expected, as well as the subsequent coalescence of SMBH binaries in a gas-rich environment \cite{Begelman1980Natur.287..307B}. Coalescence may result in a recoiling SMBH, as predicted by multiple numerical general-relativity (GR) simulations \cite{Pretorius2005PhRvL..95l1101P}. Hence, off-nucleus AGNs are probably hosted by galaxies in an early phase of galaxy merger or are ejected AGNs in case the recoiling SMBH is still active after merger. A systematic search for off-nucleus AGNs in galaxy mergers at different offsets and redshifts will help constrain the role of galaxy merger and the associated AGN fueling and feedback, while a search for recoiling SMBHs will provide insight into the distribution of mass ratios and spins in SMBH binaries prior to merger so that the GR numerical simulations are tested.

To date, the application of multiple approaches has only resulted in several hundred offset AGN candidates \cite{Reines2020ApJ...888...36R} and a few recoiling SMBH candidates \cite{Ward2021ApJ...913..102W}. Recently, adopting a novel variability-based search strategy, Ward et al. \cite{Ward2021ApJ...913..102W} identified 52 AGNs in merging galaxies and 9 recoiling SMBH candidates based on a parent sample of 5493 optically EV AGNs with flux variations over 2.5 mag in both ZTF $g-$ and $r-$bands over a 2.5-year period. Among their offset AGNs, those with available redshifts display linear separations typically larger than 2 kpc as a result of the low resolution of ZTF images. In comparison, the high-resolution multiband imaging of WFST will enable us to construct a sample of targets with smaller offsets that helps reveal the crucial phase closer to the merger event, and a more statistically complete sample allowing to test relevant physics before and after mergers is also accessible. A new method to search for off-nucleus AGNs or close AGN pairs based on their color-variation properties (e.g. the bluer-when-brighter trend) is under development. The nature of off-nucleus AGNs found by WFST will be further explored when the extremely high-resolution images from CSST become available.

\subsubsection{Strongly-lensed AGNs}\label{section_345}

When AGNs are lensed by intervening objects (galaxies in particular), multiple images may be observed. Such strongly lensed AGN systems are of fundamental importance to a number of astrophysical frontiers. They facilitate the measurement of the total-mass profile and dark-matter substructures in the lens galaxies, and are used to probe the coevolution of black holes and their hosts at cosmological distances. When the light curves of lensed AGNs are obtained, these systems can be further used to constrain the stellar initial mass function in the lens galaxies and to measure the size and temperature profile of the accretion disks surrounding BHs in the background AGNs. In addition, 
strongly-lensed AGNs with time delay measurements may deliver independent and precise measurements of the Hubble constant, a probe of particular importance to deepen the understanding of the growing tension between the $H_0$ values given by distance ladders and cosmic microwave background observations. 

The discovery of strongly lensed AGN systems traditionally relies on imaging and spectroscopy-based methods, though several variability-based methods have been developed recently \cite{Chao20b, Shu21}, which may render ongoing and upcoming time-domain surveys (e.g. ZTF, WFST, and LSST) fully exploited. To date, $\approx 200$ strongly-lensed AGN systems have been found, of which light-curve measurements for individual lensed images are available to only $\approx 30$ \footnote{\url{https://obswww.unige.ch/~millon/d3cs/COSMOGRAIL_public/code.php}}. A simulation conducted by Oguri et al. \cite{oguri10} suggests that, on average, there exist $\simeq 0.06$ galaxy-scale strongly lensed AGN systems per deg$^2$ possessing two (for two-image systems) or three (for four-image systems) lensed images brighter than $i=22$ mag. 
Therefore, we expect WFST to detect $\approx 500$ strongly lensed AGN systems, and notably, to further deliver multiband high-cadence light curves of these systems. The resultant extensive legacy dataset will potentially be a significant step forward in multiple relevant frontiers.

%%%%%%%%%%%%%%%%%%%%%%%%%%%%%%%%%%%%%%%%%%%%%%%%%%%%%%%%%%%%
%%%%%%%%%%%%%%%%%%%%%%% Solar System %%%%%%%%%%%%%%%%%%%%%%%
%%%%%%%%%%%%%%%%%%%%%%%%%%%%%%%%%%%%%%%%%%%%%%%%%%%%%%%%%%%%

\section{Asteroids and the Solar System}\label{section_4}

\subsection{Overview of NEO Science}\label{section_41}

By definition, a near-Earth Object (NEO) is any object with its perihelion q$\leq$1.3 AU and its aphelion Q$\geq$0.983 AU. Possibilities include an asteroid or a comet. NEOs may deliver information about the primordial materials of the Solar System, though a more realistic reason to construct a catalog of them as complete as possible is their potentially damaging impacts onto the Earth. Ever since the early stage of its formation, Planet Earth has been subject to NEO impacts. An exhaustive geological consensus has revealed that Cretaceous-Tertiary extinction was caused by the impact of a large asteroid or comet 65 million years ago \cite{Alvarez1980}. In July 1994, the widely observed impacts into Jupiter of the fragments of Comet Shoemaker-Levy 9 released energy equivalent to millions of megatons of TNT and generated fireballs and dark clouds on Jupiter as large as Earth. In view of the realistic threat of impacts, NEO surveys were commissioned in the late 1990s (e.g. LINEAR, NEAT, Spacewartch, CSS, Pan-STARRs \cite{Denneau2013, Magnier2020}, ATLAS, CNEOST). Knowledge on the NEO population has been accumulating for three decades, and more than 95\% of kilometer-class NEOs have been cataloged so far. The goal of LSST, NEOCam, and other next-generation sky surveys is to catalog NEOs of relatively small sizes. 

\begin{figure*}
\begin{center}
\includegraphics[width=0.8\textwidth]{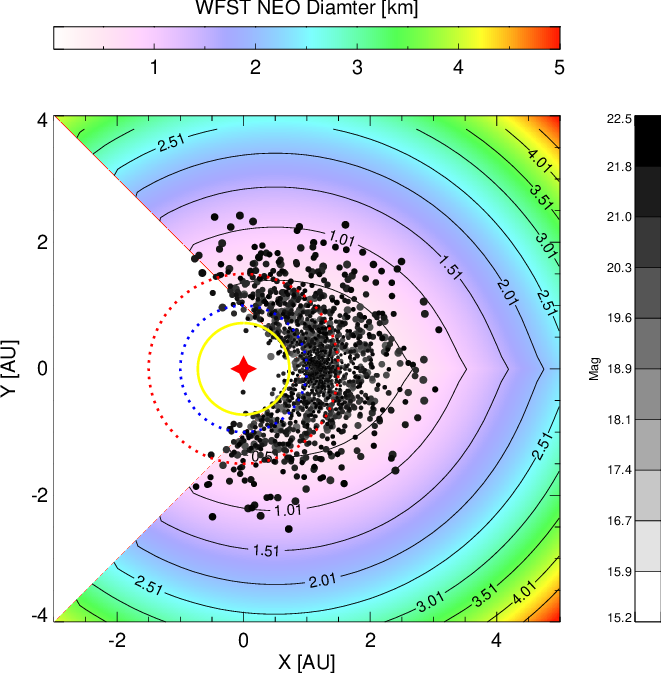}
\caption{ The size and position of near-Earth asteroids visible to WFST, the color indicates the diameter of NEA,each dot represents a NEA(form Granvik's model\cite{Granvik2018}),the dot's gray scale represents magnitude,the size of the dot indicates the diameter of the NEAs. The dotted blue line shows the orbit of the Earth, and the red asterisk at the origin represents the Sun.
\label{Fig:NEO1}}
\end{center}
\end{figure*}

Sky surveys using ground-based optical telescopes are the most efficient and systematic approach to capture NEOs. In the next decade, LSST is poised to monitor NEOs on the southern hemisphere, whereas WFST on the northern hemisphere will contribute a comprehensive catalog of NEOs at an advantage of its wide FoV. WFST will manifest itself through its ability to detect small and faint objects (r = 22.5 mag with 30-second exposure time), and its 6.5 deg$^2$ FoV that will enable frequently repeated mapping of a significant fraction of the sky to search for NEOs, and its high resolution ($0.33^{\prime\prime}$/pix) to optimize the orbital accuracy of faint NEOs. Granvik's model \cite{Granvik2018} predicts that thousands of near-Earth asteroids are readily observable by WFST every night (see: Figure \ref{Fig:NEO1}).

\begin{figure*}
\begin{center}
\includegraphics[width=0.9\textwidth]{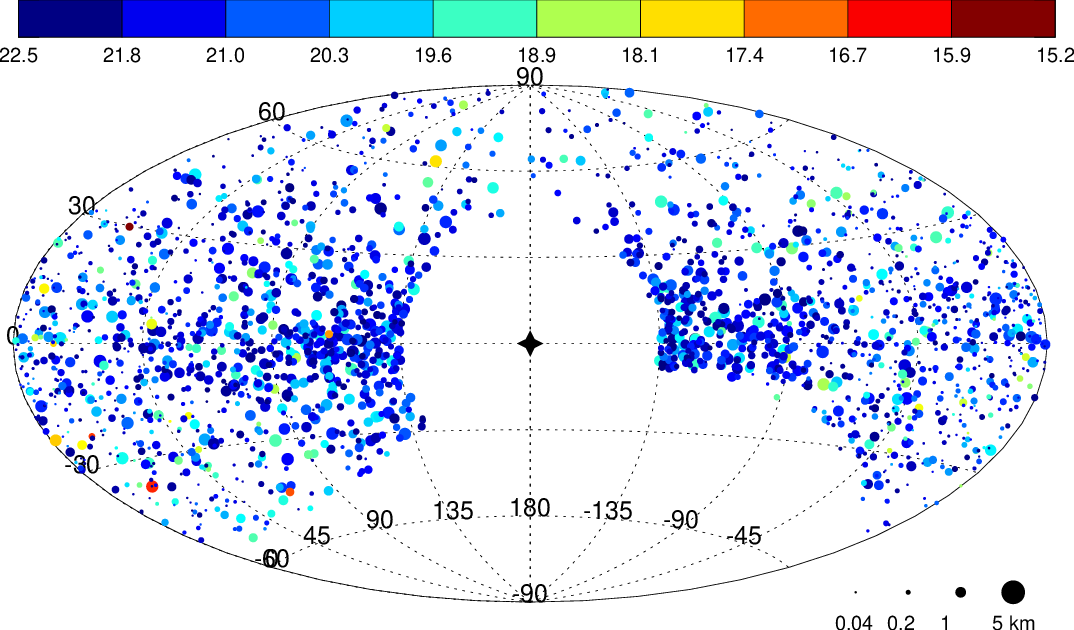}
\caption{ Ecliptic coordinate system centered on the sun, the dot's color represents magnitude,the size of the dot indicates the diameter of NEAs (form Granvik's model\cite{Granvik2018}), the star in the center represents the Sun. The 45$^\circ$ region around the Sun is considered unobservable.
\label{Fig:NEO_nearsun}}
\end{center}
\end{figure*}

Detection of NEOs on or within the Earth's orbit can be challenging for ground-based observers due to their proximity to the Sun, rendering these NEOs poorly characterized and cataloged as yet. Most of the objects that fall into this class are known as Atiras or interior-Earth objects. In general, Atiras are only observable in brief windows during evening and morning twilight. Multiple programs have surveyed Atiras, but only 28 are known to us, of which many were discovered by ZTF \cite{Bellm2019,Ye2020}. Monitoring the Atiras region may bring up additional benefits, because twilight observations at the near-Sun region (see: Figure \ref{Fig:NEO_nearsun}) will significantly increase the solar phase angle coverage of NEOs and MBAs, so that photometric models and actual detection are both improved, facilitating the discovery of Earth Trojan asteroids \cite{Wiegert2000} supposed to librate at Earth-Sun L4 and L5 Lagrange points. Dynamical simulations predict that these objects can survive on a timescale comparable to the age of the solar system, implying that an ancient population of small asteroids may exist in these regions. 

The combination of the 2-meter aperture of WFST and the excellent night sky conditions of Lenghu is advantageous to twilight observations (starting at a sun altitude of $-12$ degrees and ending at $-18$ degrees). A twilight survey is a strategy to realize the science goals described above without interfering with the operation of WFST surveys. However, we must be aware that sky background brightness at twilight\footnote{According to \cite{Patat2006}, the average of lightening or darkening is $0.23\pm 0.02$ mag/arcsec$^2$/min at twilight.} will worsen the detection limit, only a few relatively large near-Earth asteroids may be seen. Another concern during the twilight NEA survey is due to the low-orbit satellite constellation, such as Starlink. \cite{Mroz2022} estimate that once the size of the Starlink constellation reaches 10,000, virtually all ZTF images taken during twilight will be of lower quality, and the WFST twilight NEA survey likewise.

\subsection{Cometary Activity}\label{section_42}
Comets are considered to be the least modified solar system objects or ``fossils'' date back to the era of planet formation and are therefore an essential probe to the origin and evolution of the solar system. They can be classified into short-period (with orbital periods shorter than 200 years) and long-period (orbital periods over 200 years) comets. Before 2006, comets were believed to originate from two locations: most short-period comets born in the Kuiper Belt or the scattered disc, while long-period comets in the Oort Cloud. The main asteroid belt was identified as the third origin of comets in 2006 \cite{2006Sci...312..561H}. Main-belt comets are found in the main asteroid belt with orbital characteristics similar to main-belt asteroids, although tails or comas exist. The main belt comet has arisen broad interest ever since their discovery, because the existence of comets in the asteroid belt implies that water ice exists therein, an intriguing clue to tackling the origin of water on Earth \cite{2012AREPS..40..251M} and to the solar system's thermal history.  

Within 3 AU, cometary activity is consistent with the standard model, in which water ice volatilization acts as the main drive \cite{1950ApJ...111..375W}; while beyond 5 AU, volatile gas volatilization is the main cause \cite{1993PASP..105..946L}. Different modes are at work, in which dust is released from the surface of the comet nucleus as a result of the sublimation of gas ice and water ice; distant comets therefore promise to help unveil the mechanisms underlying cometary activity.

To date, only nine main-belt comets have been discovered \cite{2016ApJ...830...22H,2021ApJ...922L...9H}, thus an in-depth understanding of these objects awaits systematic searches and the accumulation of further investigations. Many international telescopes have been involved in the search for main-belt comets, such as Pan-STARRS1(PS1), Hawaii Trails Project (HTP), Canada France-Hawaii Telescope (CFHT) and Palomar Transient Factory (PTF). Among them, PS1 is more effective in discovering main belt comets (four), because PS1 has the characteristics of large field of view and the ability to detect weak objects. WFST also has performance similar to that of PS1, so it is expected to play an important role in the search and discovery of new main-belt comets. In addition, by mining comet data at different heliocentric distances in the WFST sky survey data, differences in the activity of the comet driven by sublimation of water ice and gas ice can be compared.

\subsection{Trans-Neptunian Objects and Planet Nine}\label{section_43}

Trans-Neptunian Objects (TNOs), also known as Kuiper Belt Objects (KBOs), are asteroids or dwarf planets beyond the orbit of Neptune, of which the distribution extends from about 30 AU from the Sun to nearly 1,000 AU or even further. More than 2000 of these objects have been cataloged so far, likely representing only a tiny fraction of the actual populations in this region. The Kuiper Belt is believed to be populated with millions of objects, of which hundreds of thousands are larger than 100 kilometers \cite{Morbidelli2020, Petit2011}. 

The diverse structures and characteristics of TNOs provide clues to the formation and evolution history of our solar system and offer unique information to place constraints on unknown parameters involved in planetary formation and migration simulations. TNOs are classified into several dynamical populations: resonant populations, classical belts, scattering disks, and detached objects. Comparison between different populations can shed light on their respective evolution history. The cold classical subclass, dynamically defined as TNOs with non-resonant orbits, no close encounters with Neptune, and with orbital inclinations less than 5$^\circ$, is a special population with multiple unusual physical properties (e.g. a distinctly red color, a large fraction of wide binaries, generally higher albedos, a steep slope of size distribution at large sizes) \cite{Nesvorny2018}. These unusual properties imply that this subclass may have formed or dynamically evolved in processes different from other TNOs. Furthermore, a variety of potential correlations among orbital and physical characteristics (e.g. inclination and color) await observational tests using a larger sample \cite{Marsset2019}. In particular, even the discovery of a single binary asteroid system or several high-inclination objects may impose strong constraints on planet formation and evolution theories \cite{Nesvorny2018NA}.

The Planet 9 hypothesis derives from several dynamical anomalies of known distant TNOs \cite{Batygin2016}. Distant TNOs, also known as detached objects, are far beyond the eight-planet dynamical region and may act as an indirect probe of the far reaches of the solar system. Hitherto, only 14 detached objects have been detected, of which five chaotic objects may fail to represent the dynamical statistics due to their instability \cite{Batygin2019}, necessitating a sample with higher statistical significance to help clarify the (non)existence of Planet 9. We will dig part of the WFS area to search for TNOs and Planet-9.

\section{The Milky Way and Its Satellite Dwarf Galaxies}\label{section_5}

\begin{figure*}
\begin{center}
\includegraphics[width=0.49\textwidth]{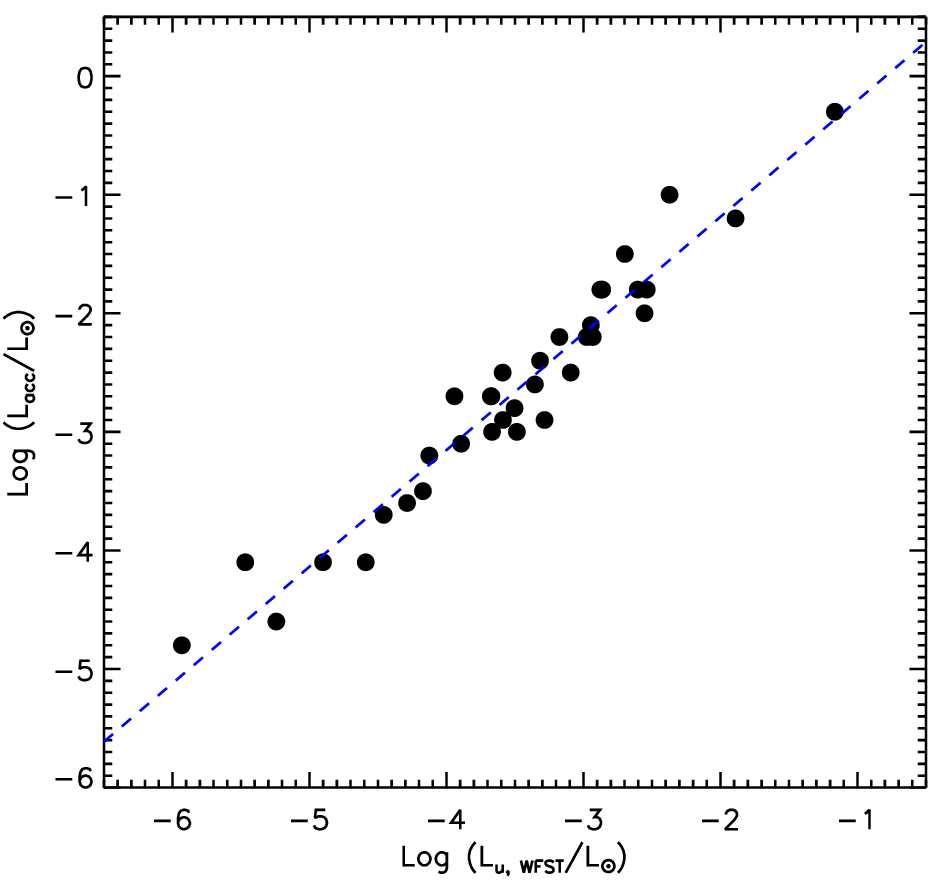}
\includegraphics[width=0.49\textwidth]{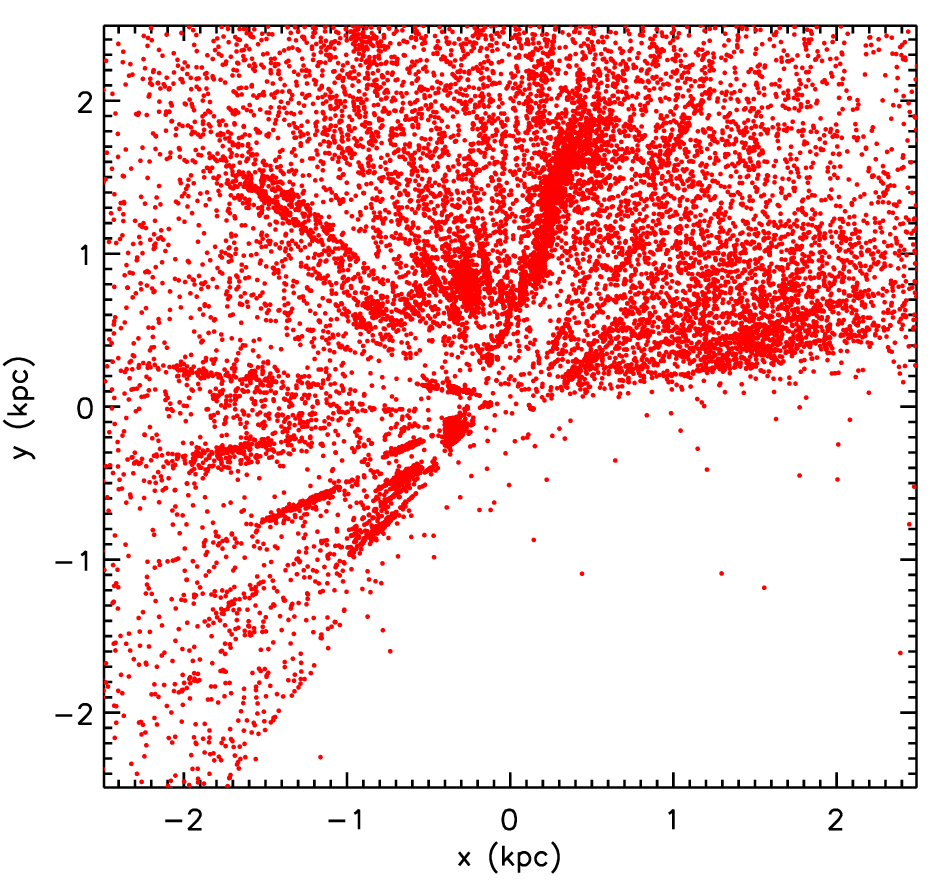}
\caption{Left: Relation between the WFST u-band brightness and the accretion luminosity. Right: A bird’s-eye view of the distribution of young stars with circumstellar disks, which can be observed with WFST, looking down on the Galactic disk with the Sun at the center.
\label{Fig:Acc}}
\end{center}
\end{figure*}

\subsection{Star Formation}\label{section_51}
\subsubsection{Young Stars}\label{section_511}
The mass accretion rate ($\dot{M}_{\rm acc}$) is a crucial parameter in modeling the evolution and dissipation of circumstellar disks and planet formation as well. Young stars commonly display accretion variability at various timescales due to different physical mechanisms, including non-steady accretion on timescales of hours, global instabilities of the magnetospheric structure on timescales of months \cite{1996A&A...314..835G,2003A&A...409..169B}. In addition, the interaction between circumstellar disks and young massive planets may induce pulsed accretion \cite{2018ApJ...853L..34B}. Pulsed accretion onto young stars also serves as a novel tool to identify young massive planets. Measurement of accretion rates of young stellar systems helps unravel the evolution of circumstellar disks in low-metallicity environments.

Magnetic pressure may expel gas from the midplane of the disk, which is funneled onto the star along the magnetic field lines. The gas flow falls onto the stellar surface at approximately a free-fall velocity, causing a strong ``accretion shock'' on the stellar surface \cite{2016ARA&A..54..135H}. Ultraviolet/optical excess emission arises when the gravitational energy of the infalling material involved in the accretion process is radiated away along with the accretion shock, manifesting itself as a direct measure of the accretion rate \cite{1998ApJ...492..323G}. WFST survey will adopt an optimized methodology by employing the $u$ band,and, in which the Balmer edge falls, and conduct $\dot{M}_{\rm acc}$ measurements. Figure~\ref{Fig:Acc} (left) depicts the relation between the WFST $u$-band brightness and the accretion luminosity for a sample of young stars in the literature, where the WFST synthetic observation is performed on their VLT/X-shooter spectra and the accretion luminosity are taken from \cite{2014A&A...561A...2A}. The tightness of the correlation promises that WFST $u$-band photometry will yield accurate measurement of accretion rates onto young stars. Using data from Gaia EDR3 and ALLWISE, we constructed a sample of over $\sim$1.8$\times$10$^{4}$ young stars surrounded by circumstellar disks observable by WFST (Figure~\ref{Fig:Acc}; right). Hence, for the first time, WFST will perform a systematic measurement of accretion rates and variability based on a large sample of young stars. 

\subsubsection{Accretion Burst Events}\label{section_512}
To date, it remains an open question how young stars gain their mass from the surrounding environment through disk accretion. Relevant models conventionally assume a steady accretion onto the central young star with a constant accretion rate \cite{shu1987}, though these models predict a significantly higher luminosity than what is observed \cite{kenyon1990}. To address this ``luminosity problem'', Kenyon\,\&\,Hartmann \cite{kenyon1995} proposed an episodic accretion scenario, under the assumption that a large fraction of disk accretion occurs during a number of short-lived bursts. Accretion bursts were first observed around low-mass young stars \cite{1996ARA&A..34..207H}, and were seen around high-mass young stars later \cite{2021ApJ...922...90C}, but it remains unclear how frequently young stars are in the state of accretion outbursts and what mechanisms drive these outbursts. 

EXors and FUors are the two types of young stars where accretion outbursts are likely ongoing. The Fuors phenomenon is the most prominent during star formation, which displays an increase of brightness by 5 magnitudes or more within a year and remains bright afterwards for decades \cite{1996ARA&A..34..207H}, while EXor outbursts occur on shorter timescales ($\sim$ years) and show lower amplitudes \cite{2014prpl.conf..387A}. It remains enigmatic whether there is a physical distinction between these two types because of the paucity of known FUors and of observations before their outbursts. Among the $\sim$1.8$\times$10$^{4}$ young stars with circumstellar disks to be monitored by WFST, we expect to detect 0.5--7 FUor outburst events each year, an estimate based on the PTF survey \cite{2015ApJ...808...68H}. Despite the fact that all-sky infrared surveys are awaited to fully characterize the evolutionary stages of these young stars, the WFST time-domain survey will significantly contribute to exploring the accretion history of young stars that are captured at different evolutionary stages.

\subsection{Mapping the Milky Way}\label{section_52}
\subsubsection{3D Dust Distribution}\label{section_521}

Dust distribution is an indispensable piece of information of Galactic science, while dust extinction is routinely invoked in astrophysical studies. A thorough dust distribution map is recovered by measuring the reddening and extinction towards a large number of stellar objects. Based on modern wide-field optical photometric and spectroscopic surveys (e.g. SDSS, Pan-STARRS1 and Gaia), the three-dimensional (3D) Galactic dust distribution has been mapped at an arcmin-scale spatial resolution, from which the structures of dust Galactic disk such as warp and spiral arms have been revealed \cite{chenbq2019,green2019}. WFST survey is at least 2--3 magnitudes deeper than Pan-STARRS1 in the $r$ band, promising to produce 3D dust maps with improved resolution and dynamical range than previous maps, so that Galactic high-density regions associated with star formation can be traced and Galactic models are better constrained. In particular, the high sensitivity and photometric accuracy of the WFST survey will allow for investigating the diffuse interstellar medium at high Galactic latitude. For instance, WFST will benefit the study of intermediate-velocity clouds (IVCs) that are considered as an inflow of gas consisting of recycled disk material and thus are believed to be connected to a Galactic fountain process \cite{rohser2016}.

\subsubsection{Stellar Clusters}\label{section_522}
 
Stellar clusters in the Milky Way serve as ideal test beds for stellar evolution from pre-main sequence to post-main sequence stages, given their ranges of age spreading over several magnitudes from a few to tens of Myr (open clusters) to a few to tens of Gyr (globular clusters) \cite{2016A&A...588A..40R,2021MNRAS.508.2688G}. The co-eval, co-spatial, and iso-metallic stellar members provide abundant clues to stellar astrophysics. As important as they are, the majority of star clusters have been relatively poorly studied because of their large distances or large angular sizes. 
  
\begin{figure*}
\begin{center}
\includegraphics[width=1.2\columnwidth]{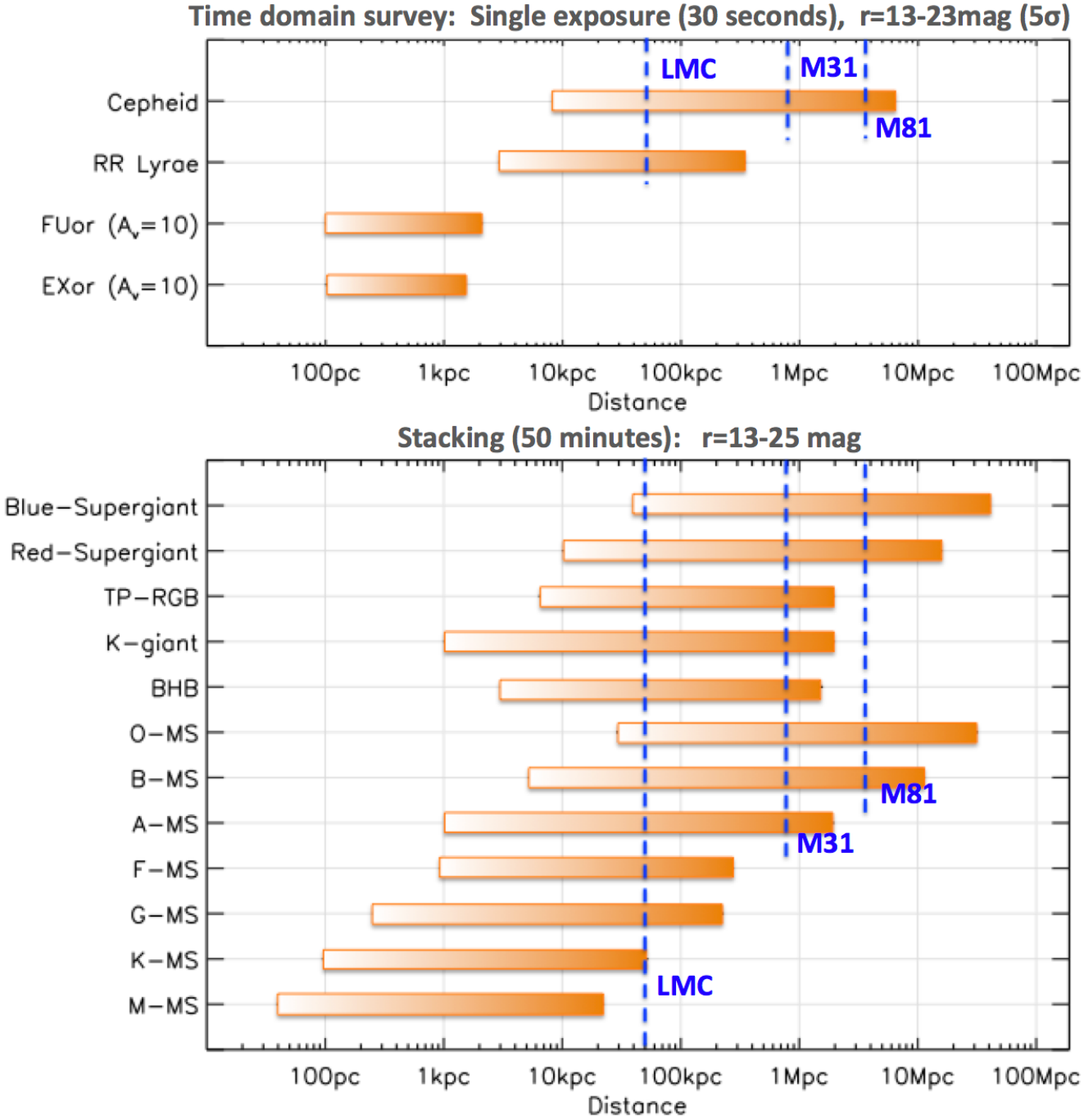}
\caption{Detectability of different types of stars vs. the distances in the WFST images with single exposure (top) and stacked images (bottom)}\label{Fig:distance}
\end{center}
\end{figure*}

\begin{figure*}
\begin{center}
\includegraphics[width=0.5\textwidth]{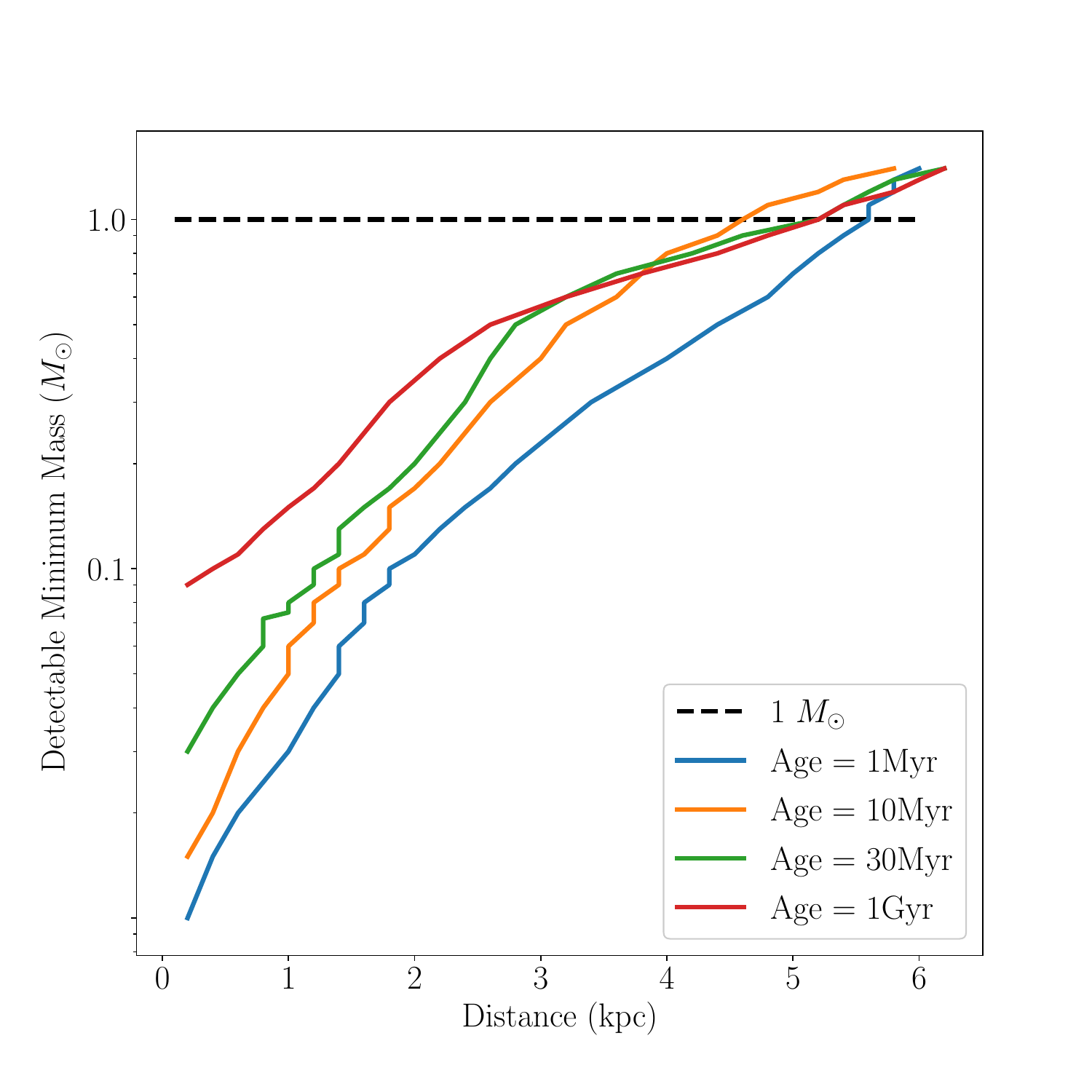}
\caption{The minimum stellar mass of the open cluster member varies with distance from near to far. The isochrones of low-mass stars at 4 ages are shown in different colors \cite{2015A&A...577A..42B}. For open clusters, an empirical relation $A_V = 1.5\,\mathrm{mag\,kpc^{-1}}$ is used in the distance module. 
\label{Fig:cluster}}
\end{center}
\end{figure*}

The detection limit of the final stacked images from the WFST survey programs in the $r$ band will reach 25 mag, 2--3 magnitude deeper than the Pan-STARRS1 survey. For open clusters confined in the galactic plane, we estimate the minimum stellar mass detectable by WFST as a function of distance (Figure~\ref{Fig:cluster}). For clusters within a 5~kpc distance from the sun, WFST can resolve them down to a stellar mass of $\lesssim1\,M_\odot$ with an appreciable completeness of the mass estimate for these open clusters. The young open clusters and the even younger embedded clusters (to be better studied in IR) represent the current star formation rate of the Milky Way \cite{2016ApJ...833..229L}. Thus, the 3D distribution of these two types of young clusters depicts the 3D star formation rate distribution of the Milky Way, which, along with the 3D molecular gas map \cite{2019ApJS..240....9S}, provides a view of the baryon cycle of the Milky Way. For nearby ($\lesssim$1\,kpc) young clusters, WFST's accuracy, sensitivity, and multi-epoch mapping will enable the detection of cluster members down to the mass level of brown dwarfs, rendering a significantly improved characterization of low-mass star formation in stellar clusters, which is of fundamental importance to deriving initial mass functions. 

The Gaia mission has released a catalog of reliable members for over 200 known clusters within 2~kpc from the Sun \cite{2018A&A...618A..93C}. These clusters range from 10~Myr to several Gyr in age \cite{2019A&A...623A.108B} and are therefore ideal calibrators for the mass–dependent relationship between stellar rotation and age. The ages of these clusters are derivable from color-magnitude fitting using the Gaia data \cite{2019A&A...623A.108B}, while the rotation periods of the individual members in a cluster are achievable from the WFST time domain surveys. A well-calibrated mass–dependent relationship between stellar rotation and age is a crucial step towards understanding the star formation history of the milky way.

\subsubsection{Structure of the Milky Way}\label{section_523}
The stellar structure of the Milky Way (MW) consists of four components: a bulge, a thin disk, a thick disk, and a diffuse stellar halo. Knowledge of the structure of the Milky Way has been rapidly increasing due to a variety of sky surveys (e.g. SDSS, Pan-STARRS, LAMOST, and Gaia) and the development of technology supporting these surveys. However, there are abundant issues about the detailed structures of the MW and their formation mechanism that remain unresolved.  

As a space-borne facility, the Gaia satellite is capable of mapping the entire sky, although its detection limit of $\sim20$~mag is insufficient for scrutinizing dense stellar fields. An investigation of the detailed MW structures and their formation mechanism necessitates a deep survey that covers a large FoV and detects a larger number of low-mass stars to large distances (cf. Figure~\ref{Fig:distance}; bottom). An accurately determined 3D distribution of MW stars and a decomposition of the MW into a number of components are crucial steps towards constraining the formation mechanism of the different MW components. The decomposition hinges on measurement of the metallicity of individual stars, which requires high sensitivity and high precision photometry in the {\em u, g, and r bands} \cite{ive2008}. Previous and current surveys with sufficient FoVs either lack a $u$-band filter (e.g. Pan-STARRS) or lack sensitivity in the $u$-band (e.g. SDSS or SkyMapper). The upcoming LSST will reach a depth of $r=27$ mag when the images are co-added, though the observations will be limited to be within the southern sky. 

The six-year WFST co-added images will reach limiting magnitudes of $u=24.6$, $g=25.2$, and $r=25.1$, which is $\sim2$ magnitude deeper than the existing SDSS data. WFST is expected to obtain high-precision multicolor measurement of nearly 5 billion MW stars and detect main sequence stars at large distances (Figure~\ref{Fig:distance}). The multi-band photometry (including $u$ band) of the WFST sky survey will enable measurement of the photometric metalicities of stars, a critical tool to distinguish halo stars from disk stars, and to accurately determine photometric distances of stars. Large metal-poor star samples will yield a metallicity distribution function (MDF) of the Galactic halo and thus constrain the chemical evolution models of the MW. With the metalicities and distances of stars in hand, we will probe the MW structure with better precision and over a broader range of distance. In particular, WFST is expected to detect several tens of more debris streams at large distances ($R>50$~ kpc) in dwarf galaxies or globular clusters.

\subsubsection{Astrometry and Variable Stars}\label{section_524}
 
WFST can survey more than 6000 square degrees per night and map the entire northern sky in each band every three nights. The six-year survey will accumulate high-quality imaging data of the northern sky in {\em u, g, r, i, z} bands at over one hundred epochs. These multi-epoch data will facilitate measurement of the proper motions ($\sigma$several mas~yr$^{-1}$) of one billion stars in the northern sky and delivery of the multidimensional information (e.g. proper motion, parallax, position and metalicity) of $\sim$100,000 nearby stars, so that the local gravitational potential field are constructed and the mass distribution and structure model of MW are constrained. WFST's sensitivity and accuracy are sufficient for detecting hypervelocity stars in the galactic halo at distances up to more than 10~kpc. The multi-band photometry will help to determine the metalicities of these hypervelocity stars, a fundamental parameter to discriminate their origin \cite{2021ApJS..252....3L}.
 
WFST time-domain survey will catalog millions of variables. Among them, RR Lyrae Stars and Classical Cepheids are two of the most important types that serve as standard candles to measure distances; eclipsing binary stars (EBs) are of significance to stellar physics; and X-ray binary systems, including high and low mass binaries, are ideal astrophysical laboratories to examine the formation and evolution of stars, compact objects, and mass transfer processes in a binary system \cite{giovannelli2003x}. 

{\bf RR Lyrae stars} are old ($>$10~Gyr), low-metallicity, horizontal-branch pulsating stars varying periodically and have been used as standard candles. Currently, the completeness of RR Lyrae star detection in existing surveys like Gaia and PanSTARRS drops to $\lesssim$ 50\% at a distance $\gtrsim$ 80 kpc \cite{2020MNRAS.496.3291M}. WFST survey 1--2 magnitude deeper than Pan-STARRS1 will significantly increase the sample size of RR Lyrae stars at a large distance in MW and in nearby dwarf spheroidal galaxies (Figure~\ref{Fig:distance}; top). 
RR~Lyrae Stars at large distances are of exceeding importance to probing the Galactic halo and the MW structure near the viral radius of the MW dark matter halo ($\sim$200--300\,kpc \cite{2019ApJ...873..118W}). Regarding the Galactic thick disk, there is ambiguity regarding whether it is a distinct component, whether it is flared or warped, and how it is related to other Galactic components (thin disk, halo, and bulge) in spatial extent, chemistry, and kinematics \cite{Mateu2018}. The deep WFST time-domain survey to search for RR~Lyrae at low Galactic latitudes, where extinction is higher than in the halo, will shed light on these puzzles. 

{\bf Classical Cepheids} are among the key standard candles to determine accurate distances within the local group. In contrast to RR~Lyrae Stars, Classical Cepheids are young stars ($\lesssim$400\,Myr). They are involved in the examination of the thin disk structure of the MW and deemed to trace the detailed morphology of the thin disk to a Galactocentric distance of $\sim$15 kpc \cite{2020ApJS..249...18C}. WFST survey promises to detect Classical Cepheids at distances over 5~Mpc (Figure~\ref{Fig:distance}; top). Construction of a WFST sample of MW Classical Cepheids will allow for depicting the Galactic structure in more detail, while the sample of Classical Cepheids in other galaxies will help tackle the intrinsic variance of Cepheid properties.
 
{\bf Eclipsing binary stars (EBs)} are indispensable for stellar physics. The accurate parameters (e.g. mass, radius, temperature, and luminosity) of the two component stars are achievable through the analysis of EBs. These parameters will impose strict constraints on stellar evolution models, especially at the low-mass end where the model is significantly uncertain. There are many open issues in eclipsing contact binaries (ECBs), such as the merging of binary stars, the evolution of their common envelope, and the short-period limit \cite{2020RAA....20..163Q}. The detection limit of the WFST survey in the $r$-band down to $\approx$23 mag in a 30-second exposure implies the discovery of faint EBs by WFST. According to the well-established period-color relationship of ECBs \cite{1967MmRAS..70..111E,2017RAA....17...87Q}, those ECBs with the shortest period possess the lowest temperature. Finding faint main-sequence ECBs will help unveil the origin of the cut-off in the period of ECBs. In particular, an ECB system with M2V+M2V components is observable within a distance of 4 kpc with a brightness of r$\approx$22 mag. 

{\bf X-ray binaries} consist of a normal star and a compact object, which is either a NS or a BH \cite{2006csxs.book.....L}. As per the mass of the optical companion, they are conventionally classified into high-mass (usually $\geq$ 10 $M_{\odot}$, \cite{2006A&A...455.1165L}) and low-mass X-ray binaries (usually $\leq$ 1,$M_{\odot}$, \cite{2007A&A...469..807L}). The two main subclasses of high-mass X-ray binaries are the supergiant X-ray binaries and the Be/X-ray binaries. To date, only 114 high-mass X-ray binaries in the MW have been cataloged, of which about 60\% are Be/X-ray binaries \cite{2006A&A...455.1165L}. In a Be/X-ray binary, the compact companion is usually an NS \cite{reig2011x}, although Be-BH binary systems also exist \cite{casares2014type}. Most Be/X-ray binaries are hard X-ray transients, usually showing two types of X-ray outbursts: Type I X-ray outbursts, of which the X-ray luminosity $L_X\sim10^{36-37}$~erg~s$^{-1}$ and the duration are the orbital period, and Type II X-ray outbursts, which are significantly brighter ($L_X>10^{37} erg s^{-1}$) and show no evident connection with the orbital period \cite{negueruela1998nature}. Long-term optical observations indicate that significant optical variations precede X-ray outbursts \cite{2012ApJ...744...37Y}, necessitating the monitoring of a sample of Be/X-ray binaries to delineate the relationship between optical variability and X-ray outbursts. Low-mass X-ray binaries are systems where an NS or BH is accreting materials from its low-mass companion donor star via a Roche lobe overflow. About 200 low-mass X-ray binaries have been cataloged in our Galaxy \cite{2007A&A...469..807L}, of which most are X-ray transients with observed outbursts. Population synthesis indicates that there are about $2.1\times10^3$ low-mass X-ray binaries with NS accretors in the MW \cite{2015A&A...579A..33V}, though the majority of them remain unexplored. The time-domain survey of WFST is expected to capture the periodic variability of the light curves and discover a remarkable number of new candidates for low-mass X-ray binaries.

\begin{figure*}[t]
\centering
\includegraphics[width=0.9\linewidth]{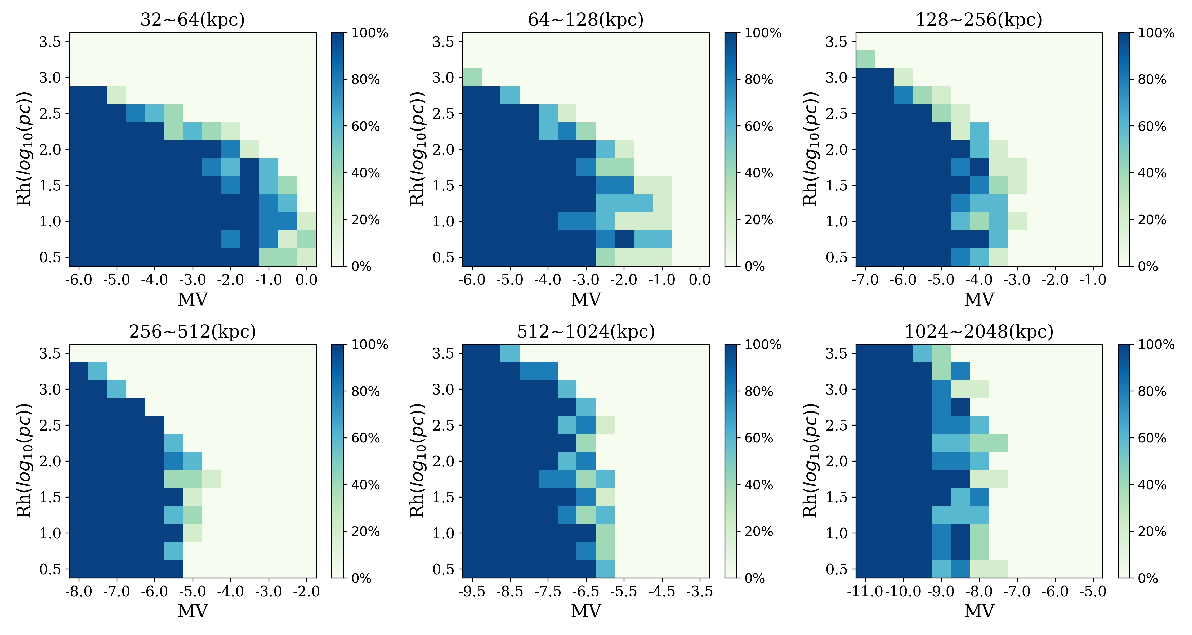}
\caption{Detection efficiency of dwarf-galaxy search. The efficiency is presented as a function of the azimuth average physical half-light radius and the absolute magnitude of the V band in various bins of the heliocentric distance. As shown in the color bar, the detection efficiency ranges from 0\% to 100 \%, which is shown in the color bar..}
\label{w_detect}
\end{figure*}

\subsection{Satellite Dwarf Galaxies in the Local Group}\label{section_53}

The dwarf galaxies surrounding the MW and M31 are the galaxies of the lowest mass observable in the Universe. These objects are of broad interest due to their astrophysical uniqueness. Bright stars in these systems can be resolved even by ground-based telescopes, rendering them ideal targets for probing the star formation history, chemical enrichment, and initial mass function in low-mass halos \cite{Weisz2014,Geha2013}. Furthermore, their abundance and spatial distribution place stringent constraints on structure formation theories on spatial scales smaller than $\sim 1$ Mpc \cite{Bullock2017}. Like their massive peers, dwarf galaxies are dominated by dark matter, which is possibly detectable through the products of its decay. In view of their physical scales and distances, dwarf galaxies in the Local Volume are ideal laboratories for the detection of dark-matter decay signals \cite{Spekkens2013}.

The search for faint dwarf galaxies in the Local Volume has been continued ever since the serendipitous discovery of the first such system in the 1930s. Only 11 MW satellite galaxies were known prior to SDSS. During the past two decades, more than 50 new MW satellites were discovered, thanks to the advent of large imaging surveys (e.g. SDSS). Most known MW satellites have luminosities comparable to those of globular clusters, though their surface brightness is remarkably lower, posing a challenge to direct identification in imaging. In practice, these galaxies are immersed in a statistical fluctuation of number density as significant as that of cataloged stellar objects.

The 6-year WFST co-added images will reach a depth of $r=25.1$ and cover $\sim$ 10000 deg$^2$ of the northern sky, well suited for the search of faint dwarf galaxies in the Local Volume. This depth is $\sim 2$ magnitude deeper than SDSS and is comparable to that of the Dark Energy Survey (DES) \cite{DES2005}. DES covers $\sim 5000 ~\rm deg^2$ of the southern sky and $\sim 20$ Milky Way satellites are found in the DES footprints. Under the simple assumption that the distribution of MW satellites is isotropic, the detection of $\sim$160 MW satellites in the full sky is expected to reach $r\sim 25$. Regarding the SDSS footprints in the northern sky, $\sim 40$ satellites are expected down to the same magnitude limit. Taking into account the classical and newly found satellite galaxies in the SDSS footprints, we expect $\sim 20$ new MW satellites to be discovered in the era of WFST. However, this number is to be treated as an upper limit because MW satellite galaxies are not observed to distribute randomly, but a trend to cluster near the Large Magellanic Cloud exists.

We conducted a simulation to analyze the capability of WFST in detecting dwarf galaxies. The detection efficiency is quantified by simulating model dwarf galaxies immersed in typical star fields, as imaged by WFST. The background MW star fields are constructed with the code {\em Galaxia} \cite{Sharma2011}. A Kroupa initial mass function \cite{Kroupa2001} is adopted for the simulated galaxies. In accordance with observations, we assume an old, metal-poor dominant stellar population that spans a stellar age range of 7--12 Gyr and a metallicity range of $\rm [Fe/H]=[-2.2,-1.5]$. We then search for dwarf galaxies using an algorithm in the literature \cite{Koposov2008}. We demonstrate the detection efficiency as a function of distance and $V-$band absolute magnitude of galaxies in Figure~\ref{w_detect}. Within the viral radius of the MW ($\sim$ 300~kpc), we conclude that galaxies with $M_{V}<-4$ are readily detectable in the stacked images of WFST. Alternatively, the detection limit within 1 Mpc is found to be $M_{V}<-6$.
 
%%%%Galaxy Formation and Cosmology%%%%%%%%%%%%%%%%%%% 
\section{Galaxy Formation and Cosmology}\label{section_6}

Modern optical imaging surveys have significantly deepened our understanding of the universe. Especially in recent years, with the help of high quality imaging of SDSS, CFHTLenS, Dark Energy Survey (DES), HSC-SSP and KiDS, we are in a position to explore the universe with unprecedented accuracy, an era known as that of precision cosmology. However, tension emerges between CMB observations and optical survey measurements, including the $\sigma_8$ tension between weak lensing and CMB, the $H_0$ tension between CMB constraints, and the strong lensing time delay / SNe Ia. It has been under suspicion whether this is due to certain hidden systematic effects or new physics beyond our knowledge. The debate arises even among research groups that pursue the same topics but using data sets from different facilities. Meanwhile, extensive efforts have been made to improve data processing pipelines to understand potential systematic effects. For example, a recent HSC-SSP shape catalog \cite{li2022PASJ} is aided by detailed simulations and systematic tests to ensure that systematic effects are under control.

Apparently, the entire collection of data sets obtained so far remains insufficient for an ultimate understanding of either new physics or systematic effects. In the near future, further progress will be made by space-based (e.g. CSST, Roman Telescope) and ground-based (e.g. LSST) facilities, which are expected to scan half of the entire sky and accomplish deep imaging down to 28 mag in the $r$-band. Once completed, the WFST multiband imaging survey will be the largest survey on the northern hemisphere overlapping with multiple spectroscopic surveys (e.g., PFS, DESI, LAMOST2, and MUST). The combination of WFST and LSST (on the southern hemisphere) will yield all-sky data, and the integration of multi-band imaging and spectroscopy promises to boost the advancement of precision cosmology. 

\subsection{Galaxy Formation}\label{section_62}

WFST shear catalog will be a key product for galaxy formation studies, delivering information about the position, shape, photometric redshift of galaxies and calibrated biases as a function of resolution and signal-to-noise ratio. Combined with preexisting and upcoming spectroscopic data available for the northern sky (e.g. SDSS, MUST and LAMOST2), it will significantly improve the accuracy of weak gravitational lensing measurements, placing more stringent constraints on theories of galaxy formation and cosmology. We elaborate science related to weak lensing analysis in three aspects, i.e. galaxy-halo connection, halo assembly effects and cluster detection. Two more important topics will also benefit from the WFST imaging surveys: $u$-band imaging will potentially facilitate the construction of large samples of $u$-band dropout galaxies and low surface brightness galaxies.

\subsubsection{Galaxy-halo Connection}\label{section_621}

Galaxies form and evolve inside dark matter halos and are affected by the large-scale environment. Exploring the connection of galaxies with their host dark matter halos and with their large-scale environment is therefore a crucial step towards the achievement of the blueprint of galaxy formation. Numerous works exist in the literature on the galaxy-halo connection based on a variety of observational measurements such as galaxy clustering and galaxy-galaxy lensing \cite{luo_gg1, luo_gg2} or the combination of the two \cite{zhang2021A&A}.

However, many questions remain unresolved. For instance, does the host halo mass depend on the galaxy properties? If this is the case, what properties of the galaxy are dominant? How are these relationships built? How do different environmental processes, an interplay of various environmental factors, affect the galaxy properties? A recent work \cite{zhang2022A&A} used a massive star-forming galaxy sample to find that about 67\% of gas has been converted to stars, which is abnormally high compared to the typical conversion fraction of 20-30\%. It remains unclear what mechanism causes this specific mass bin that bears such a high gas consumption rate.

The host halos of AGNs that are characterized by their strong central SMBH activity is another topic of interest. AGNs are different from other galaxies in the spectral energy distribution (SED) and spectral line features. Whether different types of AGNs reside in different large-scale environments remains an open question. Zhang et al. \cite{zhang2021A&A} conclude that the halo masses of AGNs are similar to those of star-forming galaxies, but are lower than the quenched control sample. However, AGNs appear to be surrounded by a larger number of satellites than star-forming galaxies, indicating an association of the AGN trigger mechanism with satellite galaxies.

\begin{figure}[H]
    \centering
    \includegraphics[scale=0.5]{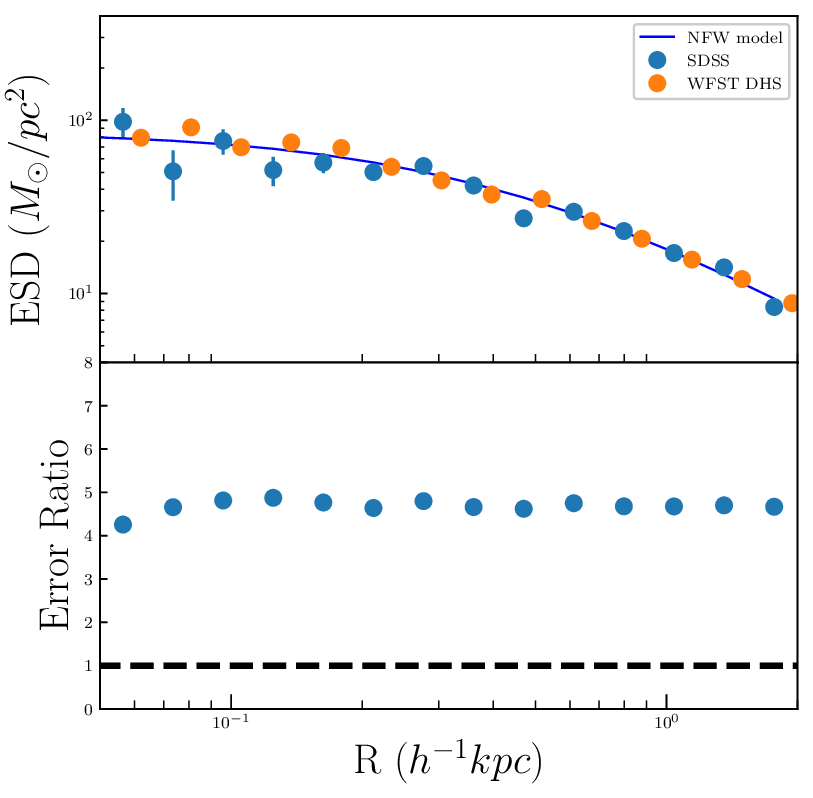}
    \caption{The comparison of the galaxy-galaxy lensing signals between SDSS shape catalog and WFST shape catalog. The lower panel is the ratio between the two errors. The error shrinks by a factor of 2.5 given the same lens sample but a different catalog.}
    \label{fig:wfst_sims_3}
\end{figure}

WFST shape catalog will manifest itself by 7 (WFS) or 20 (DHS) times deeper imaging and multiband photometry than SDSS, and the signal-to-noise ratio of Weak lensing analysis around galaxies will increase by a factor of 3, leading to an accuracy in halo mass estimation enhanced by a factor of $\sim 2$. Fig.~\ref{fig:wfst_sims_3} depicts the improved uncertainty in WFST measurement (red dots) compared to SDSS galaxy-galaxy lensing measurement (green dots). Within the halo virial radius, the WFST shape catalog alone is predicted to shrink the errorbar by a factor of 3/5 in the WFS/DHS field compared to SDSS due to number density. 

\subsubsection{Halo assembly effects}\label{section_622}
The clustering of dark-matter halos strongly depends on the halo mass, though numerical simulations have revealed dependence on other halo properties, including the formation history, the internal substructure and the spin of a halo, a higher-order effect referred to as the halo assembly bias.

Many observational efforts have been made to detect the assembly bias. For example,iyatake et al. \cite{miyatake2016PhRvL} claimed the detection of assembly bias based on RedMaPPer clusters by performing weak gravitational lensing and projected clustering analysis. However, this result is recognized by Zu et al. \cite{zu2017MNRAS} as an artifact due to projection effects of the RedMaPPer cluster members, and therefore the secondary bias is even higher than the Lambda cold dark matter (LCDM) prediction. They further predict that a 10-fold larger number of clusters with deep imaging will concretely improve detection. WFST will increase this number to about 40,200, or 4.6 times larger than the sample size of Miyatake et al. (2016), spanning the same richness range but a broader redshift range ($0.1<z<0.8$). Combining the WFST cluster sample with that of LSST is expected to suppress the uncertainty to less than 10\%, at a level comparable to the predicted LCDM assembly bias. A recent work divided 630 massive clusters into early and late formed clusters using the ELUCID simulation \cite{wang2014ApJ}, and concluded that a 4-$\sigma$ difference in clustering is detected, suggesting a real detection of assembly bias.

Clusters are not the only probe to the assembly bias. Lin et al. \cite{Lin2016ApJ} constructed low-mass samples divided into early- and late-formed galaxies to pursue this effect, who attributed null detection to their noisy measurements. McCarhty et al. \cite{McCarthy2022MNRAS} extended the previous work by analyzing a larger number of galaxies and changing the clustering estimator to the redshift space distortion (RSD), which is essentially the Legendre expansion of clustering, but velocity information is taken into account. A large amplitude of velocity bias for early-forming central galaxies is found in this work, which may originate from assembly bias, though the measurement is again overly noisy to validate their statement. Significant detection awaits the accumulation of more lens samples and deeper imaging data sets. 

In addition, the accurate determination of the halo boundaries is an important issue in galaxy formation, which are commonly defined as the radius that encloses a certain value of density contrast $\Delta$ compared to the mean/critical density of the universe. On a more physical basis, More et al. \cite{more2015ApJ...810...36M} introduced the concept of $splashback$ radius, where the accreted matter reaches its first orbital apocenter. They claim that such a radius depends on the accretion rate of the halo, whose typical value ranges from 0.8 to 1.5 virial radii. A recent work \cite{Wang2022arXiv220612163W} showed that the $splashback$ radius also depends on the major/minor axis of the local tidal tensor. We expect that the first attempt to observationally determine the $splashback$ radius will hinge on measurement of the surface number density of galaxies with spectroscopic information from existing overlapping catalog, e.g. BOSS, eBOSS, and DESI, to be followed by detecting a galaxy over density around SZ clusters as well as weak gravitational lensing. 

WFST data from the WFS and DHS fields will yield high-quality imaging and photometry to help address all of the above research topics. In particular, the shape catalog will allow us to measure the halo mass of spectroscopically selected galaxy groups, to compare the result with that of other halo mass estimation methods, and to remarkably improve the measurement of $splashback$ radius as well as the assembly bias.

\subsubsection{{\em U}-band Drop-out Galaxies at z about 2--3}\label{section_624}
Lyman Break Galaxies (LBGs) are star-burst galaxies at high redshift (for a review, see \cite{Giavalisco2002ARA&A}) that can be identified using the technique of combining $u$, $g$ and $r$ bands, as demonstrated in a recent work of joint analysis based on the CLAUDS and HSC-SSP deep imaging \cite{sawicki2019MNRAS, Thibaud2020MNRAS}. Plentiful works in the literature address the UV luminosity function (UV-LF) of LBGs that can be used to estimate the energy budget at high redshift. Since the pioneering work of Steidel et al. \cite{Steidel1995AJ}, continuous efforts have been made to analyze the UV-LF of LBGs (e.g. \cite{Reddy2009ApJ}, \cite{Cai2014ApJ}).

The selection criteria are $u-g>0.88$, $g-r<1.2$, and $u-g>1.88(g-r)+0.68$ (Equation 4 in \cite{sawicki2019MNRAS}). WFST will provide all the broad-band imaging involved with deep and wide survey regions, as mentioned in Sec.~\ref{section_2}. We estimate the number of LBGs from both the DHS fields ($\sim$1000 square degrees, down to 26.4 mag or an absolute mag of $-20.6$ at $z=3$ in $g$ and 25.9 mag in $r$) and the WFS fields (6800 square degrees, down to 25.1 mag or an absolute mag of $-21.9$ at $z=3$ in $g$, and 24.7 mag in $r$). 

We adopt the luminosity function at $z=3$ in \cite{Cai2014ApJ}, for which $M_{1350}$ monochromatic flux is used as an indicator. In the DHS region alone, we expect to detect $10^7$ galaxies (with $g$-band absolute magnitude down to -20.6, considering the completeness of 90\% for $r<25.5$). Though at an amount two orders of magnitude lower than that of the LSST survey, this catalog of galaxies will serve as a guidance for future follow-up spectroscopic surveys. LBGs at higher redshift ($z>4$) can also be selected as per {\em g, r, i} color criteria as already done in \cite{Steidel1999ApJ}, yielding another valuable legacy catalog.

\subsubsection{Low Surface Brightness Science}\label{section_625}
The low-surface-brightness (LSB) regime holds the promise to revolutionize our understanding of galaxy formation and evolution in the upcoming decade. In particular, demographics of satellites around galaxies of different morphological types and masses in the local universe will offer crucial tests of the LCDM paradigm on small scales; systematic characterization of stellar halos and tidal features in the outskirts of galaxies can provide important clues to the hierarchical assembly histories of galaxies. 

The primary focus architecture of WFST minimizes the contamination from scattered light, which is particularly desirable for LSB science. The WFST six-year co-added imaging data ($\sim$ 50 min) will reach a $r$-band 3--$\sigma$ surface brightness limit of $\sim$ 28.7 mag arcsec$^{-2}$ by averaging a 10$\times$10 arcsec$^{2}$ area, slightly deeper than the 275 deg$^{2}$ Stripe 82 field of SDSS. By scaling the results from extensive completeness simulations, we expect to achieve a detection of ordinary satellite dwarf galaxies down to an average surface brightness of $\sim$ 25.7 mag arcsec$^{-2}$ within the effective radius at a 50\% completeness limit. This corresponds to a stellar mass limit of $\sim$ 10$^{6.1\pm0.5}$ M$_{\odot}$ up to $\sim$ 60 Mpc \cite{danieli2018}. In addition, a surface brightness limit of 28.7 mag arcsec$^{-2}$ allows for detection of tidal features from galaxy merger events that happened at least $\sim$ 3-4 Gyr ago. Finally, the wide-field and homogeneous data sets from WFST will enable a robust stacking analysis of surface brightness profiles well beyond 30 mag arcsec$^{-2}$ for galaxies of different morphological types, masses, and environments, providing stringent constraints on the build-up of galaxy stellar halos in general.

Besides the combination of its sky area coverage and survey depth, an important advantage of WFST over existing optical imaging surveys, such as DES and HSC-SSP (Hyper Suprime-Cam Subaru Strategic Program), lies in its inclusion of deep $u$-band data that are indispensable in probing stellar population properties of galaxies with broad-band photometry.

\subsection{Cosmology}\label{section_63}
As tensions in cosmological parameter measurements recently emerge (e.g. $H_0$ and $S_8=\sigma_8(\Omega_m/0.3)^{0.5}$) between the CMB probe and SNe Ia, time-delay, weak lensing analysis, debates arise on whether certain hidden systematic effects are at work or new physics is in anticipation. 

In the northern sky, WFST is expected to make valuable contributions to cosmological research by detecting a large amount of SNe Ia (cf. Sect.~\ref{section_315}) and strong lensing AGN/SNe Ia (cf. Sects.~\ref{section_315} and \ref{section_34}) systems, and yielding a cluster catalog and a shape catalog. In this section, we focus on the standard cosmology constraints and cosmologies that deviates from LCDM. 

Within the framework of standard cosmology, we proceed with the analysis of cluster mass function, cosmic shear, and their combination with other measurements (e.g. clustering, cluster mass functions, time-delay, etc.). For non-standard cosmology, we focus on the constraints on modified gravity models, dark-matter particle models, among other topics.

\subsubsection{Cluster detection and cosmology}\label{section_631}
Clusters of galaxies act as a probe of cosmology and galaxy formation. The well-known Bullet cluster alone is a smoking-gun evidence of dark matter, where the spatial distribution of X-ray-emitting hot gas significantly deviates from that of dark matter inferred from weak lensing. 

Furthermore, the shock feature seen in the X-rays poses a challenge to standard cosmology in the sense that the exceedingly high collision speed of the two merging giants is difficult to realize in numerical simulations. Abell 520, another cluster intensively studied using weak-lensing techniques, poses another challenge to the classic galaxy formation paradigm in the sense that significantly fewer galaxies than theoretical prediction are found in the ``dark core'' of the cluster. Another puzzle results from the recent finding of a massive cluster (the Coma cluster) rotating at a velocity of 197 km s$^{-1}$. In summary, an appreciable number of mysteries about clusters await exploration in the future WFST survey region, e.g. XCS clusters and eFEDS clusters.

Construction of reliable cluster catalogs is a nontrivial task, especially when spectroscopic information about galaxies is absent. However, endeavors to catalog clusters using photometric information have been made (e.g. RedMaPPer and CAMIRA involving red-ridge galaxies). Yang et al. \cite{yang2021ApJ} present a novel halo-based cluster selection method (a modified approach based on \cite{yang2007}), where the adjustment is applied in the pipeline that delivers photometric data. Nevertheless, these methods suffer from the projection effect due to the accuracy of the photometric redshift, and therefore the membership estimation is biased, as described in \cite{zu2017MNRAS} and \cite{sunayama2020MNRAS}. Along with extended X-ray sources, the Sunyaev-Zeldovich (SZ) effect is another tool of cluster studies. More recently, shear maps of galaxies have been used to catalog the HSC-SSP shear map clusters. 

\begin{figure}[H]
    \centering
    \includegraphics[scale=0.4]{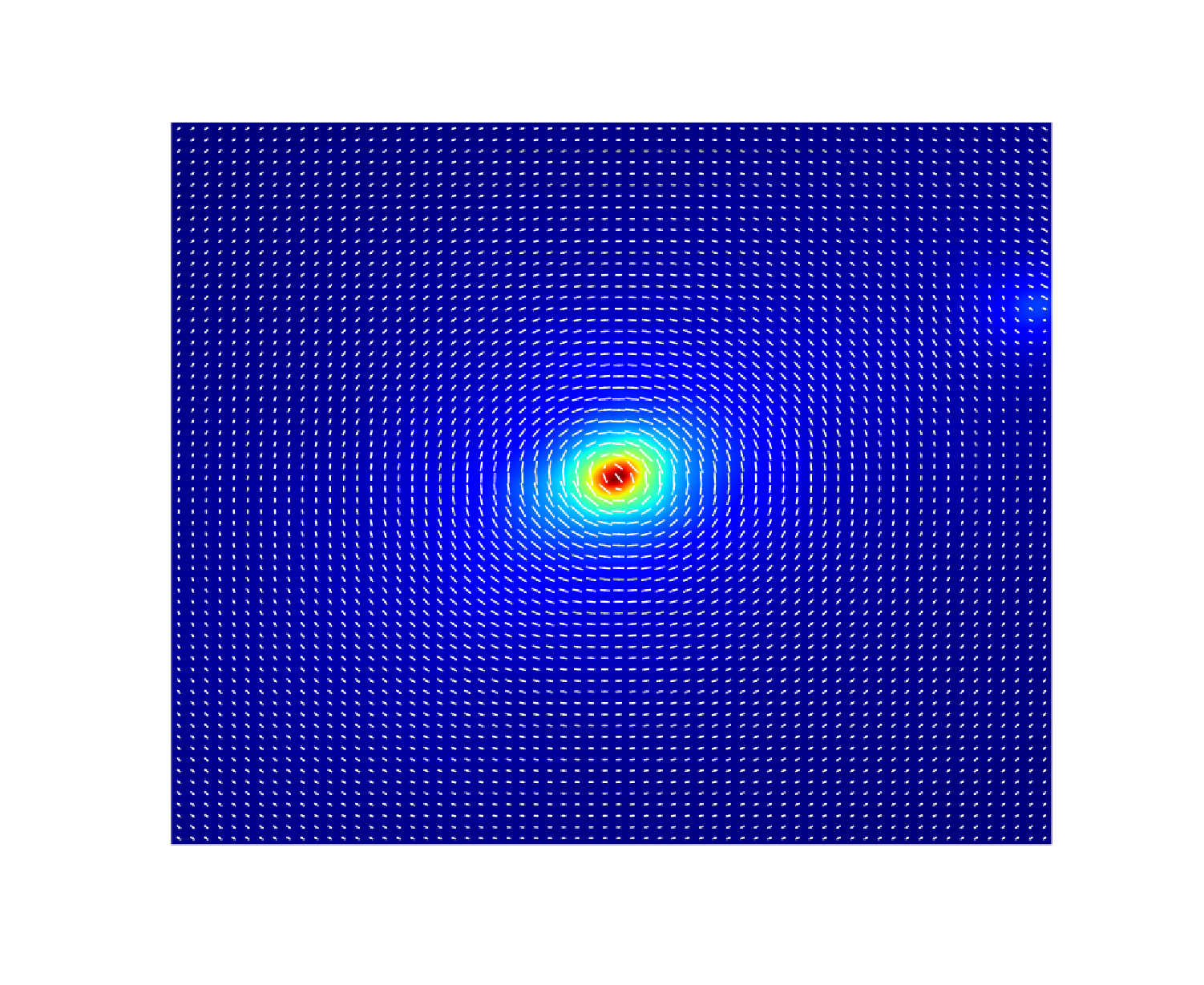}
    \caption{The shear map and kappa map of a cluster-sized dark matter halo chosen from ILLUSTRIS simulation with halo mass of $10^{14}\mathrm{h^{-1}}M_{\odot}$.}
    \label{fig:wfst_sims_4}
\end{figure}

The combination of cluster catalogs with a variety of selection methods and galaxy-galaxy lensing can yield tight constraints on observable halo mass-scaling relations. Fig.~\ref{fig:wfst_sims_4} shows the $\kappa$ (colored map) and the shear map (white ticks) of a cluster selected from the ILLUSTRIS TNG simulation.

As mentioned in the previous section on assembly bias detection, WFST will generate a photometrically selected cluster sample based on the RedMaPPer algorithm. The sample will contain more than 40,000 clusters with richness greater than 20 between $0.1<z<0.8$. The scientific goals include performing cluster mass estimation to constrain cosmology and cross-matching clusters with other observations (e.g. SZ clusters, X-ray clusters, and weak lensing mass maps). The cluster catalog together with the weak lensing shape catalog will be of appreciable value for cosmological explorations and will serve as reference data for next-generation surveys.

Once cluster catalogs from various observations are obtained, the cluster abundance and its evolution will readily constrain the fluctuation amplitude $\sigma_8$ and the parameter $\Omega_m$. The baryon fraction of the clusters can be used to estimate the ratio of the cosmic baryonic fraction $\frac{\Omega_b}{\Omega_m}$, while the core structure of the clusters is a test bed of the nature of dark matter. 

Despite the virtues of these cluster statistics, each of them has certain limitations. For instance, the systematics in converting cluster observables (X-ray luminosity, S-Z Compton parameter, richness and weak lensing) to mass may bias the scaling relation used for mass estimation. 
The construction of a reliable cluster catalog is the first step in the expedition of cluster cosmology.

\subsubsection{$3 \times 2$-point correlation functions}\label{section_632}
The digit 3 in the title of this section denotes three types of 2-point correlations employed in the statistical analysis, i.e. galaxy clustering, cosmic shear and galaxy-galaxy lensing measurements.
Cosmic shear alone is sensitive to dark matter density perturbation $\sigma_8$ and dark matter fraction $\Omega_m$. However, intrinsic alignments can bias the results. The alignment of galaxies itself is a topic of interest that addresses the misalignment between galaxies and their dark-matter halos, assuming a triaxial halo shape.

Recent weak lensing surveys (e.g. KiDS, HSC-SSP, DESc) and joint analyses combining all three surveys display a tension of 2$\sigma$ or so with CMB experiments. The WFST surveys will suppress the error by a factor of 1.3, assuming a depth similar to that of DES and an effective weak lensing area of the WFS in the northern sky.

Apart from halo masses of galaxies, cosmological constraints can also be obtained by combining clustering analysis. Leauthaud et al.  \cite{leauthaud2017} found that $\sigma_8$ predicted by weak lensing is lower than the value that fits galaxy correlation well, a discrepancy known as the "lensing is low" problem. After that, the combination of galaxy-galaxy lensing and clustering analysis becomes a standard routine to maximize the utility of different estimators, e.g. an up-to-date work using HSC-SSP data that combines galaxy-galaxy lensing and clustering. 
Also notably, Shi et al. \cite{shi2018ApJ} combine the redshift space distortion (RSD) from the SDSS DR7 spectroscopy data and galaxy-galaxy lensing and provide a tight constraint on the growth factor at $z=0.1$.

WFST WFS and DHS fields overlap with BOSS/HSC-SSP that contains spectroscopic sample with public availability, facilitating the clustering and galaxy-galaxy lensing joint analysis with LowZ and 2MASS samples. This combination will lead to an enhancement of sigal-to-noise ratio by a factor of 3.3 (WFS)/3.0 (DHS), compared to the CFHTLenS analysis and the HSC S16A shape catalog. 

\subsubsection{Joint Analysis with Other Observations}
\label{section_633}
\begin{figure}[H]
    \centering
    \includegraphics[scale=0.5]{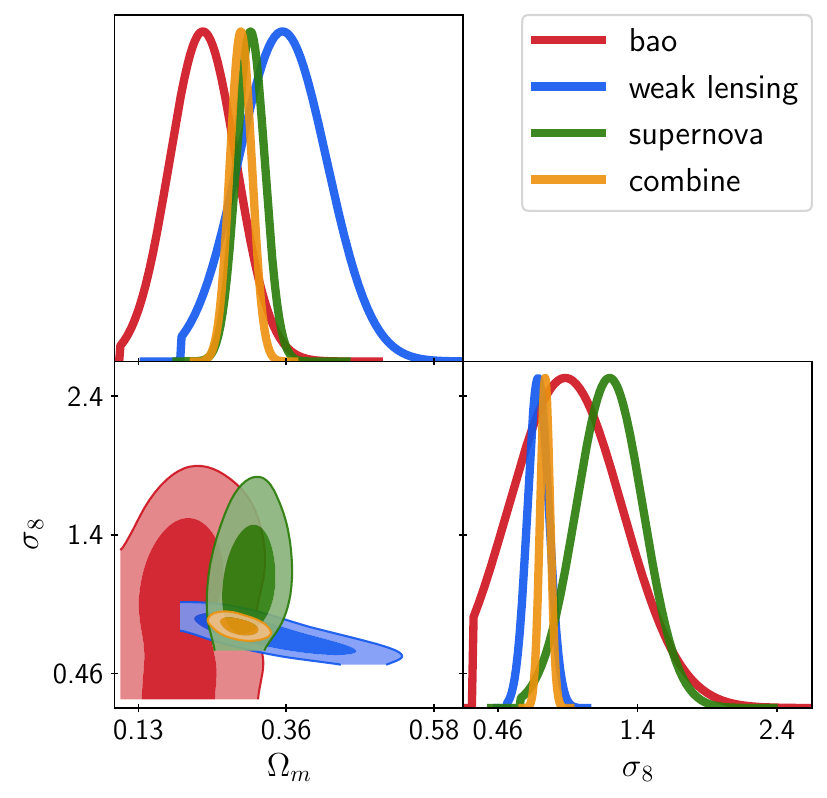}
    \caption{The joint analysis of cosmological constraints.}
    \label{fig:wfst_sims_5}
\end{figure}

Besides weak gravitational lensing, multiple cosmological probes have been employed, such as CMB radiation, baryon acoustic oscillations (BAO), and SNe Ia. Joint analysis and comparison between different probes are powerful tools, Figure~\ref{fig:wfst_sims_5} shows the joint constraints with the WFST cosmic shear measurements. In particular, comparing probes bearing different degeneracy directions for the same set of parameters may cause joint likelihood analysis to break the degeneracy between the parameters. For instance, Di Valentino et al. \cite{DiValentino2015PhRvD} combined BAO, CMB, Weak Lensing and SNe Ia analysis and extended the constraint of 6 LCDM parameters to 12 parameters by taking into account the sum of neutrino mass, the sum of neutrino species effective number, the dark energy equation of state, etc.

The commission of WFST will improve the joint analysis of SNe Ia, weak Lensing, clusters, time-dely and CMB in the northern sky, whereas the spectroscopic catalog from eBOSS and DESI survey will serve as a natural ally of WFST imaging data in relevant cosmological investigations.

\subsubsection{Non-standard Cosmology}
\label{section_634}

The tensions between the early and late probes lead to the possibility of new physics, such as the ``two dark clouds'' in the early 20th century. Maybe certain interactions between dark matter and dark energy are the cause of the low $\sigma_8$ value from the weak lensing constraint, or can we find a substitute for dark matter or dark energy? What is dark matter exactly, and can it be tested by cosmological probes like cosmic shears? Does the theory of general relativity still hold at galactic scales? 

Weak gravitational lensing can provide strong constraints on the dark side of our universe and therefore on non-standard cosmology. For instance, a recent work by Luo et al.\cite{luo2021ApJ} found that Emergent Gravity can hardly explain the difference in weak lensing signals between the blue/red galaxy sample with similar stellar mass. Zhang et al. \cite{jiajun2019ApJ} ruled out one of the interacting dark matter/energy models using weak gravitational lensing analysis based on the SDSS DR7 shear catalog. 

WFST WFS and DHS fields are estimated to contain 7 and 22 times larger density of source galaxies in the survey area similar to SDSS, resulting in a wide and deep shape catalog to improve the current constraint by a factor of 2.12 and 1.88, respectively.

\section{Summary}\label{section_7}

WFST located near the summit of the Saishiteng Mountain is a dedicated imaging facility under construction. The wide field survey (WFS) and the deep high-cadence survey (DHS) programs have been scheduled, covering a sky area of 8000 and 1000 square degrees, respectively. The unique design of the WFST survey strategy will render a $u$-band imaging depth of 26.0 mag in the deep survey, comparable to the 10-year $u$-band depth of LSST. The high cadence enables the search for multiple time domain sources such as SNe, TDEs, optical counterparts of gravitational-wave events, AGN variability, and the near-earth terrestrial objects. Stacked WFS and DHS data also facilitate cosmological investigations, such as weak/strong gravitational lensing, galaxy formation, and constraints on cosmology.

The DHS mode of WFST will produce a catalog of tens of thousands of supernovae, of which a few hundreds are expected to have early-phase observations that promise to deliver information about the progenitor systems. High-cadence $u$-band data will boost the detection of UV-luminous objects, a.k.a. FBOTs/FBUTs, as well as extreme SNe that is 100 times more luminous than SNe Ia and CCSNe. As low as the event rate of superluminous SNe, we anticipate that WFST will detect a sample of them at $z\leq 1.0$ with appreciable completeness. Furthermore, strongly lensed supernovae at high redshift can be used to constrain Hubble parameter by measuring time delays in multiple-image systems. Aware of the four preexisting ones, we estimate that about 20 such systems will be discovered in the process of the 6-year WFST survey. The combination of pre-existing and newly found strongly-lensed AGN from WFST legacy is expected to suppress the uncertainty of $H_0$ to $<1.0$\%.

A kilonova, a transient phenomenon triggered by the merging process of a NS-NS or BH-NS binary, is known as the electromagnetic counterpart of gravitational waves. This intriguing event was first confirmed by follow-up observations of GW170817/GRB170817A in 2017. Kilonovae can be used to explain the production of heavy elements through the $r$-process and to constrain $H_0$ with electromagnetic observations rendering redshift information. A kilonova is also predicted to be coupled with short GRB, despite no agreement regarding the formation of the beamed gamma-ray jets hitherto. The early optical afterglow of GRBs will remarkably help to tackle the triggering mechanism of GRBs. Meanwhile, FRBs have become a focus of time-domain astronomy since CHIME and other experiments discovered hundreds of repeating and non-repeating FRBs. WFST will provide optical information for these mysterious transients and deepen the understanding of the physics behind them. In addition, large number of high energy transients discovered by WFST will be a crucial resource for searching for the electromagnetic counterparts of high energy neutrino events; together with the carefully designed follow-up program we may unravel the origin of these mysterious particles.  

On galactic scales, another fruitful field in time-domain astronomy is attributed to TDEs, deemed a direct probe of the association of the central SMBH with AGN activity, though their rarity poses a challenge. The WFST surveys, by virtue of the large FoV and high cadence, promise to detect TDEs at a rate of hundreds per year with the redshift range to be extended to about 1.0. TDEs involving IMBHs is one of the numerous models that explain FBUTs that provide a promising way to fill or understand the gap between stellar BHs and SMBHs and to constrain the theory of seed black holes or the baby SMBHs. A grown-up SMBH residing in the galaxy center actively accretes surrounding materials to power the central engine of an AGN, another field of astrophysical importance, of which the diversity and variability have arisen of broad interest. In particular, a subclass of AGNs that exhibit extreme variability, of which more than 20\% have been confirmed as CL AGNs, remain of enigmatic physical origin. Like strongly-lensed SNe, strongly-lensed AGNs are of fundamental use to cosmological tests. WFST will discover a significant amount of the above-mentioned objects that will help exploit extremely variable AGNs as well as cosmology. 

We also assess the capability of WFST in detecting small objects in the solar system, concluding that WFST will improve both the positioning and the characterization of faint NEOs, cometary activity, and TNOs (KBOs). The dynamical anomalies of the distant TNOs hint at the existence of Planet 9. The number of known TNOs as yet is only 14, including 5 chaotic ones, necessitating a sample with higher statistical significance to facilitate a further test of the Planet 9 hypothesis. 

The stacked imaging data of WFST are a valuable legacy for exploiting the Milky Way, galaxy formation, and cosmology, of which the feasibility has been demonstrated by previous surveys. The WFST $u$-band covers the Balmer emission that serves as an indicator of the mass accretion rate, and other broad bands are required to measure the extinction. The 2-3 magnitude deeper photometry (in $r$ band) of WFST than Pan-STARRS1 will improve the 3D dust mapping of the Galaxy and help pin down the number of dwarf galaxies in the vicinity of the Milky Way, rendering a direct test of the long-standing ``missing satellite'' problem. 

Progress in the exploration of galaxy formation and cosmology is based not only on data quality but also on the amount of data. WFST will yield about 3PB imaging data in total six years, as a result of scanning 8000 square degrees in the WFS at a depth similar to that of DES, and 1000 square degrees in the DHS at a depth similar to HSC SSP. As mentioned in Sect.~\ref{section_6}, the release of the final 6-year WFST survey data promise to place remarkably improved constraints on galaxy-halo connection, cluster characterization, and cosmology. The legacy shape catalog to be combined with other survey data (e.g., KiDS, DES, HSC SSP and upcoming survey projects) or even further with BAO, SNe Ia, time-delay, and CMB measurements will facilitate a joint analysis anticipated to tighten a variety of cosmological constraints.

In summary, WFST will be located in a site with good observing conditions on the northern hemisphere. Once commissioned, this dedicated survey facility will yield massive data products that, in combination with future spectroscopic surveys of the northern sky (e.g. LAMOST II and MUST), promise to make a major step forward in time-domain astronomy that will further benefit the entire astronomical community.

%%%%%%%%%%%%%%%%%%%%%%%%%%%%%%%%%%%%%%%%%%%%%%%%%%%%%%%
%%% Acknowledgements. ??§Ý
%%%%%%%%%%%%%%%%%%%%%%%%%%%%%%%%%%%%%%%%%%%%%%%%%%%%%%%
\Acknowledgements{This work was supported by the Cyrus Chun Ying Tang Foundations, the Major Science and Technology Project of Qinghai Province (2019-ZJ-A10), the 111 Project for "Observational and Theoretical Research on Dark Matter and Dark Energy" (B23042), the National Natural Science Foundation of China (Grant Nos. 11833007, 12073078, 12173088, 12192221, 12192224, 12233008, 12273036, 12273113), and the Frontier Scientific Research Program of Deep Space Exploration Laboratory (2022-QYKYJH-HXYF-012).
ZYC and HNH acknowledge the support from the USTC Research Funds of the Double First-Class Initiative with No. YD2030002009 and Project for Young Scientists in Basic Research of the Chinese Academy of Sciences (No. YSBR-061), respectively.}

%%%%%%%%%%%%%%%%%%%%%%%%%%%%%%%%%%%%%%%%%%%%%%%%%%%%%%%
%%% Conflict of interest. ????????????
%%%%%%%%%%%%%%%%%%%%%%%%%%%%%%%%%%%%%%%%%%%%%%%%%%%%%%%
\InterestConflict{The authors declare that they have no conflict of interest.}

%%%%%%%%%%%%%%%%%%%%%%%%%%%%%%%%%%%%%%%%%%%%%%%%%%%%%%%
%%% Supplements. ????????, ????
%%%%%%%%%%%%%%%%%%%%%%%%%%%%%%%%%%%%%%%%%%%%%%%%%%%%%%%
%\Supplements{}

%%%%%%%%%%%%%%%%%%%%%%%%%%%%%%%%%%%%%%%%%%%%%%%%%%%%%%%
%%% Reference section. ?ŠÏ?????
%%% citation in the content using "some words~\cite{1,2}".
%%% ~ is needed to make the reference number is on the same line with the word before it.
%%%%%%%%%%%%%%%%%%%%%%%%%%%%%%%%%%%%%%%%%%%%%%%%%%%%%%%
%\bibliographystyle{ieeetr}
% \bibliographystyle{scpma-chenpf} %Citation order
\bibliographystyle{scpma-zycai}

%Citation order, maximum 50 authors
% \bibliographystyle{aasjournal} %Alphabetical by all authors
\bibliography{ref,ref_sn,ref_agn} %ref.bib

\end{multicols}
\end{document}